\documentclass[journal]{IEEEtran}
\IEEEoverridecommandlockouts
\usepackage{lineno}
\usepackage{cite}
\usepackage{hyperref}
\usepackage{amsmath,amssymb,amsfonts}
\usepackage{amsmath}
\usepackage{amsthm}
\DeclareMathOperator*{\argmax}{argmax}
\DeclareMathOperator*{\argmin}{argmin}
\usepackage{graphicx}
\usepackage{textcomp}
\usepackage{xcolor}
\usepackage{graphicx}
\usepackage{float}
\usepackage{subfigure}
\usepackage{amsmath}
\usepackage{amsfonts,amssymb}
\usepackage{mathrsfs}
\usepackage{mathtools}
\usepackage{algorithm}
\usepackage{algorithmicx}
\usepackage{algpseudocode}
\usepackage{bm}
\usepackage{multirow}
\usepackage{array}
\usepackage{amssymb}
\usepackage{amsmath}
\usepackage{cite}
\usepackage{url}
\usepackage{xcolor}
\usepackage{cite,graphicx,amsmath,amssymb}
\usepackage{subfigure}
\usepackage{fancyhdr}
\usepackage{mdwmath}
\usepackage{mdwtab}
\usepackage{caption}
\usepackage{amsthm}
\usepackage{setspace}
\usepackage{bm}
\usepackage{algorithm}
\usepackage{algpseudocode}
\usepackage{mathtools}
\usepackage{dsfont}
\usepackage{bbm}
\usepackage{cases}
\usepackage{stfloats}
\usepackage{bbding}
\usepackage{circledsteps}
\newtheorem{remark}{Remark}
\theoremstyle{definition}
\newtheorem{theorem}{Theorem}

\newtheorem{lemma}{Lemma}

\newtheorem{corollary}{Corollary}

\makeatletter
\newcommand{\biggg}{\bBigg@{3}}
\newcommand{\Biggg}{\bBigg@{3.5}}
\makeatother
\usepackage{pifont}
\theoremstyle{definition}
\theoremstyle{definition}
\usepackage{xcolor}
\newcommand\ytl[2]{
\parbox[b]{6em}{\hfill{\color[RGB]{76,102,67}\bfseries #1}~$\cdots\cdots$~}\makebox[0pt][c]{$\bullet$}\vrule\quad \parbox[c]{14cm}{\vspace{4pt}\color[RGB]{212,120,40}\raggedright #2.\\[7pt]}\\[-3pt]}
\makeatletter
\begin{document}
\title{The Road to Next-Generation Multiple Access:\\A 50-Year Tutorial Review}
\author{Yuanwei Liu, \IEEEmembership{Fellow, IEEE,} Chongjun Ouyang, \IEEEmembership{Member, IEEE,}\\ Zhiguo Ding, \IEEEmembership{Fellow, IEEE,} and Robert Schober, \IEEEmembership{Fellow, IEEE}\\
\vspace{0.2cm}
\emph{(Invited Paper)}

\thanks{Yuanwei Liu is with the Department of Electrical and Electronic Engineering, The University of Hong Kong, Hong Kong (email: yuanwei@hku.hk).}
\thanks{Chongjun Ouyang is with the School of Electrical and Electronic Engineering, College of Engineering and Architecture, University College Dublin, Dublin, D04 V1W8 Ireland, and also with the School of Electronic Engineering and Computer Science, Queen Mary University of London, E1 4NS London, U.K.}
\thanks{Zhiguo Ding is with the Department of Computer and Information Engineering, Khalifa University, Abu Dhabi, United Arab Emirates, and Department of Electronic and Electrical Engineering, The University of Manchester, M1 9BB Manchester, U.K. (e-mail: zhiguo.ding@ku.ac.ae).}
\thanks{Robert Schober is with the Institute for Digital Communications, Friedrich-Alexander-Universität Erlangen-Nürnberg, 91054 Erlangen, Germany (e-mail: robert.schober@fau.de).}}
\maketitle
\begin{abstract}
The evolution of wireless communications has been significantly influenced by remarkable advancements in multiple access (MA) technologies over the past five decades, shaping the landscape of modern connectivity. Within this context, a comprehensive tutorial review is presented, focusing on representative MA techniques developed over the past 50 years. The following areas are explored: \romannumeral1) The foundational principles and information-theoretic capacity limits of power-domain non-orthogonal multiple access (NOMA) are characterized, along with its extension to multiple-input multiple-output (MIMO)-NOMA. \romannumeral2) Several MA transmission schemes exploiting the spatial domain are investigated, encompassing both conventional space-division multiple access (SDMA)/MIMO-NOMA systems and near-field MA systems utilizing spherical-wave propagation models. \romannumeral3) The application of NOMA to integrated sensing and communications (ISAC) systems is studied. This includes an introduction to typical NOMA-based downlink/uplink ISAC frameworks, followed by an evaluation of their performance limits using a mutual information (MI)-based analytical framework. \romannumeral4) Major issues and research opportunities associated with the integration of MA with other emerging technologies are identified to facilitate MA in next-generation networks, i.e., next-generation multiple access (NGMA). Throughout the paper, promising directions are highlighted to inspire future research endeavors in the realm of MA and NGMA.
\end{abstract}
\section{Introduction}
Since the demonstration of the feasibility of wireless communications in the late 19th century, wireless communication technologies have rapidly developed and evolved across five generations. These advancements have provided diverse means of communications, profoundly altering human life and society. Throughout this evolution, multiple access (MA) has stood as a fundamental building block of wireless networks. MA refers to techniques that allow multiple users or devices to simultaneously share a finite amount of radio resources (e.g., frequency, time, code, etc.). It enables multiple users to transmit and receive data over the same network while minimizing disruption, thereby facilitating efficient interference management and resource allocation. Existing MA schemes are broadly classified into \emph{orthogonal} transmission strategies and \emph{non-orthogonal} transmission strategies \cite{islam2016power,ding2017application,ding2017survey,liu2017nonorthogonal}, depending on whether multiple users are allocated the same frequency/time/code resource.

Next-generation 6G wireless networks are envisioned to meet more stringent requirements than their predecessors. These include peak data rates on the order of terabits per second, extremely low latency, ultra-high reliability, and massive connectivity \cite{zhang20196g,you2021towards}. Additionally, they are expected to connect heterogeneous types of devices, provide ubiquitous coverage, integrate diverse functionalities, and support native intelligence \cite{saad2019vision,wang2023road}. In the pursuit of these ambitious goals, MA will undoubtedly play a pivotal role, evolving from conventional \emph{communications-oriented MA} to advanced MA schemes known as \emph{next-generation multiple access (NGMA)} \cite{liu2022developing,liu2022evolution}.
\subsection{A Brief History of Multiple Access}
Before delving into further investigations, we provide a brief overview of the historical development of MA.

We begin by examining orthogonal multiple access (OMA) technologies. The roots of MA trace back to British Patent No. 7777 awarded in 1900 to Marconi for the invention of the ``Tuned Circuit'' \cite{marcon1900inprovements}, which laid the foundation for frequency-division multiple access (FDMA). In the digital domain, two primary technologies have emerged for MA: time-division multiple access (TDMA) and code-division multiple access (CDMA). TDMA saw initial use in the 1960s for geosynchronous satellite networks \cite{campanella1984satellite}. On the other hand, CDMA has a distinct pedigree, dating back to the 1950s \cite{scholtz1982origins}. Unlike TDMA/FDMA, CDMA isolates users' signals not by allocating separate time or frequency slots but by employing unique associated codes. These codes, when decoded, ideally retrieve the original desired signal while eliminating the effects of other users' coded signals.

An extension of FDMA is orthogonal frequency-division multiplexing (OFDM), initially conceptualized by Chang at Bell Labs in 1966 \cite{chang1966synthesis} and practically implemented by Zimmerman and Kirsch in 1967 \cite{zimmerman1967gsc}. OFDM utilizes multiple subcarriers on adjacent frequencies, which are engineered to be orthogonal to each other. Consequently, there is no crosstalk between subcarriers, which enables overlapping channels and enhances spectral efficiency. The multiuser variant of OFDM, known as orthogonal frequency-division multiple access (OFDMA), allocates subsets of subcarriers to individual users. OFDMA has gained widespread adoption in prominent standards such as IEEE 802.16a/d/e \cite{weinstein2009history}.

These OMA strategies have found extensive application in practical wireless communication systems. For instance, FDMA was prevalent in the first generation (1G), TDMA dominated the second generation (2G) \cite{steele1999mobile}, CDMA with orthogonal codes characterized the third generation (3G) \cite{gilhousen1991capacity}, and OFDMA emerged as a key technology in the fourth generation (4G) \cite{li2013ofdma}. In these systems, users are assigned orthogonal frequency, time, or code resources, facilitating efficient communications.

\begin{figure}[!t]
    \centering
    \subfigure[Multiple Access Channel (MAC).]{
        \includegraphics[width=0.45\textwidth]{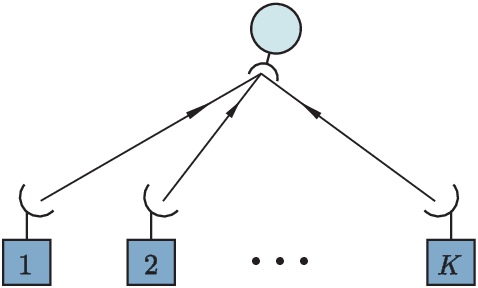}
        \label{Figure: MA_MAC_System}	
    }
    \subfigure[Broadcast Channel (BC).]{
        \includegraphics[width=0.45\textwidth]{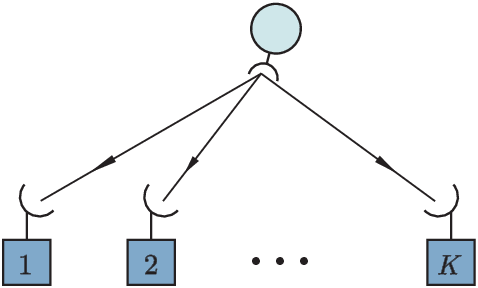}
        \label{Figure: MA_BC_System}
    }
    \caption{Illustration of MAC and BC.}
    \label{Figure: MA_BC_MAC_System}
\end{figure}

Another line of research has explored non-orthogonal transmission strategies, where multiple users are served simultaneously within a single frequency/time/code resource block. In this context, two types of channels are of particular interest: multiple access channels (MACs) and broadcast channels (BCs), representing uplink and downlink transmissions, respectively (refer to {\figurename} {\ref{Figure: MA_BC_MAC_System}}\footnote{{\figurename} {\ref{Figure: MA_MAC_System}} and {\figurename} {\ref{Figure: MA_BC_System}} illustrate two key scenarios: multiple terminals simultaneously communicating with a common station, and a common station transmitting messages simultaneously to multiple terminals. These are referred to as the MAC (multiple access channel) and BC (broadcast channel), respectively. It is important to note that both scenarios rely on the fact that the wireless channel is inherently a broadcast medium. From this perspective, using the term ``broadcast channel'' to describe the downlink case shown in {\figurename} {\ref{Figure: MA_BC_System}} could potentially lead to misunderstanding. Referring to it as a ``multiplexing scenario'' might be more accurate. However, the terms ``MAC'' and ``BC'' have been used in our field for over 50 years to describe these scenarios. Since our goal is to highlight key milestones and important results in MA technologies over the past five decades, we have chosen to retain the traditional terms ``MAC'' and ``BC'' for consistency with the established literature.}). Both cases require effective methods to mitigate inter-user interference (IUI).

The concept of MAC was initially hinted at by Shannon in his seminal paper ``\emph{Two-Way Communication Channels}'' \cite{shannon1961two}. In 1961, in paragraph 17 of \cite{shannon1961two}, Shannon mentioned, ``\emph{In another paper we will discuss the case of a channel with two or more terminals having inputs only and one terminal with an output only, a case for which a complete and simple solution of the capacity region has been found.}'' However, Shannon never published this finding before his passing. It remains unknown which solution Shannon had in mind when he wrote this, but it took a decade before other researchers addressed this problem. In 1971, Ahlswede presented his solution at the \emph{Tsahkadsor Symposium} \cite{ahlswede1971multi}, which utilized successive interference cancellation (SIC) decoding to tackle the IUI and obtained the MAC capacity region for discrete memoryless channels (DMCs). This work is considered to be one of the earliest major advancements in non-orthogonal transmission since Shannon's paper \cite{shannon1961two}.

On the downlink side, in 1972, Cover pioneered the concept of BC and conjectured that the capacity region of degraded BCs could be achieved through superposition coding (SPC) and SIC decoding \cite{cover1972broadcast}. In 1973, Bergmans proved that any rate tuples within the SPC-SIC rate region are attainable \cite{bergmans1973random}. There followed a year of rigorous efforts to establish the converse---proving that any rate tuples lying outside the SPC-SIC rate region are unattainable. Correspondences were exchanged among Wyner, Bergmans, and Gallager. At the end of the year, Wyner received proofs of the converse from Bergmans \cite{bergmans1974simple} and Gallager \cite{gallager1974capacity}. Among them, Bergmans asserted that the capacity region of Gaussian BCs precisely coincides with the SPC-SIC rate region.

In 1984, Winter introduced the utilization of antenna arrays with linear minimum mean-squared error (LMMSE) combining to mitigate IUI in MACs \cite{winters1984optimum}, marking the inception of space-division multiple access (SDMA). The capacity region of uplink Gaussian SDMA channels was subsequently delineated by Cheng and Verdú \cite{cheng1993gaussian}. Although the capacity region of the multiple-antenna BC remains elusive, Vishwanath \emph{et al.} demonstrated in 2003 \cite{vishwanath2003duality} that the sum-rate capacity of the multiple-antenna Gaussian BC can be attained via dirty paper coding (DPC), with Weingarten \emph{et al.} confirming in 2006 \cite{weingarten2006capacity} that DPC indeed achieves the entire capacity region. Since then, multiple-antenna non-orthogonal transmission has garnered significant attention, with SDMA emerging as a cornerstone technology of the fifth generation (5G) wireless networks \cite{agiwal2016next}.

In 2013, Saito \emph{et al.} \cite{saito2013non} introduced the concept of non-orthogonal multiple access (NOMA) and demonstrated its potential for simultaneous transmission of multiple users’ signals by leveraging SPC and SIC. Their work highlighted the improved system throughput and user fairness achievable over a single-antenna channel compared to conventional OMA techniques. Often referred to as power-domain NOMA, this approach represents an extension of Cover's pioneering work \cite{cover1972broadcast}. Due to its promising spectral efficiency, NOMA has become a focal point of research over the past decade. In 2022, Liu \emph{et al.} \cite{liu2022developing} presented the first comprehensive overview of NGMA, which explored NOMA's evolutionary trajectory to meet the demands of future wireless networks. In 2023, NOMA was highlighted in the International Mobile Telecommunications-2030 Framework, which stated: ``\emph{for MA, technologies including NOMA and grant-free MA are expected to be considered to meet future requirements}'' \cite{recommendation2023framework,ding2024next}.

For reference, the milestones of MA are succinctly presented in the timeline illustrated in Table \ref{Table: timeline}. The historical review presented above indicates a clear evolutionary trend in MA technologies, viz. shifting from orthogonal to non-orthogonal transmission strategies. While OMA techniques have traditionally been favored for their simplicity, they often suffer from poor resource efficiency \cite{liu2017nonorthogonal}. As we move towards the era of 6G networks, there is a growing reliance on non-orthogonal transmission strategies for MA \cite{liu2022developing,liu2022evolution}. This evolution reflects the ongoing quest for enhanced spectral efficiency and improved performance in modern wireless communication systems.

\begin{table*}[!t]
\caption{Timeline of Multiple Access Milestones}
\centering
\begin{minipage}[t]{0.92\linewidth}
\color{gray}
\rule{\linewidth}{1pt}
\ytl{1900}{Marconi was awarded British Patent No. 7777, which laid the foundation for FDMA \cite{marcon1900inprovements}}
\ytl{1950}{The earliest versions of CDMA were reported \cite{scholtz1982origins}}
\ytl{1960}{TDMA saw initial use for geosynchronous satellite networks \cite{campanella1984satellite}}
\ytl{1961}{Shannon hinted at the concept of MAC in his paper ``\emph{Two-Way Communication Channels}'' \cite{shannon1961two}}
\ytl{1966}{Chang conceptualized OFDM at Bell Labs \cite{chang1966synthesis}}
\ytl{1967}{Zimmerman and Kirsch presented the first practical implementation of OFDM \cite{zimmerman1967gsc}}
\ytl{1971}{Ahlswede obtained the MAC capacity region for DMCs via SIC decoding \cite{ahlswede1971multi}}
\ytl{1972}{Cover pioneered the concept of BC and conjectured that the capacity region of degraded BCs could be achieved through SPC and SIC decoding \cite{cover1972broadcast}}
\ytl{1973}{Bergmans proved that any rate tuples within the SPC-SIC rate region are attainable \cite{bergmans1973random}}
\ytl{1974}{Bergmans asserted that the capacity region of Gaussian BCs precisely coincides with the SPC-SIC rate region \cite{bergmans1974simple}}
\ytl{1984}{Winters utilized antenna arrays with LMMSE combining to mitigate IUI in MACs \cite{winters1984optimum}}
\ytl{1993}{Cheng and Verdú delineated the capacity region of uplink Gaussian SDMA channels \cite{cheng1993gaussian}}
\ytl{2003}{Vishwanath \emph{et al.} demonstrated that the sum-rate capacity of the multiple-antenna Gaussian BC can be attained via DPC \cite{vishwanath2003duality}}
\ytl{2006}{Weingarten \emph{et al.} confirming that DPC indeed achieves the entire capacity region \cite{weingarten2006capacity}}
\ytl{2013}{Saito \emph{et al.} introduced the concept of power-domain NOMA \cite{saito2013non}}
\ytl{2022}{The first overview paper on NGMA was presented \cite{liu2022developing}}
\ytl{2023}{NOMA was highlighted in the International Mobile Telecommunications-2030 Framework \cite{recommendation2023framework}}
\rule{\linewidth}{1pt}%
\end{minipage}%
\label{Table: timeline}
\end{table*}
\subsection{Motivation and Contributions}
It has been 50 years since Cover \cite{cover1972broadcast} and Ahlswede \cite{ahlswede1971multi} pioneered the use of SPC and SIC to address IUI and achieve the capacity region of BC and MAC. The introduction of these fundamental techniques represents a watershed moment in the development of MA and will continue to play a pivotal role in unlocking the potential of NGMA systems. In light of this, we provide a 50-year tutorial review on MA technologies, with a specific focus on their non-orthogonal components in both the power domain and spatial domain. Furthermore, we will explore their integration with other emerging technologies in the 6G landscape. Our aim is not to delve into the details of OMA technologies, such as FDMA, TDMA, CDMA, and OFDMA, and their variants\footnote{It is worth noting that code-domain NOMA can be considered a variant of CDMA, which is thus not the focus of this tutorial review. Code-domain NOMA shares similarities with CDMA or multi-carrier CDMA, except for its preference for low-density sequences or non-orthogonal sequences with low cross-correlation. For further details, we refer to the overview papers \cite{dai2018survey,wu2018comprehensive,budhiraja2021systematic,deolia2022code}.}, as these have already been extensively discussed and standardized for practical implementation over the past 70 years.

In recent years, numerous excellent overview, survey, and tutorial papers have illuminated the landscape of MA, particularly focusing on non-orthogonal transmission strategies. The authors of \cite{islam2016power,ding2017application,ding2017survey} reviewed the fundamental aspects of power-domain NOMA. These works encompassed a range of topics including capacity analysis, power allocation, user fairness, and user-pairing, with the main emphasis on single-antenna NOMA. A succinct overview of multiple-antenna NOMA was provided in \cite{liu2018multiple}. Expanding upon this foundation, a comprehensive survey of power-domain NOMA was presented by the authors of \cite{liu2017nonorthogonal}, elaborating on theoretical NOMA principles, the design of multiple-antenna-aided NOMA, and its integration potential with 5G and future applications. In parallel, the authors of \cite{vaezi2019interplay,maraqa2020survey} explored the integration of power-domain NOMA with cutting-edge technologies such as massive multiple-input multiple-output (MIMO), millimeter-wave (mmWave)/terahertz (THz) communications, visible light communications, cognitive radio, and unmanned aerial vehicle (UAV) communications.

Contrasting these power-domain NOMA-centric works, the authors of \cite{mao2022rate} provided a comprehensive survey of MA technologies focusing on the spatial domain, particularly highlighting rate-splitting multiple access (RSMA). Additionally, the evolution of MA towards NGMA was delineated by the authors of \cite{liu2022developing}, who outlined the concept of NGMA for 6G and beyond, along with several promising research directions. Further elaboration of promising application scenarios, emerging requirements, and potential techniques within the realm of NGMA was offered in \cite{liu2022evolution}. Moreover, a concise magazine paper \cite{mu2023noma} introduced the application of NOMA in integrated sensing and communications (ISAC), a recently emerging technology garnering significant attention from academia and industry.

While the aforementioned contributions have addressed general concepts or specific aspects of MA technologies, several crucial areas remain unexplored. These include the development of novel MA frameworks, a review of the information-theoretic limits and analytical foundations of MA, and the exploration of newly emerged applications in wireless networks. Additionally, a comprehensive historical overview of MA milestones, combined with a detailed tutorial for newcomers, has been lacking. Motivated by these gaps, this tutorial review aims to provide a comprehensive perspective that covers the fundamental theories of MA technologies---exploiting both the power and spatial domains---their diverse applications, and a forward-looking exploration of the future of MA. For clarity, we present a detailed comparison between existing contributions and the content of this paper in Table \ref{tab: Section: Comparison of this work with existing}. 

As shown in Table \ref{tab: Section: Comparison of this work with existing}, our paper overlaps with the work in \cite{liu2022evolution} in some areas. This overlap arises because both \cite{liu2022evolution} and our work discuss the developments of MA technologies. However, while \cite{liu2022evolution} focuses on presenting a survey of the available literature, our goal is to provide a comprehensive tutorial review, offering step-by-step teaching materials on key milestones and important results. Therefore, the objectives of our paper and \cite{liu2022evolution} are fundamentally different. In terms of content, our work emphasizes the theoretical foundations and principles of MA, whereas \cite{liu2022evolution} places more focus on applications, including existing implementations and future prospects. Moreover, our paper covers several important aspects not addressed in \cite{liu2022evolution}, such as the historical development of MA and its information-theoretic foundations. Additionally, \cite{liu2022evolution} does not delve into recent advancements, such as MA in near-field communications (NFC) and ISAC, both of which are thoroughly covered in our work. Finally, since conventional applications of MA, such as power-domain MA, are detailed in \cite{liu2022evolution}, we do not delve deeply into these topics. In summary, our paper differs significantly from \cite{liu2022evolution} in both its fundamental objectives and content.

\begin{table*}[!t]
\centering
\caption{Comparison of this work with existing overview/survey/tutorial papers on MA. Here, $\bigstar$ and $\blacksquare$ refer to ``reviewed in detail'' and ``mentioned but not considered in detail'', respectively. The existing papers are listed in chronological order.}
\label{tab: Section: Comparison of this work with existing}
\resizebox{0.99\textwidth}{!}{
\begin{tabular}{|ll|l|l|l|l|l|l|l|l|l|l|l|l|}
\hline
\multicolumn{2}{|l|}{\textbf{Existing Works}}                                                    & {\cite{islam2016power}} & {\cite{ding2017application}} & {\cite{ding2017survey}} & {\cite{liu2017nonorthogonal}} & {\cite{liu2018multiple}} & {\cite{vaezi2019interplay}} & {\cite{maraqa2020survey}} & {\cite{mao2022rate}} & {\cite{liu2022developing}} & {\cite{liu2022evolution}} & {\cite{mu2023noma}} & This work \\ \hline
\multicolumn{2}{|l|}{\textbf{Type}}                                                              & Survey        &  Overview       &  Survey       &  Survey       &   Overview      & Survey        & Survey        &  Tutorial Review       &   Overview     &   Survey        &     Overview     &    Tutorial Review                                                  \\ \hline
\multicolumn{2}{|l|}{\textbf{Year}}                                                              & 2016        &  2017       &  2017       &  2017       &   2018      & 2019        & 2020        &  2022       &   2022     &   2022        &     2023     &    2024                                                  \\ \hline
\multicolumn{2}{|l|}{\textbf{Timeline}}                                                          &         &         &         &   $\bigstar$      &         &         &         &         &         & &          &  $\bigstar$                                                    \\ \hline
\multicolumn{1}{|l|}{\multirow{4}{*}{\textbf{Power-Domain MA}}}   & \textbf{Information-Theoretic Limits} &   $\blacksquare$      &   $\blacksquare$      &   $\blacksquare$      & $\blacksquare$        &  $\blacksquare$       &  $\blacksquare$       &         &         &         &          &          &        $\bigstar$                                              \\ \cline{2-14} 
\multicolumn{1}{|l|}{}                                   & \textbf{Foundational Principles}      &  $\bigstar$       &   $\bigstar$      &  $\bigstar$       &  $\bigstar$       &   $\bigstar$      &  $\blacksquare$       &    $\blacksquare$     &         &         &   $\bigstar$       &          &          $\bigstar$                                            \\ \cline{2-14} 
\multicolumn{1}{|l|}{}                                   & \textbf{Applications}                 &  $\bigstar$       &   $\blacksquare$      & $\bigstar$        &   $\bigstar$      &   $\blacksquare$ &   $\bigstar$      &   $\bigstar$      &         &         &   $\bigstar$       &          &           $\blacksquare$                                           \\ \cline{2-14} 
\multicolumn{1}{|l|}{}                                   & \textbf{MIMO-NOMA}                    &   $\blacksquare$      &   $\blacksquare$      &  $\blacksquare$       &  $\bigstar$       &    $\bigstar$     &         &         &         &         &    $\bigstar$      &          &            $\bigstar$                                          \\ \hline
\multicolumn{1}{|l|}{\multirow{4}{*}{\textbf{Spatial-Domain MA}}} & \textbf{Foundational Principles}      &         &         &         &  $\blacksquare$       &         &         &         &    $\blacksquare$     &         &    $\blacksquare$      &          &            $\bigstar$                                          \\ \cline{2-14} 
\multicolumn{1}{|l|}{}                                   & \textbf{Applications and Variants}    &         &         &         &  $\bigstar$       &         &  $\bigstar$       &  $\bigstar$       &    $\blacksquare$     &         &    $\blacksquare$      &          &            $\bigstar$                                          \\ \cline{2-14} 
\multicolumn{1}{|l|}{}                                   & \textbf{RSMA}                         &         &         &         &         &         &         &         &    $\bigstar$     &         &    $\blacksquare$      &          &            $\blacksquare$                                          \\ \cline{2-14} 
\multicolumn{1}{|l|}{}                                   & \textbf{Near-Field MA}                &         &         &         &         &         &         &         &         &         &          &          &              $\bigstar$                                        \\ \hline
\multicolumn{1}{|l|}{\multirow{2}{*}{\textbf{MA for ISAC}}}       & \textbf{NOMA-Based Framework}         &         &         &         &         &         &         &         &         &   &          $\blacksquare$& $\bigstar$         &              $\bigstar$                                        \\ \cline{2-14} 
\multicolumn{1}{|l|}{}                                   & \textbf{Performance Analysis}         &         &         &         &         &         &         &         &         &         &          &          &              $\bigstar$                                      \\ \hline
\multicolumn{1}{|l|}{\multirow{4}{*}{\textbf{NGMA}}}              & \textbf{New Applications}             &         &         &         &         &         &         &         &         &   $\bigstar$      &    $\bigstar$      &          &              $\bigstar$                                        \\ \cline{2-14} 
\multicolumn{1}{|l|}{}                                   & \textbf{New Techniques}               &         &         &         &         &         &         & $\blacksquare$        &         &   $\bigstar$      &     $\bigstar$     &          &                $\bigstar$                                      \\ \cline{2-14} 
\multicolumn{1}{|l|}{}                                   & \textbf{New Tools}                    &         &         &         &         &         &         &         &         &   $\bigstar$      &    $\bigstar$      &          &                 $\bigstar$                                     \\ \cline{2-14} 
\multicolumn{1}{|l|}{}                                   & \textbf{New Requirements}             &         &         &         &         &         &         &         &         &     $\bigstar$    &    $\bigstar$      &          &                $\blacksquare$                                       \\ \hline
\end{tabular}}
\end{table*}

We anticipate that our work will offer valuable insights into the design of NGMA systems and serve as a primer on MA for graduate students and researchers seeking to gain a fundamental understanding of MA technologies. The key contributions of our work are outlined as follows.
\begin{enumerate}
  \item[1)] \emph{\textbf{Exploration of Information-Theoretic Limits of Power-Domain NOMA:}} We conduct a comprehensive examination of the information-theoretic boundaries of power-domain NOMA. This investigation encompasses an analysis of capacity-achieving coding/decoding strategies and their associated capacity regions for both scalar and vector channels. Emphasis is placed on elucidating the pivotal role of SPC and SIC decoding. Furthermore, we extend our scrutiny to encompass MIMO-NOMA systems, offering insights into the performance of a unified cluster-free MIMO-NOMA framework.
  \item[2)] \emph{\textbf{Overview of MA Transmission Schemes in the Spatial Domain:}} We provide a detailed overview of MA transmission schemes operating in the spatial domain. This examination begins with an assessment of SDMA systems in the context of linear beamforming. Subsequently, we study SDMA and MIMO-NOMA as well as their variants in traditional far-field communications (FFC) using planar-wave propagation models. Furthermore, we analyze the performance of spatial-domain MA in NFC, considering scenarios where the base station (BS) is equipped with an extremely-large array and employing spherical-wave propagation models.
  \item[3)] \textbf{\emph{Integration of NOMA with ISAC Systems:}} We study the integration of NOMA with ISAC systems. First, we introduce four novel NOMA-based frameworks designed to facilitate the practical implementation of ISAC. Subsequently, we present an analytical ISAC framework based on mutual information (MI), which enables the evaluation of the sensing and communication performance of the considered NOMA-based ISAC frameworks.
  \item[4)] \textbf{\emph{Identification of Research Opportunities in NGMA:}} We identify promising avenues for research on NGMA in three principal directions: new application scenarios, new techniques, and new tools. For each direction, we select a key technology as a case study, namely semantic communications for application scenarios, reconfigurable intelligent surfaces (RISs) for techniques, and machine learning (ML) for tools. Along with each direction, we discuss potential solutions and avenues for future exploration.
\end{enumerate}

\begin{figure}[!t]
  \centering
  \includegraphics[width=0.45\textwidth]{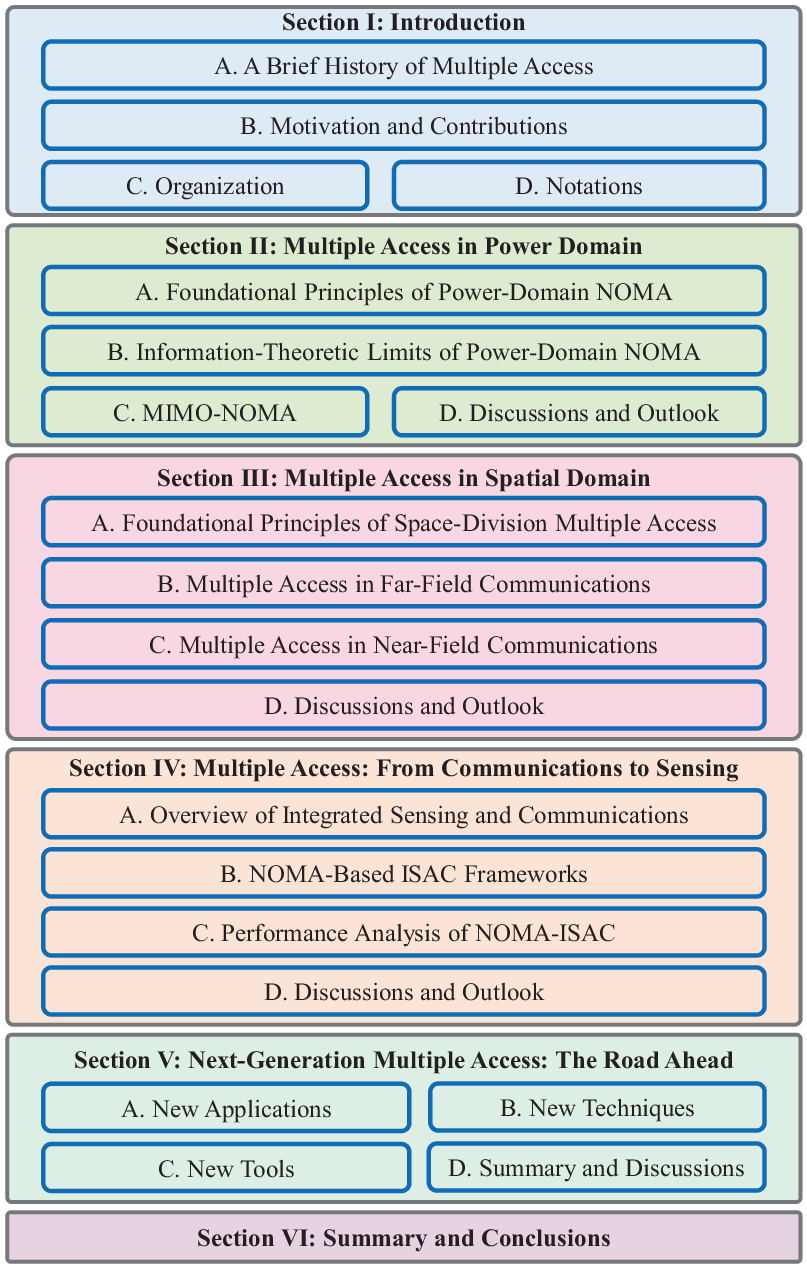}
  \caption{Condensed overview of this tutorial review on MA.}
  \label{Figure: Outline}
\end{figure}

\subsection{Organization}
The remainder of this paper is structured as follows. Specifically, the four main contributions mentioned earlier are addressed in Section \ref{Section: Power Domain}, Section \ref{Section: Spatial Domain}, Section \ref{Section: MA for ISAC}, and Section \ref{Section: MA for NGMA} of our work, respectively. In Section \ref{Section: Power Domain}, we explore the information-theoretic limits of power-domain NOMA, with a focus on channel coding/decoding and the capacity region. Section \ref{Section: Spatial Domain} investigates MA transmission schemes exploiting the spatial domain, considering both conventional SDMA systems and near-field MA systems. Section \ref{Section: MA for ISAC} studies the integration of NOMA into ISAC, concentrating on the fundamental uplink and downlink transmission frameworks. Section \ref{Section: MA for NGMA} investigates the integration of MA with other emerging 6G wireless technologies. Finally, Section \ref{Section: Conclusions} concludes the paper. {\figurename} {\ref{Figure: Outline}} illustrates the organizational structure of the paper, and Table \ref{tab:LIST OF ACRONYMS} presents key abbreviations used throughout this treatise.

\begin{table*}[!t]
\caption{List of Acronyms}
\label{tab:LIST OF ACRONYMS}
\centering
\resizebox{0.99\textwidth}{!}{
\begin{tabular}{|l||l|l||l|}
\hline
ADMA 	&	Angle Division Multiple Access	&	NFC	&	Near-Field Communications	\\ \hline
APES 	&	Amplitude and Phase Estimation	&	NGMA	&	Next-Generation Multiple Access 	\\ \hline
AWGN	&	Additive White Gaussian Noise	&	NLoS	&	Non-Line-of-Sight 	\\ \hline
BC	&	Broadcast Channels	&	NOMA	&	Non-orthogonal Multiple Access	\\ \hline
BDMA	&	Beam Division Multiple Access	&	NU	&	Near User	\\ \hline
BS	&	Base Station	&	OFDM	&	Orthogonal Frequency-Division Multiplexing	\\ \hline
CAP	&	Continuous-Aperture	&	OFDMA	&	Orthogonal Frequency-Division Multiple Access	\\ \hline
C-C	&	Communications-Centric	&	OMA	&	Orthogonal Multiple Access	\\ \hline
CDMA	&	Code-Division Multiple Access	&	OSAC	&	Orthogonal Sensing and Communications	\\ \hline
CL	&	Centralized Learning	&	QoS	&	Quality-of-Service	\\ \hline
CR	&	Communication Rate 	&	RCS	&	Radar Cross Section	\\ \hline
CRB	&	Cramér-Rao Bound	&	RDMA	&	Range Division Multiple Access	\\ \hline
C-SIC	&	Communications-Centric Successive Interference Cancellation	&	RF	&	Radio Frequency	\\ \hline
CU	&	Communication User	&	RIS	&	Reconfigurable Intelligent Surface	\\ \hline
DFSAC	&	Dual-Function Sensing and Communications	&	RL	&	Reinforcement Learning	\\ \hline
DFT	&	Discrete Fourier Transform	&	RSMA	&	Rate-Splitting Multiple Access	\\ \hline
DL	&	Deep Learning	&	RZF	&	Regularized Zero-Forcing	\\ \hline
DMC	&	Discrete Memoryless Channel	&	S\&C	&	Sensing and Communications	\\ \hline
DoA	&	Direction of Arrival	&	S-C	&	Sensing-Centric	\\ \hline
DoF	&	Degree-of-Freedom	&	SCA	&	Successive Convex Approximation	\\ \hline
DPC	&	Dirty Paper Coding	&	SDM	&	Spatial Division Multiplexing	\\ \hline
ELAA	&	Extremely Large Aperture Array 	&	SDMA	&	Space-Division Multiple Access	\\ \hline
EM	&	Electromagnetic	&	SDP	&	Semi-Definite Programming	\\ \hline
eMBB	&	Enhanced Mobile Broadband	&	SDR	&	Software-Defined Radio	\\ \hline
FDMA	&	Frequency-Division Multiple Access	&	SIC	&	Successive Interference Cancellation	\\ \hline
FFC	&	Far-Field Communications	&	SINR	&	Signal-to-Interference-Plus-Noise Ratio	\\ \hline
FL	&	Federated Learning	&	SISO	&	Single-Input Single-Output	\\ \hline
FP	&	Fractional Programming	&	SLNR	&	Signal-to-Leakage-Plus-Noise Ratio	\\ \hline
FU	&	Far User	&	SPC	&	Superposition Coding	\\ \hline
GSVD	&	Generalized Singular Value Decomposition	&	SPD	&	Spatially Discrete	\\ \hline
HB	&	Hybrid Beamforming	&	SR	&	Sensing Rate	\\ \hline
IFI	&	Inter-Functionality Interference	&	SRE	&	Smart Radio Environment	\\ \hline
ISAC	&	Integrated Sensing and Communications	&	S-SIC	&	Sensing-Centric Successive Interference Cancellation	\\ \hline
ISCC	&	Integrated Sensing, Communications, and Computation	&	ST	&	Simultaneous Triangularization	\\ \hline
IUI	&	Inter-User Interference	&	STAR	&	Simultaneously Transmitting and Reflecting	\\ \hline
LMMSE	&	Linear Minimum Mean-Squared Error	&	SU	&	Single-User	\\ \hline
LoS	&	Line-of-Sight	&	SWIPT	&	Simultaneous Wireless Information and Power Transfer	\\ \hline
MA	&	Multiple Access	&	TDMA 	&	Time-Division Multiple Access	\\ \hline
MAC	&	Multiple Access Channel	&	THz	&	Terahertz	\\ \hline
MD	&	Minimum Distance	&	TIN	&	Treating Interference as Noise	\\ \hline
MI	&	Mutual Information	&	UAV	&	Unmanned Aerial Vehicle	\\ \hline
MIMO	&	Multiple-Input Multiple-Output	&	ULA	&	Uniform Linear Array	\\ \hline
MISO	&	Multiple-Input Single-Output	&	URLLC	&	Ultra-Reliable Low-Latency Communication	\\ \hline
ML	&	Machine Learning	&	V-BLAST	&	Vertical-Bell Laboratories Layered Space-Time	\\ \hline
MMSE	&	Minimum Mean-Squared Error	&	WMMSE	&	Weighted Minimum Mean-Squared Error	\\ \hline
mMTC	&	Massive Machine-Type Communication	&	ZF	&	Zero-Forcing	\\ \hline
mmWave	&	Millimeter-Wave	&	1G	&	First Generation	\\ \hline
MRC	&	Maximal-Ratio Combining	&	2G	&	Second Generation	\\ \hline
MRT	&	Maximal-Ratio Transmission	&	3G	&	Third Generation	\\ \hline
MSE	&	Mean Squared Error	&	4G	&	Fourth Generation	\\ \hline
MUSIC	&	Multiple Signal Classification	&	5G	&	Fifth Generation	\\ \hline
\end{tabular}}
\end{table*}
\subsection{Notations}
Throughout this paper, scalars, vectors, and matrices are denoted by non-bold, bold lower-case, and bold upper-case letters, respectively. For the matrix $\mathbf{A}$, $[\mathbf{A}]_{i,j}$, ${\mathbf{A}}^{\mathsf{T}}$, ${\mathbf{A}}^{*}$, and ${\mathbf{A}}^{\mathsf{H}}$ denote the $(i,j)$th entry, transpose, conjugate, and conjugate transpose of $\mathbf{A}$, respectively. For a square matrix $\mathbf{B}$, ${\mathbf{B}}^{\frac{1}{2}}$, ${\mathbf{B}}^{-1}$, ${\mathsf{tr}}(\mathbf{B})$, and $\det(\mathbf{B})$ denote the principal square root, inverse, trace, and determinant of $\mathbf{B}$, respectively. The notation $[\mathbf{a}]_i$ denotes the $i$th entry of vector $\mathbf{a}$, and ${\mathsf{diag}}\{\mathbf{a}\}$ returns a diagonal matrix whose diagonal elements are entries of $\mathbf{a}$. The notations $\lvert a\rvert$ and $\lVert \mathbf{a} \rVert$ denote the magnitude and norm of scalar $a$ and vector $\mathbf{a}$, respectively. The $N\times N$ identity matrix is denoted by ${\mathbf{I}}_N$. The matrix inequalities ${\mathbf{A}}\succeq{\mathbf{0}}$ and ${\mathbf{A}}\succ{\mathbf{0}}$ imply that $\mathbf{A}$ is positive semi-definite and positive definite, respectively. The sets $\mathbbmss{Z}$, $\mathbbmss{R}$, and $\mathbbmss{C}$ stand for the integer, real, and complex spaces, respectively, and notation $\mathbbmss{E}\{\cdot\}$ represents mathematical expectation. The MI between random variables $X$ and $Y$ conditioned on $Z$ is denoted by $I\left(X;Y|Z\right)$. The Kronecker product and the ceiling function are represented by $\otimes$ and $\lceil \cdot \rceil$, respectively. Finally, ${\mathcal{CN}}({\bm\mu},\mathbf{X})$ is used to denote the circularly-symmetric complex Gaussian distribution with mean $\bm\mu$ and covariance matrix $\mathbf{X}$.

\section{Multiple Access in Power Domain}\label{Section: Power Domain}
In this section, we provide a comprehensive review of power-domain NOMA (for short NOMA), wherein multiple users share the same orthogonal time-frequency-code-space resource block \cite{ding2017application}. NOMA offers a larger rate region compared to OMA techniques such as FDMA, TDMA, and OFDMA. This section is divided into three parts. In Section \ref{Section: Foundational Principles of Power-Domain NOMA}, the foundational principles of NOMA are presented, detailing the capacity-achieving encoding and decoding schemes for several typical NOMA channels. In Section \ref{Section: Information-Theoretic Limits of Power-Domain NOMA}, the information-theoretic capacity limits achieved by these coding schemes are discussed. These results are further extended to MIMO-NOMA using linear beamforming schemes in Section \ref{Section: MIMO-NOMA}. It is worth mentioning that the information-theoretic results reported in this section can be considered the pinnacle of MA, as they laid the foundation for the development of MA technologies over the past years.
\subsection{Foundational Principles of Power-Domain NOMA}\label{Section: Foundational Principles of Power-Domain NOMA}
The central idea behind power-domain NOMA lies in the efficient utilization of a designated resource block to accommodate multiple users. This involves the implementation of \emph{transmitter-side SPC} and \emph{receiver-side SIC decoding}. This marks a significant departure from traditional OMA as well as from code-domain NOMA approaches. 

In the following pages, we explain the foundational principles of power-domain NOMA by examining its key features. To enhance clarity, our discussion focuses on discrete memoryless channels, allowing for concrete descriptions of coding schemes and associated capacity regions. 
\subsubsection{Shannon Channels, Multiple Access Channels, and Broadcast Channels}\label{Section: Shannon Channels, Multiple Access Channels, and Broadcast Channels}
By definition, NOMA involves multiple users. In the scenario where $K$ users communicate simultaneously with a receiver, as depicted in {\figurename} \ref{Figure: MA_MAC_System}, this configuration is referred to as \emph{MAC}. If we reverse the network and have one transmitter broadcasting independent messages simultaneously to $K$ users, as shown in {\figurename} \ref{Figure: MA_BC_System}, it becomes the \emph{BC}. MAC and BC correspond to uplink and downlink NOMA transmission, respectively. Before delving further into these multiuser NOMA models, let us commence our exploration with the single-user case to introduce some fundamental information-theoretic preliminaries. Our focus is on the single-user DMC (discrete memoryless channel), also known as the \emph{Shannon channel}. This choice is motivated by two key reasons. Firstly, DMC allows us to provide concrete descriptions of the coding schemes and associated capacity regions. Secondly, many channels discussed in the current NOMA literature are essentially special cases of the DMC, such as Gaussian channels.

\begin{figure}[!t]
  \centering
  \includegraphics[width=0.45\textwidth]{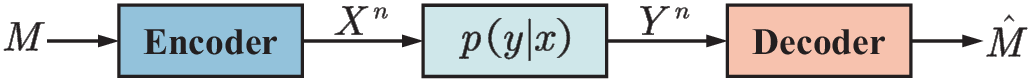}
  \caption{Illustration of the single-user DMC, where $p(y|x)$ denotes the transition probability density function of the channel for each channel use.}
  \label{Figure: DMGC_SU_Coding}
\end{figure}

$\bullet$ \emph{Shannon Channels}: We begin the theoretical discourse by revisiting Shannon's original channel capacity theorem. The fundamental Shannon model for a communication channel is the DMC with a channel transition probability density function $p(y|x)$, an input alphabet $\mathcal{X}$, and an output alphabet $\mathcal{Y}$, as illustrated in {\figurename} {\ref{Figure: DMGC_SU_Coding}}. The term \emph{discrete memoryless} implies that if a sequence ${{X}}^{n}=[{\mathsf{x}}_1,{\mathsf{x}}_2,\ldots,{\mathsf{x}}_n]\in{\mathcal{X}}^{n}$ is transmitted, the received sequence $Y^n=[{\mathsf{y}}_1,{\mathsf{y}}_2,\ldots,{\mathsf{y}}_n]\in{\mathcal{Y}}^{n}$ is generated according to $\prod_{i=1}^{n}p({\mathsf{y}}_i|{\mathsf{x}}_i)$. 

For any block length $n$, a $(2^{n {\mathsf{R}}},n)$ code for the Shannon channel is characterized by a message set ${\mathscr{M}}=\{1,\ldots,2^{\lceil n {\mathsf{R}}\rceil}\}$, a codebook $\mathcal{C}$ comprising $2^{\lceil n {\mathsf{R}}\rceil}$ \emph{codewords} of the form $X^n=[{\mathsf{x}}_1,{\mathsf{x}}_2,\ldots,{\mathsf{x}}_n]\in{\mathcal{X}}^{n}$, an encoding function $f_{\rm{en}}:{\mathscr{M}}\mapsto{\mathcal{C}}$ mapping each message index $M\in{\mathscr{M}}$ onto a unique codeword within $\mathcal{C}$, and a decoding function $f_{\rm{de}}:{\mathcal{Y}}^{n}\mapsto{\mathscr{M}}$. The error probability is given by $\epsilon=\Pr(f_{\rm{de}}(Y^n)=\hat{M}\ne M)$. A coding rate $\mathsf{R}$ is \emph{achievable} if there exists a sequence of $(2^{n {\mathsf{R}}},n)$ codes with rate \emph{approaching} $\mathsf{R}$ and \emph{vanishing} $\epsilon$, i.e., $\lim_{n\rightarrow\infty}\epsilon=0$. The capacity $\mathsf{C}$ of the DMC is the supremum of the set of achievable rates. 

In 1948, Shannon showed that the capacity $\mathsf{C}$ of the DMC is given as follows \cite{shannon1948mathematical}.
\vspace{-5pt}
\begin{theorem}\label{Shannon_Capacity_Theorem}
(\textbf{Shannon’s Capacity Theorem}) The capacity of the DMC is given by
\begin{align}\label{Channel_Capacity_SU_DM}
\mathsf{C}=\max\nolimits_{p(x)}I(X;Y),
\end{align}
where $p(x)$ denotes the input distribution of each letter $\mathsf{x}_i$.
\end{theorem}
\vspace{-5pt}
\begin{IEEEproof}
The proof involves demonstrating \cite{el1980multiple}:
\begin{enumerate}
  \item[(\romannumeral1)] \emph{Achievability} of any ${\mathsf{R}}<{\mathsf{C}}$, i.e., that there exists a sequence of $(2^{n {\mathsf{R}}},n)$ codes such that $\lim_{n\rightarrow\infty}\epsilon=0$,
  \item[(\romannumeral2)] and a \emph{converse} showing that given any sequence of $(2^{n {\mathsf{R}}},n)$ codes with $\lim_{n\rightarrow\infty}\epsilon=0$ then ${\mathsf{R}}<{\mathsf{C}}$.
\end{enumerate}
\subsubsection*{Achievability}
This can be proven using \emph{random coding} and \emph{joint typicality decoding}. Assume maximization is achieved for a distribution $p^{\star}(x)$. \emph{Randomly} generate $2^{\lceil n {\mathsf{R}}\rceil}$ \emph{independent identically distributed (i.i.d.)} sequences $X^n(M)$ for $M\in{\mathscr{M}}$, according to the probability distribution $\prod\nolimits_{i=1}^{n}p^{\star}(\mathsf{x}_i)$. The chosen codebook is revealed to both the encoder and the decoder before transmission begins. Decoding is accomplished by a \emph{joint typical decoder} \cite{cover1999elements}, ensuring that $\lim_{n\rightarrow\infty}\epsilon=0$. 
\subsubsection*{Converse}The converse of Theorem \ref{Shannon_Capacity_Theorem} can be proved by using Fano's inequality and the data processing inequality. For further details, interested readers are referred to \cite{cover1999elements}.
\end{IEEEproof}
The arguments in the proof of achievability also reveal that a possible approach to constructing a \emph{capacity-achieving encoder-decoder pair} is to utilize \emph{random coding along with joint typical decoding}.

%In 1948, upon the publication of Shannon's groundbreaking work, the proofs of the capacity theorem were initially considered heuristic. The first rigorous proof, presented by Feinstein \cite{feinstein1954new} eight years later, diverged significantly from Shannon's original concept of random coding. Subsequent proofs by Wolfowitz \cite{wolfowitz1957coding}, Fano \cite{fano1961transmission}, Gallager \cite{gallager1965simple}, and various others pursued distinct approaches. The proof sketched in Theorem \ref{Shannon_Capacity_Theorem} aligns with the conceptual foundation outlined by Shannon \cite{el1980multiple}. It is presented as a problem in Gallager's work \cite{gallager1968information} and in Forney's unpublished class notes \cite{forney1972inf}.

We proceed by specializing the DMC to a discrete-time Gaussian scalar channel and provide its channel capacity.

$\bullet$ \emph{Gaussian Scalar Channels}: The discrete-time additive white Gaussian noise (AWGN) channel is modeled as follows:
\begin{equation}
Y=X+Z,
\end{equation}
where $Z\sim{\mathcal{CN}}(0,N)$ is the additive noise with noise variance $N$. The input sequence $X^{n}=[{\mathsf{x}}_1,\ldots,{\mathsf{x}}_n]$ adheres to a power constraint $\frac{1}{n}\sum\nolimits_{i=1}^{n}\lvert {\mathsf{x}}_i\rvert^2\leq P$. The Shannon capacity $\mathsf{C}$, obtained by maximizing $I(X;Y)$ over all random variables $X$ such that ${\mathbbmss{E}}\{\lvert X\rvert^2\}\leq P$, is given by \cite{tse2005fundamentals}
\begin{equation}\label{AWGN_Discrete_Capacity}
\mathsf{C}=\log_2(1+P/N)~{\rm{bits/transmission}}.
\end{equation}
Moreover, the coding and decoding procedure for achieving a coding rate $\mathsf{R}<\mathsf{C}$ in the Gaussian channel is outlined as follows.
\subsubsection*{Gaussian Random Coding}
The codebook is generated by randomly selecting $2^{\lceil n {\mathsf{R}}\rceil}$ i.i.d. $X^n$ sequences. Each sequence is composed of $n$ independent complex Gaussian random variables each subject to $\mathcal{CN}(0,P)$. Then, the encoding function $f_{\rm{en}}(\cdot)$ is defined by establishing a random one-to-one mapping from the message set to the Gaussian codebook.
\subsubsection*{Minimum Distance (MD) Decoding}
Given ${Y}^n=f_{\rm{en}}(M)+Z^n$ with $Z^n$ denoting the sequences of noise samples, the receiver identifies the index $\hat{M}$ of the codeword closest to $Y^n$, i.e.,
\begin{align}\label{Gaussian_P2P_ML}
\hat{M}=f_{\rm{de}}({Y}^n)=\argmin\nolimits_{m\in{\mathscr{M}}}\lVert{Y}^n- f_{\rm{en}}(m)\rVert.
\end{align}
This MD decoding scheme for the Gaussian channel is essentially equivalent to joint typical decoding. MD decoding is also known as nearest neighbour decoding and is equivalent to maximum likelihood decoding.

Having elucidated the channel capacity of the single-user DMC, we will now transition to the multiuser channel, i.e., NOMA transmission.

\begin{figure}[!t]
  \centering
  \includegraphics[width=0.45\textwidth]{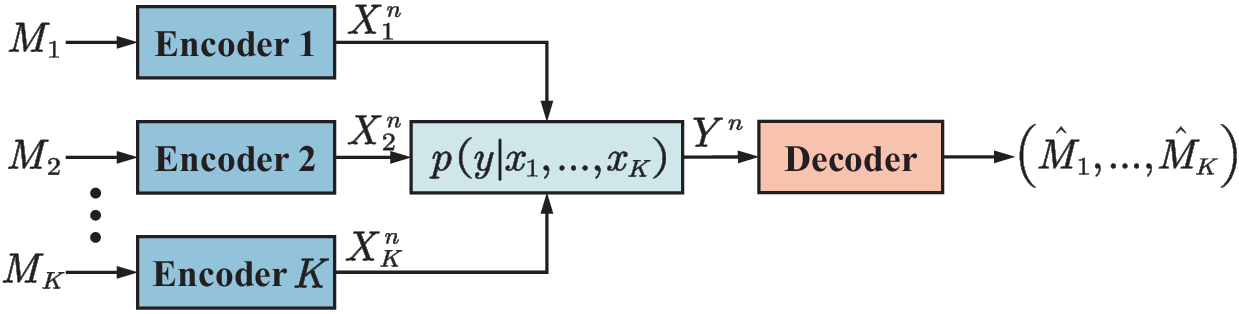}
  \caption{Illustration of the discrete memoryless MAC, where $p(y|x_1,\ldots,x_K)$ denotes the transition probability density function of the channel for each channel use, $\mathcal{X}_k$ is the input alphabet of user $k$, and $\mathcal{Y}$ is the output alphabet.}
  \label{Figure: DMGC_MAC_Coding}
\end{figure}

$\bullet$ \emph{Multiple Access Channel (Uplink NOMA)}: One of the most thoroughly understood multiuser channels is the MAC (i.e., uplink NOMA), as depicted in {\figurename} {\ref{Figure: DMGC_MAC_Coding}}, where $K\geq2$ users each simultaneously attempt to send a message to a common receiver. These users must contend not only with the receiver noise but also with interference from one another.

A $((2^{n {\mathsf{R}}_1},\ldots,2^{n {\mathsf{R}}_K}),n)$ code for the discrete memoryless MAC is characterized by $K$ message sets ${\mathscr{M}}_k=\{1,\ldots,2^{\lceil n {\mathsf{R}}_k\rceil}\}$ for $k\in{\mathcal{K}}=\{1,\ldots,K\}$, $K$ codebooks $\{{\mathcal{C}}_k\}_{k=1}^{K}$ each comprising $2^{\lceil n {\mathsf{R}}_k\rceil}$ \emph{codewords} of the form $X_k^{n}=[{\mathsf{x}}_{k,1},\ldots,{\mathsf{x}}_{k,n}]\in{\mathcal{X}}_k^n$, $K$ encoding function $f_{\rm{en}}^{(k)}:{\mathscr{M}}_k\mapsto{\mathcal{C}}_k$, and a decoding function $f_{\rm{de}}:{\mathcal{Y}}^{n}\mapsto{\mathscr{M}}_1\times\ldots\times{\mathscr{M}}_K$. The error probability is given by $\epsilon=\Pr(f_{\rm{de}}(Y^n)\ne (M_1,\ldots,M_K))$. A $K$-tuple of rates $\bm{\mathsf{R}}=({\mathsf{R}}_1,\ldots,{\mathsf{R}}_K)$ is \emph{achievable} if there exists a sequence of $((2^{n {\mathsf{R}}_1},\ldots,2^{n {\mathsf{R}}_K}),n)$ codes with rates approaching $\bm{\mathsf{R}}$ and vanishing $\epsilon$. The MAC \emph{capacity region} is the closure of the set of all achievable rate tuples, which was established by Ahlswede \cite{ahlswede1971multi} and Liao \cite{liao1972coding,liao1972multiple}.
\vspace{-5pt}
\begin{theorem}\label{Shannon_MAC_Capacity_Theorem}
The capacity region of the discrete memoryless MAC is given by the convex hull of the union of the sets
\begin{equation}\label{Channel_Capacity_MAC_DM}
\begin{split}
\Big\{\bm{\mathsf{R}}:\sum\nolimits_{k\in{\mathscr{K}}}{\mathsf{R}}_k\leq
I(X(\mathscr{K});Y|X({\mathscr{K}}^{\rm{c}})),\forall {\mathscr{K}}\subseteq{\mathcal{K}}\Big\},
\end{split}
\end{equation}
where the MI expressions are computed with respect to the joint distribution $p(y|x_1,\ldots,x_K)\prod_{k=1}^{K}p_k(x_k)$. Here, $X(\mathscr{K})=\{X_k:k\in{\mathscr{K}}\}$ for any set of indices $\mathscr{K}$, ${\mathscr{K}}^{\rm{c}}\in{\mathcal{K}}\backslash \mathscr{K}$, $\prod_{k=1}^{K}p_k(x_k)$ is the input distribution, and $p(y|x_1,\ldots,x_K)$ is the channel transition probability density function.
\end{theorem}
\vspace{-5pt}
\begin{IEEEproof}
The proof of Theorem \ref{Shannon_MAC_Capacity_Theorem} involves establishing the achievability and converse.
\subsubsection*{Achievability}
The MAC capacity region exhibits a polymatroidal structure in $K$-dimensional space, whose achievability can be demonstrated through the utilization of \emph{point-to-point random coding} and \emph{joint typical decoding}. 

Fix $\prod_{k=1}^{K}p_k(x_k)$. For each user $k\in{\mathcal{K}}$, \emph{randomly} generate $2^{\lceil n {\mathsf{R}}_k\rceil}$ i.i.d. sequences $X_k^n(M_k)$ for $M_k\in{\mathscr{M}}_k$, according to the probability distribution $\prod\nolimits_{i=1}^{n}p_k(\mathsf{x}_{k,i})$. 

Joint typical decoding can be realized in either a parallel (simultaneous decoding) or sequential (SIC decoding) manner. Simultaneous decoding searches all the message tuples within ${\mathscr{M}}_1\times\ldots\times{\mathscr{M}}_K$ to declare that a message tuple $(\hat{M}_1,\ldots,\hat{M}_K)$ has been sent, realizing each point in the capacity region. In contrast, SIC decoding decodes messages transmitted by distinct users in a sequential manner, wherein each successfully decoded message serves as auxiliary \emph{a priori} information for the subsequent decoding process. SIC decoding can realize the $K!$ \emph{corner points of the capacity region} and the remaining region is established through \emph{time-sharing among these corner points} \cite{el2011network}. The error probability satisfies $\lim_{n\rightarrow\infty}\epsilon=0$ for both simultaneous and SIC decoding strategies.
\begin{figure}[!t]
  \centering
  \includegraphics[width=0.45\textwidth]{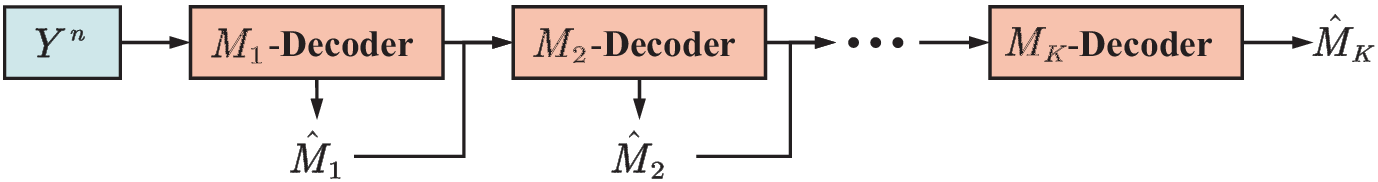}
  \caption{Illustration of SIC decoding for decoding order $1\rightarrow2\rightarrow\ldots\rightarrow K$.}
  \label{Figure: DM_MAC_SIC}
\end{figure}
\subsubsection*{Converse}
The proof is based on Fano’s inequality and the data processing inequality; see \cite{el2011network} for more details.
\end{IEEEproof}
For illustration, the SIC precoding procedure is shown in {\figurename} {\ref{Figure: DM_MAC_SIC}}. As can be observed, SIC decoding reduces a complex multiuser detection problem to a series of single-user detection steps \cite{guess1996multiuser}. Besides, the complexity associated with simultaneous decoding scales exponentially with $K$, while the complexity of SIC decoding scales linearly with $K$. Given a decoding order $\pi(1)\rightarrow\ldots\rightarrow\pi(K)$, the corner point of the capacity region achieved by SIC decoding is characterized by
\begin{align}
\bm{\mathsf{R}}:{\mathsf{R}}_{\pi(k)}\leq
I(X_{\pi(k)};Y|X_{\pi(k+1)},\ldots,X_{\pi(K)}),\forall k\in{\mathcal{K}}.
\end{align}

In 1971, Ahlswede \cite{ahlswede1971multi} characterized the discrete memoryless MAC capacity region for two (and three) users. In the same year, van der Meulen \cite{van1971discrete} proposed inner and outer bounds for the MAC capacity region. Subsequently, Liao \cite{liao1972coding,liao1972multiple} provided other simple characterizations, and Ahlswede \cite{ahlswede1974capacity} extended these findings. However, it was not clear how many input distributions needed consideration in computing each point on the region boundary. Slepian and Wolf \cite{slepian1973coding} addressed this by utilizing an auxiliary random variable with bounded cardinality to characterize the capacity region of the two-user MAC. This method was later extended by Csiszár and Körner \cite{csiszar2011information}. The original proof of achievability involved SIC decoding \cite{ahlswede1971multi}, with simultaneous decoding first introduced by El Gamal and Cover \cite{el1980multiple}. The strong converse was established by Dueck \cite{dueck1981strong}. The capacity region of the Gaussian MAC was contributed by Cover \cite{cover1975some} and Wyner \cite{wyner1974recent}. The polymatroidal structure of the MAC capacity region was studied by Han \cite{te1979capacity} and Tse and Hanly \cite{tse1998multiaccess}. For those interested in the early literature on the MAC, comprehensive surveys can be found in \cite{van1977survey,van1985recent}.
\begin{figure}[!t]
  \centering
  \includegraphics[width=0.45\textwidth]{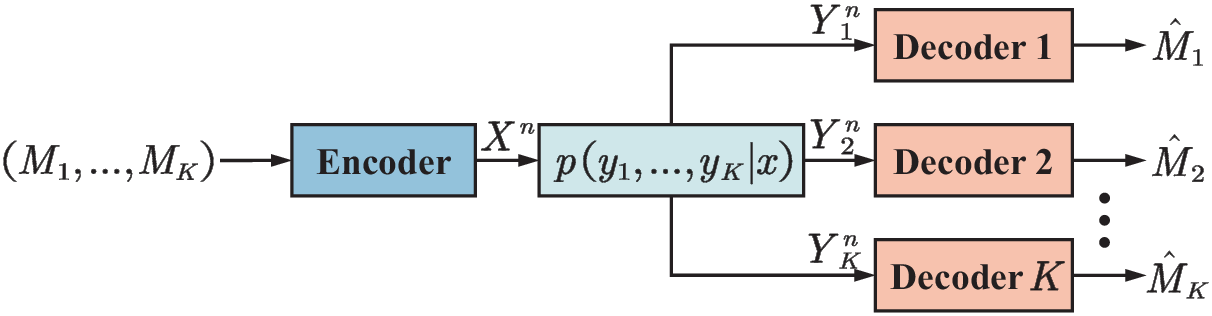}
  \caption{Illustration of the discrete memoryless BC, where $p(y_1,\ldots,y_K|x)$ denotes the transition probability density function of the channel for each channel use, $\mathcal{X}$ is the input alphabet, and $\mathcal{Y}_k$ is the output alphabet of user $k$.}
  \label{Figure: DMGC_BC_Coding}
\end{figure}

$\bullet$ \emph{Broadcast Channel (Downlink NOMA)}: The BC (i.e., downlink NOMA) comprises one transmitter and $K\geq2$ receivers, as illustrated in {\figurename} {\ref{Figure: DMGC_BC_Coding}}. A $((2^{n {\mathsf{R}}_1},\ldots,2^{n {\mathsf{R}}_K}),n)$ code for the discrete memoryless BC is characterized by $K$ message sets ${\mathscr{M}}_k=\{1,\ldots,2^{\lceil n {\mathsf{R}}_k\rceil}\}$ for $k\in{\mathcal{K}}$, a codebook ${\mathcal{C}}$ comprising $2^{\sum_{k=1}^{K}\lceil n {\mathsf{R}}_k\rceil}$ \emph{codewords} of the form $X^{n}=[{\mathsf{x}}_{1},\ldots,{\mathsf{x}}_{n}]\in{\mathcal{X}}^n$, an encoding function $f_{\rm{en}}:{\mathscr{M}}_1\times\ldots\times{\mathscr{M}}_K\mapsto{\mathcal{C}}$, and $K$ decoding functions $f_{\rm{de}}^{(k)}:{\mathcal{Y}}_k^{n}\mapsto{\mathscr{M}}_k$ for $k\in{\mathcal{K}}$. The error probability for user $k$ is given by $\epsilon_k=\Pr(f_{\rm{de}}^{(k)}(Y_k^n)\ne M_k)$. A $K$-tuple of rates $\bm{\mathsf{R}}=({\mathsf{R}}_1,\ldots,{\mathsf{R}}_K)$ is \emph{achievable} if there exists a sequence of $((2^{n {\mathsf{R}}_1},\ldots,2^{n {\mathsf{R}}_K}),n)$ codes with rates approaching $\bm{\mathsf{R}}$ and vanishing $\epsilon_k$ for $k\in{\mathcal{K}}$. The BC \emph{capacity region} is the closure of the set of all achievable rate tuples. Unlike the MAC, the capacity region of the BC has not been fully characterized, and remains an open problem. Marton \cite{marton1979coding} proposed a general achievable BC rate region, yet its status as the actual capacity region remains unverified. Despite this, various coding schemes and corresponding inner bounds on the capacity region have been suggested, proving to be tight in certain scenarios, such as the degraded BC.
\subsubsection*{Degraded Broadcast Channel}
\begin{figure}[!t]
  \centering
  \includegraphics[width=0.4\textwidth]{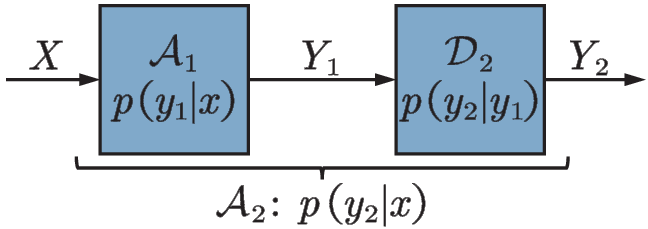}
  \caption{${\mathcal{A}}_2$ is a degraded version of ${\mathcal{A}}_1$, where $p(\cdot|\cdot)$ denotes the transition probability density function of the channel for each channel use.}
  \label{Figure: Def_Deg}
\end{figure}

In a degraded BC, one receiver is statistically stronger than the other \cite{bergmans1973random}. We define a channel ${\mathcal{A}}_2$ as a degraded version of a channel ${\mathcal{A}}_1$ if there exists a third channel ${\mathcal{D}}_2$ such that ${\mathcal{A}}_2$ can be represented as the cascade of ${\mathcal{A}}_1$ and ${\mathcal{D}}_2$, as illustrated in {\figurename} {\ref{Figure: Def_Deg}}. This degradation concept is attributed to Shannon \cite{shannon1958note}.
\begin{figure}[!t]
  \centering
  \includegraphics[width=0.45\textwidth]{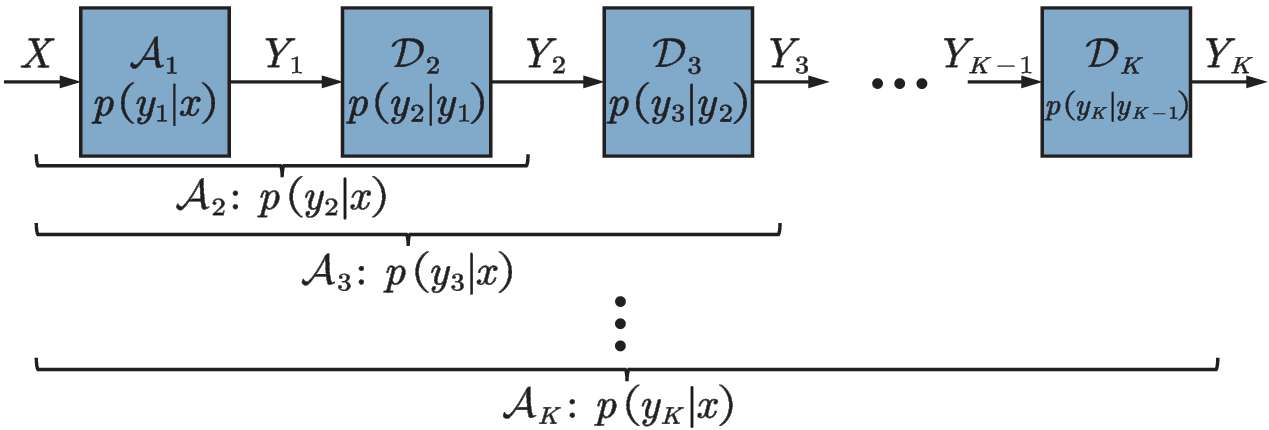}
  \caption{Illustration of the degraded BC, where $p(\cdot|\cdot)$ denotes the transition probability density function of the channel for each channel use.}
  \label{Figure: DM_BC_Deg}
\end{figure}
A degraded BC can be represented as a cascade formed by the best channel ${\mathcal{A}}_{1}$, followed by successive degrading channels ${\mathcal{D}}_2,\ldots,{\mathcal{D}}_K$, as depicted in {\figurename} {\ref{Figure: DM_BC_Deg}}, where $X\rightarrow Y_1\rightarrow\ldots\rightarrow Y_K$ forms a Markov chain. The capacity region of the degraded BC was established by Bergmans \cite{bergmans1973random} and Gallager \cite{gallager1974capacity}, rooted in enabling a ``stronger'' user to recover the message of the ``weaker'' user.
\vspace{-5pt}
\begin{theorem}\label{Shannon_BC_Capacity_Theorem}
Consider a discrete memoryless degraded BC, where $X\rightarrow Y_1\rightarrow\ldots\rightarrow Y_K$ forms a Markov chain. The capacity region is given by the closure of the union of the sets
\begin{equation}\label{Channel_Capacity_BC_DM}
\begin{split}
\{{\bm{\mathsf{R}}}:{\mathsf{R}}_k\leq I(X_k;Y_k|X_{k+1}),\forall k\in{\mathcal{K}}\}
\end{split}
\end{equation}
over all input probability distributions $p_{1|2}(x|x_2)\times p_{2|3}(x_2|x_3)\times\ldots \times p_{K-1|K}(x_{K-1}|x_{K})\times p_K(x_{K})$. Here, $X=X_1$ and $X_{K+1}=\varnothing$.
\end{theorem}
\vspace{-5pt}
\begin{IEEEproof}
The proof of Theorem \ref{Shannon_BC_Capacity_Theorem} involves establishing the achievability and converse.
\begin{figure}[!t]
  \centering
  \includegraphics[width=0.4\textwidth]{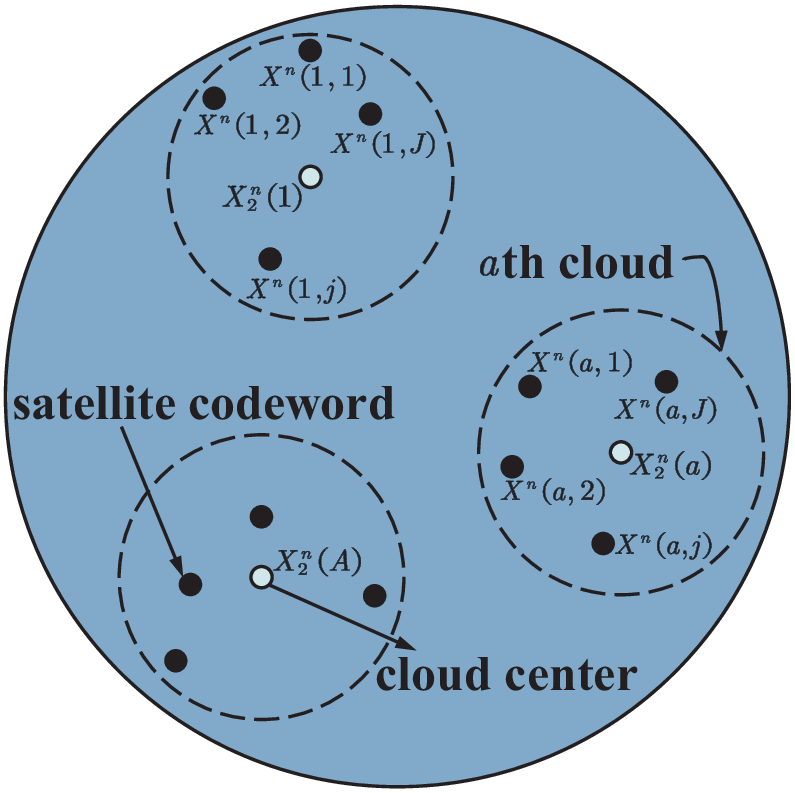}
  \caption{Illustration of the clouds and cloud centers of a broadcast code, where $A=2^{\lceil n {\mathsf{R}}_2\rceil}$ and $J=2^{\lceil n {\mathsf{R}}_1\rceil}$.}
  \label{Figure: SC_Cloud}
\end{figure}
\subsubsection*{Achievability}
Achievability is demonstrated through \emph{SPC} and \emph{joint typical decoding}.

For a two-user degraded BC, where user 1 is statistically stronger than user 2, SPC is implemented in three steps: \romannumeral1) fixing $p(x_2)p_{1|2}(x|x_2)$, \romannumeral2) randomly generating $2^{\lceil n {\mathsf{R}}_2\rceil}$ i.i.d. $X_2^n$ sequences (cloud centers) with $n$ letters independently drawn according to $p(x_2)$, and \romannumeral3) for each cloud center, appending $2^{\lceil n {\mathsf{R}}_1\rceil}$ i.i.d. $X^n$ sequences (satellites) with $n$ letters independently drawn according to $p_{1|2}(x|x_2)$ conditioned on the cloud center $X_2^{n}$. Here, $X_2^{n}(M_2)$ plays the role of the cloud center identifiable to both users, while $X^n(M_1,M_2)$ is the $M_1$th satellite codeword in the $M_2$th cloud, as depicted in {\figurename} {\ref{Figure: SC_Cloud}}. The above procedure extends directly to the $K$-user case, as shown in {\figurename} {\ref{Figure: SC_DG_BC}}. 
\begin{figure*}[!b]
  \centering
  \includegraphics[width=0.95\textwidth]{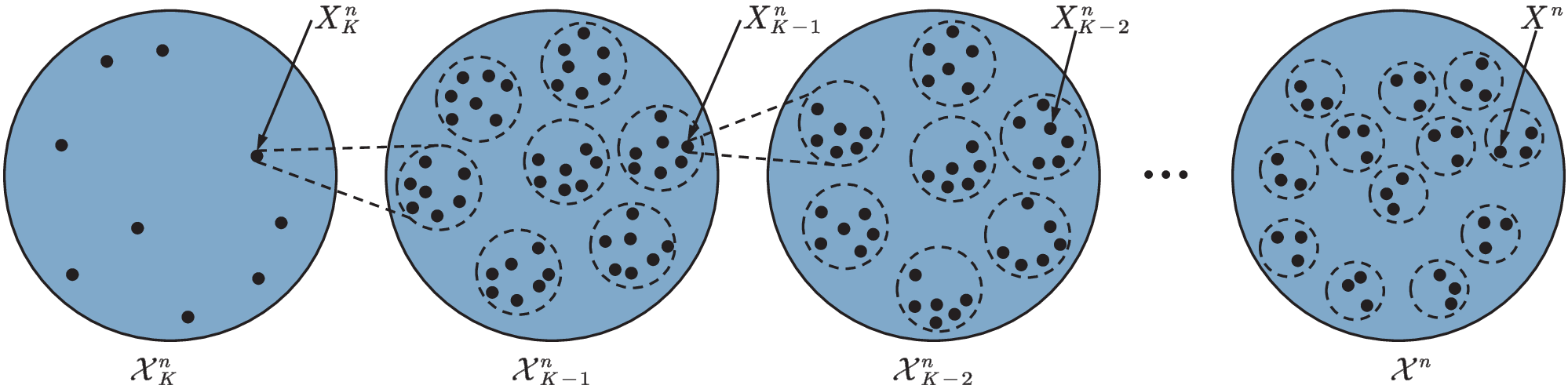}
  \caption{Illustration of SPC for the degraded BC, where $\mathcal{X}_k$ denotes the alphabet of $X_k$ for $k=2,\ldots,K$.}
  \label{Figure: SC_DG_BC}
\end{figure*}
\begin{figure}[!t]
  \centering
  \includegraphics[width=0.45\textwidth]{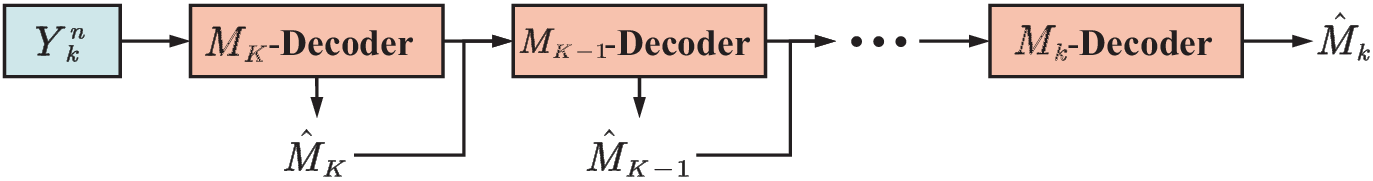}
  \caption{Illustration of SIC decoding for the degraded BC.}
  \label{Figure: SIC_DG_BC}
\end{figure}

The capacity region is established by enabling the stronger user to decode both his own message and the messages of other weaker users. Specifically, for user $k$, the decoder aims to recover $(M_k,M_{k+1},\ldots,M_K)$. Similar to the MAC, joint typical decoding at user $k$ can be realized through either simultaneous decoding or SIC decoding (shown in {\figurename} {\ref{Figure: SIC_DG_BC}}). The error probability for both simultaneous and SIC decoding is proven to satisfy $\lim_{n\rightarrow\infty}\epsilon=0$ \cite{el2011network}.
\subsubsection*{Converse}
The converse proof leverages Mrs. Gerber’s lemma \cite{wyner1973theorem} and the entropy power inequality \cite{shannon1948mathematical}. Further details are provided in \cite{el2011network}.
\end{IEEEproof}
The capacity region characterized in Theorem \ref{Shannon_BC_Capacity_Theorem} was conjectured by Cover \cite{cover1972broadcast}, proved to be achievable by Bergmans \cite{bergmans1973random}, and the converse was established by Bergmans \cite{bergmans1974simple}, Gallager \cite{gallager1974capacity}, and Wyner \cite{wyner1973theorem}. Bergmans' proof specifically applied to the Gaussian channel, while Gallager's proof does not apply to the Gaussian channel with a power constraint, and Wyner's proof applied to the binary symmetric BC \cite{wyner1973theorem}. The strong converse for the degraded BC capacity region was established by Ahlswede and Korner \cite{ahlswede1975source}, and further extended by Willems \cite{willems1990maximal}. To gain a comprehensive understanding of the BC, the surveys provided by van der Meulen \cite{van1977survey,van1981recent} and Cover \cite{cover1998comments} are recommended.
\subsubsection{Key Technologies of Power-Domain NOMA}\label{Section: Key Technologies of Power-Domain NOMA}
Section \ref{Section: Shannon Channels, Multiple Access Channels, and Broadcast Channels} has emphasized the critical role of SPC and SIC in downlink and uplink NOMA. These foundational principles act as the linchpin for unlocking the capacity potential of these channels, forming the core tenets of power-domain NOMA techniques. The literature review in Section \ref{Section: Shannon Channels, Multiple Access Channels, and Broadcast Channels} reveals that these techniques are not novel; their exploration predates their formal integration into the NOMA framework. The subsequent sections provide a concise review of SPC/SIC.
\subsubsection*{Superposition Coding}
The ingenious concept of SPC was initially proposed by Cover in 1972 \cite{cover1972broadcast}, based on which the OMA rate region was proven to be a subset of the NOMA capacity region. At its core, SPC involves encoding a message for a user experiencing suboptimal channel conditions at a lower rate and subsequently ``superimposing'' the signal of a user with superior channel conditions. This fundamental idea has permeated various communication channels, including interference channels \cite{carleial1978interference}, relay channels \cite{cover1979capacity}, MACs \cite{grant2001rate}, and wiretap channels \cite{csiszar1978broadcast}. While these theoretical explorations laid a robust foundation, the transition of SPC from theory to practice necessitated additional research efforts \cite{zhang2010unified,vanka2012superposition,xiong2015open}. Vanka \emph{et al.} \cite{vanka2012superposition} developed an experimental platform utilizing a software-defined radio (SDR) system to scrutinize the performance of SPC. Another open-source SDR-based prototype was reported in \cite{xiong2015open}.
\subsubsection*{Successive Interference Cancellation}
While SPC constructs capacity-achieving codes for downlink NOMA, SIC decoding has emerged as a crucial element applicable to both downlink and uplink NOMA. The significance of SIC decoding for achieving channel capacity was demonstrated by Ahlswede in 1971 \cite{ahlswede1971multi} for MACs (uplink NOMA) and later by Bergmans in 1973 \cite{bergmans1973random} for degraded BCs (downlink NOMA). Diverging from simultaneous decoding approaches that consider all users concurrently, SIC decoding operates iteratively, presenting a notable reduction in receiver hardware complexity \cite{andrews2005interference,hanzo2007ofdm}. The adoption of SIC decoding represents a distinctive feature distinguishing NOMA from OMA, although it is noted that SIC had been considered in earlier 3G and 4G research phases within the realms of interference cancellation and receiver designs \cite{li2010mimo}. Beyond theoretical investigations, SIC has found extensive applications in practical systems, such as CDMA \cite{patel1994analysis}, SDMA \cite{hanzo2007ofdm}, and Vertical-Bell Laboratories Layered Space-Time (V-BLAST) \cite{wolniansky1998v}. Furthermore, SIC has been harnessed across diverse scenarios, including multiuser MIMO networks \cite{gelal2012topology}, multihop networks \cite{jiang2012squeezing}, random access systems \cite{xu2013decentralized}, and in large-scale networks modeled by stochastic geometry \cite{lee2012spectrum}. A pivotal aspect validating the practical utility of SIC is its successful integration into commercial systems, such as IEEE 802.15.4. Additionally, real-world applications leverage SIC-aided spatial division multiplexing (SDM) detectors in SDM-assisted OFDM systems \cite{hanzo2005ofdm}. 
\subsection{Information-Theoretic Limits of Power-Domain NOMA}\label{Section: Information-Theoretic Limits of Power-Domain NOMA}
Having established the foundational principles of power-domain NOMA, our exploration now shifts to the examination of its information-theoretic capacity limits, with a specific focus on discrete-time Gaussian channels. Throughout our discussion, the channels are assumed to be time-invariant and known to both the transmitters and the receiver. 
\subsubsection{Gaussian Scalar Channels}\label{Section: Single-Input Single-Output (SISO) Case}
We begin by considering the Gaussian scalar channel, i.e., the single-input single-output (SISO) channel. The superiority of NOMA over OMA was established in Cover's seminal paper in 1972, which states that downlink NOMA contains the achievable rate region of OMA as a subset \cite{cover1972broadcast, tse2005fundamentals, goldsmith2005wireless}. Motivated by these findings, we now explore the information-theoretic limits of the Gaussian scalar BC, i.e., downlink NOMA.

\begin{figure}[!t]
  \centering
  \includegraphics[width=0.45\textwidth]{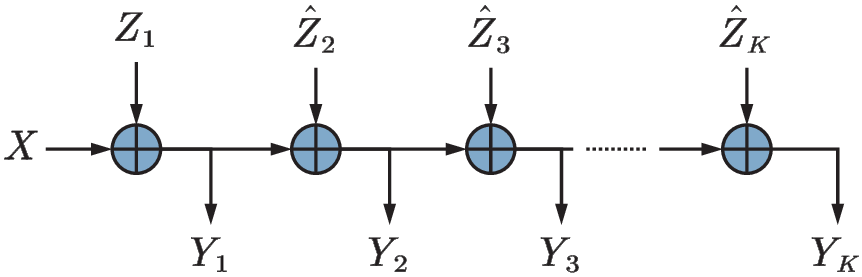}
  \caption{Representation of the discrete-time Gaussian BC as a degraded BC, where $\hat{Z}_k\sim{\mathcal{CN}}(N_k-N_{k-1})$ for $k\in{\mathcal{K}}$.}
  \label{Figure: GBC_Deg}
\end{figure}

\begin{figure*}[!t]
  \centering
  \includegraphics[width=0.95\textwidth]{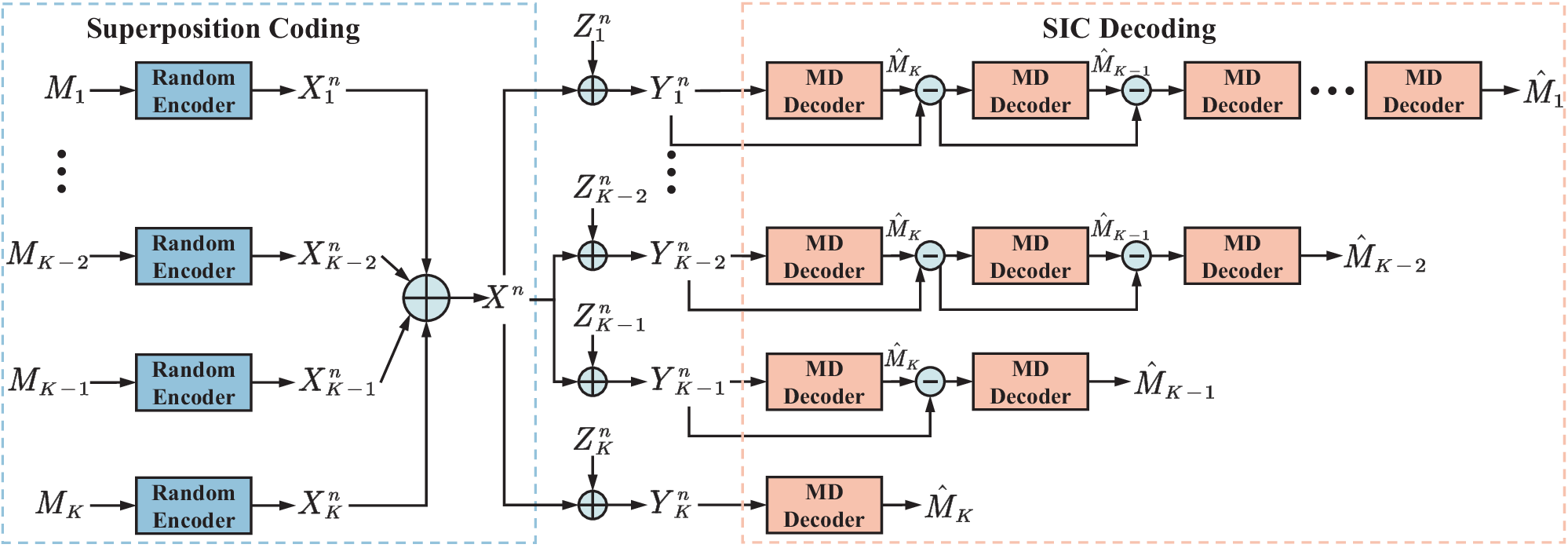}
  \caption{Illustration of SPC and SIC decoding for the Gaussian BC.}
  \label{Figure: SC_SIC_DG_GBC}
\end{figure*}

$\bullet$ \emph{Gaussian Scalar Broadcast Channel (Downlink NOMA)}: The discrete-time Gaussian scalar BC is represented by
\begin{equation}
Y_k=X+Z_k,~k\in{\mathcal{K}},
\end{equation}
where $Z_k\sim{\mathcal{CN}}(0,N_k)$ denotes AWGN. The sequence $X^n=({\mathsf{x}}_{1},\ldots,{\mathsf{x}}_{n})$ adheres to a power constraint $\frac{1}{n}\sum\nolimits_{i=1}^{n}\lvert {\mathsf{x}}_{i}\rvert^2\leq P$. Without loss of generality, we assume that the noise powers satisfy $N_1\leq N_2\ldots\leq N_K$. Consequently, the Gaussian BC can be viewed as a degraded channel, as illustrated in {\figurename} {\ref{Figure: GBC_Deg}}. 
The capacity region of the Gaussian scalar BC was established by Bergmans \cite{bergmans1974simple} in 1974, and is characterized by the set of rate tuples $({\mathsf{R}}_1,\ldots,{\mathsf{R}}_K)$ such that
\begin{align}\label{Gaussian_BC_Capacity_Region}
{\mathsf{R}}_k\leq
{\mathsf{C}}\left(\frac{\alpha_k P}{N_k + \sum_{k'<k}\alpha_{k'}P}\right),~k\in{\mathcal{K}}
\end{align}
with $\alpha_k\geq 0$, $\sum_{k=1}^{K}\alpha_k=1$, and ${\mathsf{C}}(x)=\log_2(1+x)$. The SPC and SIC strategy to achieve a coding rate tuple $({\mathsf{R}}_1,\ldots,{\mathsf{R}}_K)$ in the Gaussian BC is illustrated in {\figurename} {\ref{Figure: SC_SIC_DG_GBC}}. Notice that the SPC and SIC decoding for Gaussian channels can be implemented via \emph{mathematical addition and subtraction of codewords}\footnote{This principle applies specifically to Gaussian channels. However, for general discrete memoryless channels, the simplification of SPC and SIC through mathematical addition and subtraction of codewords is not possible.}. Concerning the sum-rate capacity, it is obtained by assigning all power to the strongest user. Thus, achieving the sum-rate capacity does not lead to user fairness \cite{el2011network}.

\begin{figure}[!t]
  \centering
  \includegraphics[width=0.45\textwidth]{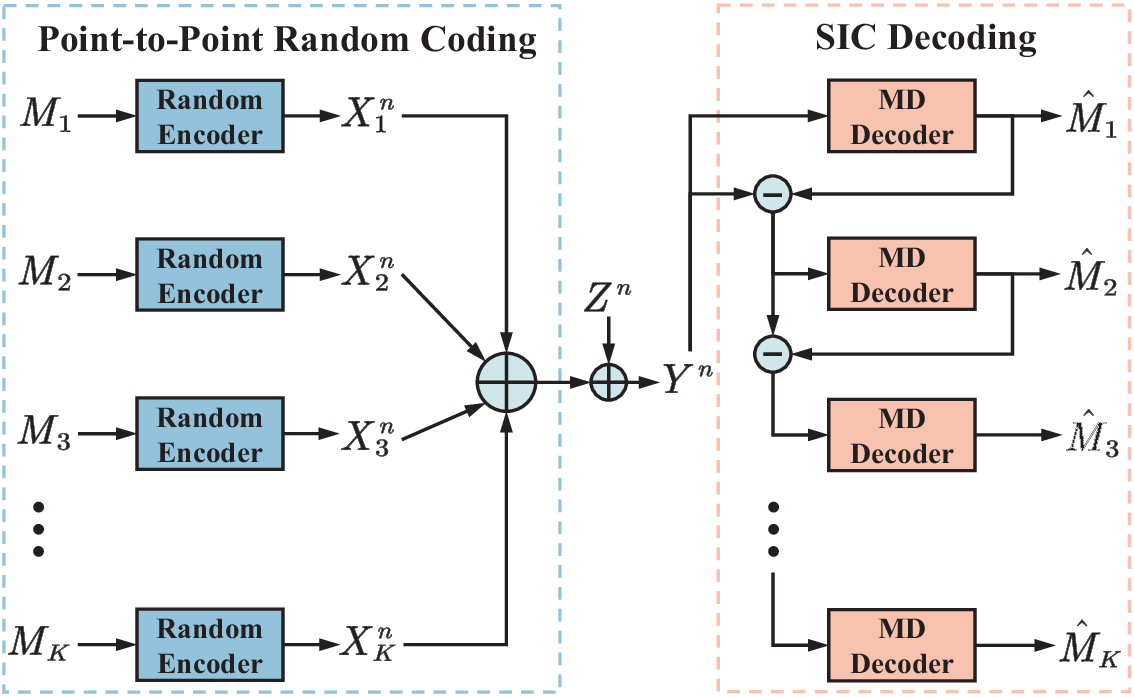}
  \caption{Illustration of point-to-point random coding and SIC decoding for the Gaussian MAC, where the SIC decoding order is $1\rightarrow2\rightarrow\ldots\rightarrow K$.}
  \label{Figure: SIC_DG_GMAC}
\end{figure}

$\bullet$ \emph{Gaussian Scalar Multiple Access Channel (Uplink NOMA)}:
The discrete-time Gaussian MAC is modeled as follows:
\begin{equation}
Y=\sum\nolimits_{k=1}^{K}X_k+Z,
\end{equation}
where $Z\sim{\mathcal{CN}}(0,N)$ is AWGN. The sequence $X_k^n=({\mathsf{x}}_{k,1},\ldots,{\mathsf{x}}_{k,n})$ adheres to a power constraint $\frac{1}{n}\sum\nolimits_{i=1}^{n}\lvert {\mathsf{x}}_{i,k}\rvert^2\leq P_k$. The capacity region is the set of rate tuples $({\mathsf{R}}_1,\ldots,{\mathsf{R}}_K)$ such that
\begin{align}\label{Gaussian_MAC_Capacity_Region}
\sum\nolimits_{k\in{\mathscr{K}}}{\mathsf{R}}_k\leq
{\mathsf{C}}\Big(\sum\nolimits_{k\in{\mathscr{K}}}P_k/N\Big),~\forall {\mathscr{K}}\subseteq{\mathcal{K}}.
\end{align}
Moreover, the point-to-point Gaussian random coding and SIC decoding procedure for achieving a coding rate tuple $({\mathsf{R}}_1,\ldots,{\mathsf{R}}_K)$ is depicted in {\figurename} {\ref{Figure: SIC_DG_GMAC}}. Regarding the sum-rate capacity, it is achieved when each user transmits at full power. Note that the $K!$ corner points of the capacity region yields the same sum-rate regardless of the decoding order \cite{el2011network}.
\subsubsection{Gaussian Vector Channels}
We now explore the capacity limits of Gaussian vector channels, i.e., multiuser multiple-input single-output (MU-MISO) and MU-MIMO channels. In the following discussion, we consider a narrowband single-cell multiuser system, where a BS with $N_{\mathsf{BS}}$ antennas serves $K$ users. Each user $k\in{\mathcal{K}}$ is equipped with $N_k$ antennas, and the user $k$-to-BS channel matrix is denoted as ${\mathbf{H}}_k\in{\mathbbmss{C}}^{N_{\mathsf{BS}}\times N_k}$.

$\bullet$ \emph{Gaussian Vector Multiple Access Channel (Uplink NOMA)}: Under our specified system configuration, the memoryless Gaussian vector MAC can be represented as follows:
\begin{align}
{\mathbf{y}}=\sum\nolimits_{k=1}^{K}{\mathbf{H}}_k\mathbf{x}_k+{\mathbf{n}},
\end{align}
where $\mathbf{x}_k\in{\mathbbmss{C}}^{N_k\times1}$ is the input vector signal for user $k\in{\mathcal{K}}$, ${\mathbf{y}}\in{\mathbbmss{C}}^{N_{\mathsf{BS}}\times1}$ is the output vector signal, and ${\mathbf{n}}\sim{\mathcal{CN}}({\mathbf{0}},\sigma^2{\mathbf{I}}_{\mathsf{BS}})$ is the AWGN vector with noise power $\sigma^2$. Let ${\bm\Sigma}_k={\mathbbmss{E}}\{{\mathbf{x}}_k{\mathbf{x}}_k^{\mathsf{H}}\}\in{\mathbbmss{C}}^{N_k\times N_k}$ be the covariance matrices of $\mathbf{x}_k$, subject to the power constraint ${\mathsf{tr}}({\bm\Sigma}_k)\leq P_k$.

For the Gaussian vector MAC, the input distribution that achieves the capacity region is a Gaussian distribution with ${\mathbf{x}}_k\sim{\mathcal{CN}}({\mathbf{0}},{\bm\Sigma}_k)$. Given $({\bm\Sigma}_1,\ldots,{\bm\Sigma}_K)$, the achievable region is expressed as follows:
\begin{equation}\label{Gaussian_Vector_MAC_Rate_Region}
\begin{split}
&\{({\mathsf{R}}_1,\ldots,{\mathsf{R}}_K):\sum\nolimits_{k\in{\mathscr{K}}}{\mathsf{R}}_k\\
&\leq\log_2\det\Big({\mathbf{I}}+\frac{1}{\sigma^2}\sum\nolimits_{k\in{\mathscr{K}}}{\mathbf{H}}_k{\bm\Sigma}_k{\mathbf{H}}_k^{\mathsf{H}}\Big),\forall {\mathscr{K}}\subseteq{\mathcal{K}}\},
\end{split}
\end{equation}
which corresponds to a $K$-dimensional polyhedron. The capacity region is the union of all such polyhedrons, considering all covariance matrices satisfying the trace constraints. SIC decoding achieves the corner points of each polyhedron, where users' signals are successively decoded and subtracted from the received signal. Concerning the sum-rate capacity, it is defined as follows:
\begin{equation}\label{Gaussian_Vector_MAC_Sum_Rate}
\begin{split}
{\mathsf{C}}_{\mathsf{MAC}}=&\max_{\{{\bm\Sigma}_k\}_{k=1}^{K}:{\bm\Sigma}_k\succeq{\mathbf{0}},{\mathsf{tr}}({\bm\Sigma}_k)\leq P_k}\\
&\log_2\det\Big({\mathbf{I}}+\frac{1}{\sigma^2}\sum\nolimits_{k=1}^{K}{\mathbf{H}}_k{\bm\Sigma}_k{\mathbf{H}}_k^{\mathsf{H}}\Big),
\end{split}
\end{equation}
which can be efficiently solved by the iterative water-filling method \cite{yu2004iterative}.

$\bullet$ \emph{Gaussian Vector Broadcast Channel (Downlink NOMA)}: Without loss of generality, the uplink and downlink channels are assumed to be reciprocal. Thus, the memoryless Gaussian vector BC can be represented as follows:
\begin{align}
{\mathbf{y}}_k={\mathbf{H}}_k^{\mathsf{H}}\mathbf{x}+{\mathbf{n}}_k,~k\in{\mathcal{K}},
\end{align}
where $\mathbf{x}\in{\mathbbmss{C}}^{N_{\mathsf{BS}}\times1}$ is the transmitted vector signal, ${\mathbf{y}}_k\in{\mathbbmss{C}}^{N_{k}\times1}$ is the received vector signal at user $k\in{\mathcal{K}}$, and ${\mathbf{n}}_k\sim{\mathcal{CN}}({\mathbf{0}},\sigma_k^2{\mathbf{I}}_{N_k})$ is the AWGN vector with noise power $\sigma_k^2$. In the case of a single-antenna transmitter, the Gaussian BC becomes a degraded BC, allowing the users to be unequivocally ranked by their channel strength, and the capacity region is well-known \cite{bergmans1974simple}. However, when the transmitter has multiple antennas, the Gaussian BC is generally non-degraded, as users receive signals of different strengths from distinct transmit antennas. Hence, achieving the capacity region through SPC-SIC becomes challenging. Although the capacity region for general non-degraded BCs is unknown, it has been proved that the capacity region of the Gaussian vector BC is achieved via DPC \cite{costa1983writing}. 

DPC operates under the premise that if the BS possesses perfect, non-causal knowledge of the additive Gaussian interference in the channel, then the channel's capacity remains unaffected, similar to a scenario with no interference or where the user also has knowledge of the interference. DPC enables the BS to ``pre-subtract'' known interference without increasing the transmit power. In the Gaussian BC, DPC is applied when choosing codewords for different users. The BS first picks a codeword (i.e., ${\mathbf{x}}_1$) for user 1. The BS then chooses a codeword for user 2 (i.e., $\mathbf{x}_2$) with full (noncausal) knowledge of the codeword intended for user 1. Therefore, the codeword of user 1 can be ``pre-subtracted'' such that user 2 does not see the codeword intended for user 1 as interference. Similarly, the codeword for user 3 is chosen such that user 3 does not see the signals intended for users 1 and 2 (i.e., ${\mathbf{x}}_1+{\mathbf{x}}_2$ ) as interference. This process continues for all users. 

Given a preset DPC encoding order $\pi(1)\rightarrow\ldots\rightarrow\pi(K)$, the achievable rate tuple $({\mathsf{R}}_1,\ldots,{\mathsf{R}}_K)$ is given by
\begin{equation}
\begin{split}
{\mathsf{R}}_{\pi(k)}&=\log_2\det\Big({\mathbf{I}}_{N_k}+\mathbf{H}_{\pi(k)}^{\mathsf{H}}{\bm\Xi}_{\pi(k)}\mathbf{H}_{\pi(k)}\\
&\times\Big(\sigma_k^2{\mathbf{I}}_{N_k}+\sum\nolimits_{k'>k}{\bm\Xi}_{\pi(k')}\Big)^{-1}\Big),
\end{split}
\end{equation}
where ${\bm\Xi}_k\succeq{\mathbf{0}}$ is a semi-definite covariance matrix used to generate the Gaussian codebook of user $k$, subject to the sum-power constraint $\sum_{k=1}^{K}{\mathsf{tr}}({\bm\Xi}_k)\leq P$. The capacity region is defined as the convex hull of the union of all such rate tuples over all feasible covariance matrices $\{{\bm\Xi}_k\}_{k=1}^{K}$ and over all permutations of the encoding order. Regarding the sum-rate capacity, it can be expressed as follows:
\begin{equation}\label{Gaussian_Vector_BC_Sum_Rate}
\begin{split}
{\mathsf{C}}_{\mathsf{BC}}=\max_{\{\pi(k)\}_{k=1}^{K},\{{\bm\Sigma}_k\}_{k=1}^{K}:{\bm\Sigma}_k\succeq{\mathbf{0}},\sum_{k=1}^{K}{\mathsf{tr}}({\bm\Xi}_k)\leq P}\sum\nolimits_{k=1}^{K}{\mathsf{R}}_{\pi(k)}.
\end{split}
\end{equation}
This non-convex maximization problem can be efficiently solved by employing the BC–MAC duality \cite{vishwanath2003duality} along with the sum-power iterative water-filling algorithm \cite{jindal2005sum}.

The Gaussian vector channel was first studied by Telatar \cite{telatar1999capacity} and Foschini \cite{foschini1996layered}. The capacity region of the Gaussian vector MAC was established by Cheng and Verdú \cite{cheng1993gaussian}. The iterative water-filling algorithm was developed by Yu \emph{et al.}. Caire and Shamai \cite{caire2003achievable} proved that DPC achieves the sum-rate capacity of the two-user MISO BC. The sum-rate optimality of DPC was extended to the MU-MISO channel by Viswanath and Tse \cite{viswanath2003sum} and to the more general case by Vishwanath \emph{et al.} \cite{vishwanath2003duality} using the BC–MAC duality and Yu and Cioffi \cite{yu2004sum} using a minimax argument. The equivalence between the DPC rate region and the capacity region of the Gaussian vector BC was established by Weingarten \emph{et al.} \cite{weingarten2006capacity} using the technique of channel enhancement. Surveys of the literature on Gaussian vector channels can be found in \cite{biglieri2007mimo,goldsmith2003capacity,heath2018foundations}.
\vspace{-5pt}
\begin{remark}
In this paper, the term ``NOMA'' specifically refers to technologies that employ linear encoding schemes. While DPC is an important capacity-achieving non-orthogonal transmission strategy for MA, it relies on non-linear encoding and is therefore not classified as NOMA. We discuss DPC here to emphasize that the capacity region of Gaussian vector BCs is well-established. Generally, NOMA using SPC and SIC only achieves the capacity of Gaussian vector BCs in special cases, such as those discussed in\cite{chen2016application}. 
\end{remark}
\vspace{-5pt}

\begin{table*}[!t]
\centering
\caption{Summary of capacity results for downlink (DL) and uplink (UL) NOMA for discrete memoryless, Gaussian scalar, and Gaussian vector channels.}
\label{tab: Section: Information-Theoretic Limits of Power-Domain NOMA}
\resizebox{0.99\textwidth}{!}{
\begin{tabular}{|l|l|lll|l|}
\hline
\multicolumn{1}{|c|}{\multirow{3}{*}{\textbf{Direction}}} & \multicolumn{1}{c|}{\multirow{3}{*}{\textbf{Scenarios}}} & \multicolumn{3}{c|}{\textbf{Capacity Region}}                                                                            & \multicolumn{1}{c|}{\multirow{3}{*}{\textbf{Sum-Rate Capacity}}} \\ \cline{3-5}
\multicolumn{1}{|c|}{}                           & \multicolumn{1}{c|}{}                           & \multicolumn{2}{c|}{\textbf{Achievability}}                             & \multicolumn{1}{c|}{\multirow{2}{*}{\textbf{Converse}}} & \multicolumn{1}{c|}{}                                   \\ \cline{3-4}
\multicolumn{1}{|c|}{}                           & \multicolumn{1}{c|}{}                           & \multicolumn{1}{c|}{\textbf{Coding}} & \multicolumn{1}{c|}{\textbf{Decoding}}    & \multicolumn{1}{c|}{}                          & \multicolumn{1}{c|}{}                                   \\ \hline
DL                                               & DMC MU-SISO                                      & \multicolumn{3}{l|}{Unknown}                                                                                    & Unknown                                                    \\ \hline
DL                                               & Degraded DMC MU-SISO                             & \multicolumn{1}{l|}{SPC plus random coding}     & \multicolumn{1}{l|}{SIC Decoding}         & {\cite{ahlswede1975source}}                                        & Exhaustive Search                                                     \\ \hline
UL                                               & DMC MU-SISO                                         & \multicolumn{1}{l|}{Point-to-point
random coding} & \multicolumn{1}{l|}{SIC plus time-sharing}         & {\cite{ahlswede1971multi}}                                          & Exhaustive Search                                                     \\ \hline
DL                                               & Gaussian MU-SISO                                & \multicolumn{1}{l|}{SPC plus random coding}     & \multicolumn{1}{l|}{SIC Decoding}         & {\cite{bergmans1974simple}}                                        & Allocate all power to the strongest user                                                       \\ \hline
UL                                               & Gaussian MU-SISO                                & \multicolumn{1}{l|}{Point-to-point
random coding} & \multicolumn{1}{l|}{SIC plus time-sharing}         & {\cite{cover1975some,wyner1974recent}}                                        & Each user transmits at full power                                                       \\ \hline
DL                                               & Gaussian MU-MIMO                                & \multicolumn{1}{l|}{DPC}    & \multicolumn{1}{l|}{MD Decoding} & {\cite{weingarten2006capacity}}                                        & BC–MAC duality plus iterative water-filling                                                       \\ \hline
UL                                               & Gaussian MU-MIMO                                & \multicolumn{1}{l|}{Point-to-point
random coding} & \multicolumn{1}{l|}{SIC plus time-sharing}         & {\cite{cheng1993gaussian}}                                        & Iterative water-filling                                                       \\ \hline
\end{tabular}}
\end{table*}

\subsubsection*{Summary}
Numerous numerical findings in existing literature demonstrate that the capacity region of OMA is encompassed by that of NOMA for both uplink and downlink transmissions \cite{liu2017nonorthogonal,liu2022evolution,dai2018survey}. Table \ref{tab: Section: Information-Theoretic Limits of Power-Domain NOMA} provides a summary of the prevailing contributions to the information-theoretic capacity limits of power-domain NOMA\footnote{For discrete memoryless channels, an efficient algorithm for computing the sum-rate capacity is lacking, as the corresponding problem is generally non-convex \cite{el2011network}.}. The results in Table \ref{tab: Section: Information-Theoretic Limits of Power-Domain NOMA} suggest that, for all the considered uplink NOMA channels, SIC decoding is crucial for achieving the capacity region. In contrast, for downlink NOMA channels, SPC-SIC can only achieve the capacity region of degraded discrete memoryless channels, including Gaussian MU-SISO channels. In the case of Gaussian MU-MIMO channels, the realization of the capacity region is credited to the utilization of DPC. While DPC can be heuristically viewed as a ``pre-SIC'' coding scheme, it has been demonstrated that the achievable region with SIC decoding is included in the DPC region \cite{jafar2004phantomnet}. Although DPC is a potent capacity-achieving scheme, its practical implementation proves challenging. Consequently, non-DPC downlink NOMA\footnote{As mentioned earlier, DPC is a capacity-achieving but non-linear encoding scheme, and thus does not fall within the scope of NOMA. The motivation for employing MIMO-NOMA lies in the pursuit of low-complexity transmission strategies that can closely approach the capacity limits achieved by DPC.} transmission schemes hold practical significance. This motivates research on power-domain MIMO-NOMA, incorporating downlink beamforming and SPC-SIC to approach the capacity limits achieved by DPC.
\subsection{MIMO-NOMA}\label{Section: MIMO-NOMA} 
The focus in this subsection is on the downlink MIMO-NOMA. In contrast to SISO-NOMA, where transceivers are single-antenna devices, MIMO-NOMA enhances the capabilities of both the BS and users by deploying multiple antennas. This enhancement necessitates the design of transmit/receive beamformers at the BS and/or users \cite{liu2018multiple,huang2018signal,tian2019multiple}, which is elaborated on below. On the one hand, the beamformer plays a crucial role in shaping the signal-to-interference-plus-noise ratio (SINR) for each user in MIMO-NOMA. On the other hand, in MIMO-NOMA, where each user receives signals of varying strengths from distinct transmit antennas, the effective channel gain of each user is influenced by both the wireless propagation environment and the associated transmit/receive beamformer, which further influences the SIC decoding order. This complexity contrasts with SISO-NOMA, where the decoding order is solely determined by ordering the user channel gains. These considerations have motivated a large body of works on the optimization and analysis of MIMO-NOMA. Specifically, two primary beamforming strategies have emerged for MIMO-NOMA: \emph{beamformer-based MIMO-NOMA} and \emph{cluster-based MIMO-NOMA} \cite{liu2018multiple}. The key difference between these two structures is whether each beamformer serves multiple users or one user.

\begin{figure*}[!t]
    \centering
    \subfigure[SDMA.]{
        \includegraphics[width=0.315\textwidth]{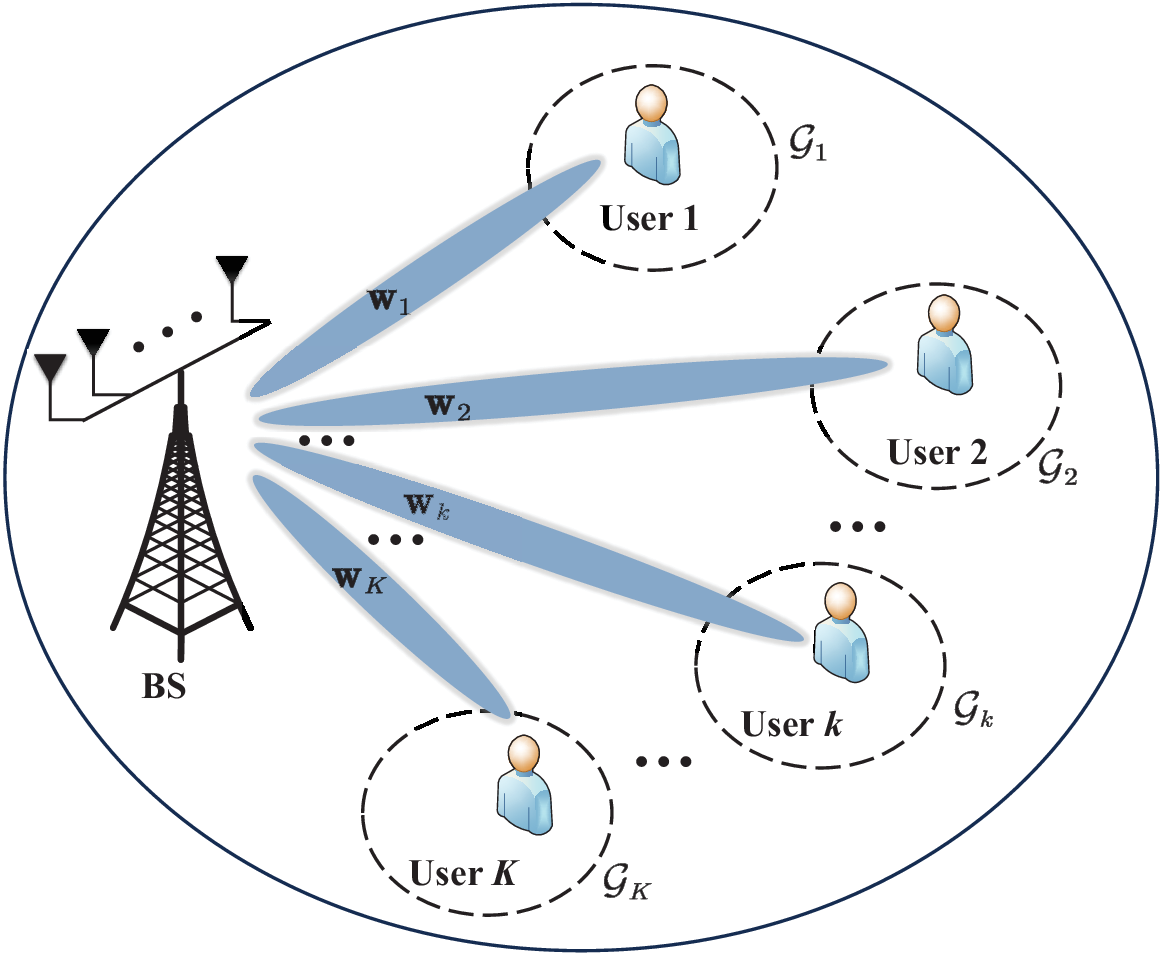}
        \label{Figure: SDMA_System}	
    }
    \subfigure[Beamformer-Based NOMA.]{
        \includegraphics[width=0.315\textwidth]{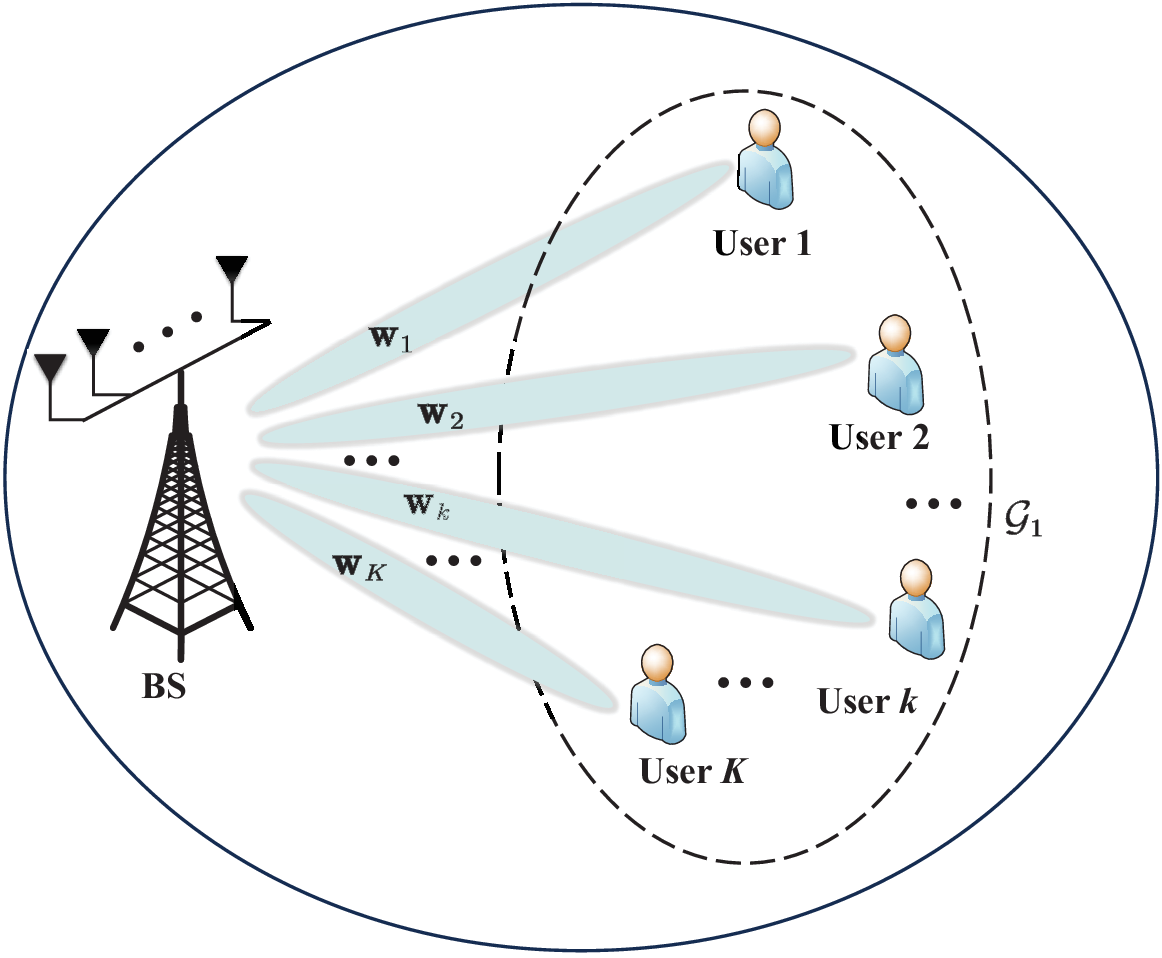}
        \label{Figure: BB_NOMA_System}
    }
    \subfigure[Cluster-Based NOMA.]{
        \includegraphics[width=0.315\textwidth]{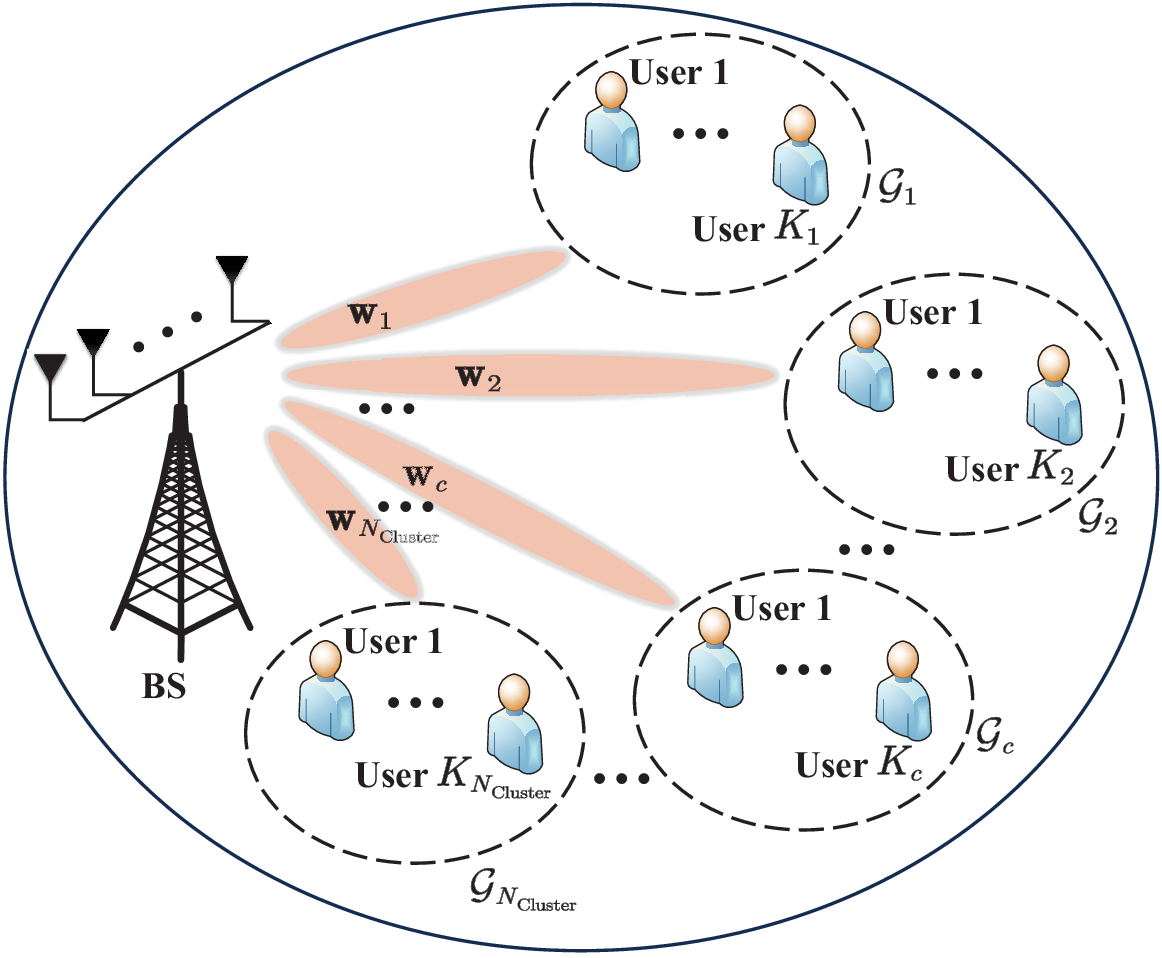}
        \label{Figure: CB_NOMA_System}	
    }
    \caption{Illustration of SDMA, beamformer-based MIMO-NOMA, and cluster-based MIMO-NOMA.}
\end{figure*}

\subsubsection{Beamformer-Based MIMO-NOMA}
Beamformer-based MIMO-NOMA ({\figurename} {\ref{Figure: BB_NOMA_System}}) is inspired by SDMA ({\figurename} {\ref{Figure: SDMA_System}}), both allocating a linear beamformer to each user\footnote{For simplicity of presentation, we focus on the case where each user is allocated only one data stream.}. However, SIC decoding is not integrated into SDMA. To illustrate this, consider a $K$-user MISO channel, where the BS is equipped with $N_{\mathsf{BS}}$ antennas. The signal transmitted by the BS for both NOMA and SDMA can be expressed as follows:
\begin{align}\label{MISO_NOMA_Bemformer_Signal}
{\mathbf{x}}=\sum\nolimits_{k=1}^{K}{\mathbf{w}}_k\sqrt{p_k}s_k,
\end{align} 
where ${\mathbf{w}}_k\in{\mathbbmss{C}}^{N_{\mathsf{BS}}\times1}$ is the normalized beamformer for user $k$, $p_k$ is user $k$'s transmit power, and $s_k\in{\mathbbmss{C}}$ is the associated normalized signal. In the absence of SIC decoding in SDMA, the transmission rate of user $k$ is given by
\begin{align}\label{SDMA_User_k_Rate}
{\mathsf{R}_k}=\log_2\left(1+\frac{\lvert{\mathbf{h}}_k^{\mathsf{H}}{\mathbf{w}}_k\rvert^2p_k}{\sum_{k'\ne k}\lvert{\mathbf{h}}_k^{\mathsf{H}}{\mathbf{w}}_{k'}\rvert^2p_{k'}+\sigma_k^2}\right),
\end{align}
where ${\mathbf{h}}_k\in{\mathbbmss{C}}^{N_{\mathsf{BS}}\times1}$ is the channel vector, and $\sigma_k^2$ is the noise power. SDMA beamformers are designed to enhance the desired message power at each user while mitigating IUI to enhance individual SINR as well as overall system throughput. 

Contrastingly, in MIMO-NOMA, the SIC decoding order is crucial. Let the binary variable $\alpha_{k,k'}\in\{0,1\}$, $\forall k\ne k'$, specify the SIC decoding order of users $k$ and $k'$. $\alpha_{k,k'}=0$ means that user $k$ will carry out SIC to first decode user $k'$'s signal; otherwise, $\alpha_{k,k'}=1$. Accordingly, we have the condition $\alpha_{k,k'}+\alpha_{k',k}=1$ since it is in general impossible to mutually carry out SIC at both users. Adhering to SIC principles, the rate at which user $k''$ decodes user $k$'s message is given by
\begin{align}
{\mathsf{R}_{k'',k}}=\log_2\left(1+\frac{\lvert{\mathbf{h}}_{k''}^{\mathsf{H}}{\mathbf{w}}_{k}\rvert^2p_k}{\sum_{k'\ne k}\alpha_{k,k'}
\lvert{\mathbf{h}}_{k''}^{\mathsf{H}}{\mathbf{w}}_{k'}\rvert^2p_{k'}+\sigma_{k''}^2}\right),
\end{align}
and the transmission rate of user $k$ is given by
\begin{align}\label{NOMA_User_k_Rate}
{\mathsf{R}_{k,k}}=\log_2\left(1+\frac{\lvert{\mathbf{h}}_{k}^{\mathsf{H}}{\mathbf{w}}_{k}\rvert^2p_k}{\sum_{k'\ne k}\alpha_{k,k'}
\lvert{\mathbf{h}}_{k}^{\mathsf{H}}{\mathbf{w}}_{k'}\rvert^2p_{k'}
+\sigma_k^2}\right).
\end{align} 
Successful SIC is guaranteed only if the condition 
\begin{align}\label{SIC_Ordering_Condition}
{\mathsf{R}_{k'',k}}>{\mathsf{R}_{k,k}},\forall k''\ne k,\alpha_{k'',k}=0
\end{align}
is satisfied. This emphasizes that NOMA beamformers must not only improve the SINR but also ensure the success of SIC, distinguishing it from SDMA. A comparison of \eqref{SDMA_User_k_Rate} with \eqref{NOMA_User_k_Rate} reveals that NOMA results in less IUI than SDMA, and this benefit extends to both underloaded/critically loaded and overloaded regimes.

Despite the above advantage, the design of NOMA beamforming, whether for a fixed SIC decoding order or jointly optimized with the SIC ordering, poses a challenging task, necessitating the development of new methodologies. Assuming a fixed SIC order, various efficient NOMA beamforming algorithms have been proposed to improve the sum-rate \cite{hanif2015minorization}, minimum power consumption \cite{chen2016optimal}, outage probability \cite{chen2016application}, and energy efficiency \cite{al2019energy}, among others. Notably, the authors of \cite{chen2016optimal} established a quasi-degraded channel condition for a two-user MISO-NOMA channel, which was later extended to the multiuser case \cite{zhu2020optimal}. Under this condition, an optimal beamforming strategy was formulated to minimize the transmit power under user rate constraints, demonstrating a performance comparable to the optimal DPC-based scheme. Additionally, some researchers explored the joint optimization of NOMA beamforming and SIC ordering; see \cite{zakeri2021robust} for more details. The previously mentioned works assumed a single data stream is allocated to each user. To overcome this limitation, a generalized singular value decomposition (GSVD) beamforming-based framework for a two-user MIMO-NOMA system was reported. This framework utilizes GSVD to decompose MIMO channels into multiple SISO channels \cite{ma2016general,chen2018asymptotic}. This approach effectively reduces the computational complexity of user detection while necessitating that each user employs the same number of antennas as the BS and requires matrix inversions. As a further improvement, a simultaneous triangularization (ST)-based beamforming method was proposed \cite{krishnamoorthy2020precoder,krishnamoorthy2021uplink}. This method employs ST to decompose MIMO channels, eliminating the requirements on specific antenna numbers and matrix inversion operations.
\subsubsection{Cluster-Based MIMO-NOMA}
Another popular beamforming strategy for MIMO-NOMA is termed cluster-based MIMO-NOMA ({\figurename} {\ref{Figure: CB_NOMA_System}}). In this approach, users are divided into distinct clusters/groups, each characterized by similar spatial attributes. Within a cluster, all users employ a common beamformer. Through the application of well-designed transmit and receive beamforming techniques, inter-cluster interference can be effectively eliminated or suppressed. Simultaneously, intra-cluster interference is mitigated through SIC decoding. The transmitted signal in cluster-based MIMO-NOMA is given by
\begin{align}
{\mathbf{x}}=\sum\nolimits_{c=1}^{N_{\mathsf{Cluster}}}{\mathbf{w}}_c\sum\nolimits_{k=1}^{K_c}\sqrt{p_{c,k}}s_{c,k},
\end{align}
where $N_{\mathsf{Cluster}}>1$ is the number of clusters, ${\mathbf{w}}_c\in{\mathbbmss{C}}^{N_{\mathsf{BS}}\times1}$ is the normalized beamformer for cluster $c$, $s_{c,k}\in{\mathbbmss{C}}$ is the normalized signal dedecated to user $k$ in cluster $c$ (denoted as user ${\mathcal{U}}_k^{c}$), and $p_{c,k}$ is the associated transmit power. Let ${\mathcal{G}}_c$ denote the set of users in cluster $c=1,\ldots,N_{\mathsf{cluster}}$ subject to $\sum_{c=1}^{N_{\mathsf{cluster}}}\lvert{\mathcal{G}}_c\rvert=K$, where $\lvert{\mathcal{G}}_c\rvert$ calculates the cardinality of ${\mathcal{G}}_c$. Let the binary variable $\alpha_{k,k'}^{c}\in\{0,1\}$, $\forall k\ne k'$, $c=1,\ldots,N_{\mathsf{Cluster}}$, specify the SIC decoding order of users ${\mathcal{U}}_k^{c}$ and ${\mathcal{U}}_{k'}^{c}$. $\alpha_{k,k'}^{c}=0$ means that user ${\mathcal{U}}_k^{c}$ will carry out SIC to first decode user ${\mathcal{U}}_{k'}^{c}$'s signal; otherwise, $\alpha_{k,k'}^{c}=1$. Accordingly, we have
the condition $\alpha_{k,k'}^{c}+\alpha_{k',k}^{c}=1$ since it is in general impossible to mutually carry out SIC at both users.

Let ${\mathbf{h}}_{c,k}\in{\mathbbmss{C}}^{N_{\mathsf{BS}}\times1}$ denote the user ${\mathcal{U}}_k^{c}$-to-BS channel vector. Then, the received signal of user ${\mathcal{U}}_k^{c}$ can be written as follows:
\begin{equation}
\begin{split}
y_{c,k}&={\mathbf{h}}_{c,k}^{\mathsf{H}}{\mathbf{w}}_c\sqrt{p_{c,k}}s_{c,k}+{\underbrace{{\mathbf{h}}_{c,k}^{\mathsf{H}}{\mathbf{w}}_c\sum\nolimits_{k'\ne k}\sqrt{p_{c,k'}}s_{c,k'}}_{\mathsf{intra-cluster~interference}}}\\
&+{\underbrace{\sum\nolimits_{c'\ne c}{\mathbf{h}}_{c,k}^{\mathsf{H}}{\mathbf{w}}_{c'}\sum\nolimits_{k=1}^{K_{c'}}\sqrt{p_{c',k}}s_{c',k}}_{\mathsf{inter-cluster~interference}}}+n_{c,k},
\end{split}
\end{equation}
where $n_{c,k}\sim{\mathcal{CN}}(0,\sigma_{c,k}^2)$ is AWGN with noise power $\sigma_{c,k}^2$. Upon observing $y_{c,k}$, user ${\mathcal{U}}_k^{c}$ will employ SIC decoding to detect the message sent by user ${\mathcal{U}}_{k'}^{c}$ with $\alpha_{k,k'}^{c}=0$ for $k'\ne k$, and then remove the corresponding message from its observation. The remaining intra-cluster interference after SIC can be expressed as follows:
\begin{align}
{\mathbf{h}}_{c,k}^{\mathsf{H}}{\mathbf{w}}_c\sum\nolimits_{k'\ne k}\alpha_{k,k'}^{c}\sqrt{p_{c,k'}}s_{c,k'}.
\end{align}
As a result, the achievable rate of user ${\mathcal{U}}_k^{c}$ is given by
\begin{equation}\label{NOMA_Cluster_User_k_Rate}
\begin{split}
{\mathsf{R}_k}=\log_2\left(\!\!1+\frac{p_{c,k}\lvert{\mathbf{h}}_{c,k}^{\mathsf{H}}{\mathbf{w}}_{c}\rvert^2}{\lvert{\mathbf{h}}_{c,k}^{\mathsf{H}}{\mathbf{w}}_{c}\rvert^2\overline{p}_{c,k}
+\sum\limits_{c'\ne c}\lvert{\mathbf{h}}_{c,k}^{\mathsf{H}}{\mathbf{w}}_{c'}\rvert^2\overline{p}_{c'}+\sigma_{c,k}^2}\!\!\right),
\end{split}
\end{equation}
where $\overline{p}_{c,k}=\sum_{k'\ne k}\alpha_{k,k'}^{c}p_{c,k'}$ and $\overline{p}_{c'}=\sum_{k=1}^{K_{c'}}p_{c',k}$. It is worth noting that successful SIC is assured under a rate condition akin to \eqref{SIC_Ordering_Condition}, considering both intra-cluster and inter-cluster interference. Compared to beamformer-based MIMO-NOMA, cluster-based MIMO-NOMA exhibits reduced channel ordering complexity since SIC decoding is confined to individual clusters. Consequently, this strategy is particularly well-suited for scenarios characterized by extensive user overload.

One straightforward approach to beamformer design for each cluster is applying the zero-forcing (ZF) criterion to eliminate inter-cluster interference. Under this criterion, the beamformer ${\mathbf{w}}_{c}$ for each cluster $c$ is designed as follows:
\begin{equation}\label{Cluster_NOMA_ZF_MISO}
\left\{
\begin{array}{ll}
{\mathbf{h}}_{c,k}^{\mathsf{H}}{\mathbf{w}}_{c}\ne 0             & k=1,\ldots,K_c\\
{\mathbf{h}}_{c',k}^{\mathsf{H}}{\mathbf{w}}_{c}=0           & c'\ne c,~k=1,\ldots,K_{c'}
\end{array} \right..
\end{equation}
Consider an ideal case where the spatial channel vectors of all users within each cluster are parallel to each other, i.e., matrix $[{\mathbf{h}}_{c,1} \ldots {\mathbf{h}}_{c,K_c}]\in{\mathbbmss{C}}^{N_{\mathsf{BS}}\times K_c}$ has rank $1$ for $c=1,\ldots,N_{\mathsf{cluster}}$. In this case, the condition $N_{\mathsf{BS}}\geq N_{\mathsf{cluster}}$ has to be satisfied to ensure the feasibility of \eqref{Cluster_NOMA_ZF_MISO}. On the contrary, in the worst-case scenario, the spatial channel vectors of all users within each cluster are mutually independent, i.e., matrix $[{\mathbf{h}}_{c,1} \ldots {\mathbf{h}}_{c,K_c}]\in{\mathbbmss{C}}^{N_{\mathsf{BS}}\times K_c}$ has full column rank for $c=1,\ldots,N_{\mathsf{cluster}}$. In this case, the condition $N_{\mathsf{BS}}\geq \max_{c}\sum_{c'\ne c}K_{c'}$ has to be satisfied to ensure the feasibility of the beamformer design. This is because we need to solve a system of $\sum_{c'\ne c}K_{c'}$ linear equations to obtain ${\mathbf{w}}_c$.

If each user has multiple antennas, say $N_{\mathsf{U}}$, ZF equalization can be implemented at each user to eliminate inter-cluster interference. Let ${\mathbf{H}}_{c,k}\in{\mathbbmss{C}}^{N_{\mathsf{BS}}\times N_{\mathsf{U}}}$ and ${\mathbf{v}}_{c,k}\in{\mathbbmss{C}}^{N_{\mathsf{U}}\times1}$ denote the user $k$-to-BS channel matrix and the detection vector of user ${\mathcal{U}}_k^{c}$, respectively. Then, the equalizer ${\mathbf{v}}_{c,k}$ for each user ${\mathcal{U}}_k^{c}$ should be designed as follows:
\begin{equation}\label{Cluster_NOMA_ZF_MIMO_1}
\left\{
\begin{array}{ll}
{\mathbf{v}}_{c,k}^{\mathsf{H}}{\mathbf{H}}_{c,k}^{\mathsf{H}}{\mathbf{w}}_{c}\ne 0             & \\
{\mathbf{v}}_{c,k}^{\mathsf{H}}{\mathbf{H}}_{c,k}^{\mathsf{H}}{\mathbf{w}}_{c'}=0           & c'\ne c
\end{array} \right.,
\end{equation}
which can be realized if $N_{\mathsf{U}}\geq N_{\mathsf{Cluster}}$ \cite{ding2015application}. To further reduce the antenna number at each user, a signal alignment-based MIMO-NOMA framework was introduced, which yields $N_{\mathsf{U}}\geq \frac{N_{\mathsf{Cluster}}}{2}$ \cite{ding2016general}. In all cases, ZF beamforming or equalization completely cancels the inter-cluster interference, transforming the MIMO-NOMA system into several parallel SISO-NOMA systems with lower design complexity. Theoretical analyses suggest that cluster-based NOMA outperforms its OMA counterpart in terms of system throughput \cite{zeng2017sum,zeng2017capacity}. Besides, it is emphasized that cluster-based NOMA can enhance user fairness \cite{liu2018multiple}. Several user clustering algorithms, along with power allocation schemes, were proposed in \cite{liu2016fairness}, striking a balanced tradeoff between throughput and user fairness.

The primary limitation of the aforementioned scheme is constrained by the required relationship between $N_{\mathsf{U}}$, $N_{\mathsf{BS}}$, and $N_{\mathsf{Cluster}}$. This has motivated another cluster-based MIMO-NOMA design which tolerates a certain level of inter-cluster interference \cite{ali2016non}. The key advantage of this inter-cluster interference-tolerant design lies in its ability to operate without imposing constraints on the numbers of antennas at the BS and users. In this direction, several beamforming and user clustering methods have been proposed; for detailed discussions, we refer to \cite{ali2016non,cui2018outage,sun2018joint} and relevant works.
\subsubsection{From Conventional NOMA to Cluster-Free NOMA}\label{Section: From Conventional NOMA to Cluster-Free NOMA}
Beamformer-based MIMO-NOMA assigns all users to a single cluster, potentially giving rise to an overuse of SIC. Specifically, in scenarios where the users' channels exhibit low correlation, resulting in minimal IUI at each user, both the need for user ordering and SIC become negligible \cite{liu2022evolution}. In such cases, SDMA outperforms NOMA. Cluster-based MIMO-NOMA offers a partial solution to the overuse of SIC by assigning users to different clusters. This approach supports a large number of users with moderate SIC complexity. However, it relies on the assumption that users within the same cluster exhibit high channel correlations, while users in different clusters experience low channel correlations. This assumption may not always hold due to the inherent randomness of wireless channels. Both beamformer- and cluster-based MIMO-NOMA approaches are scenario-centric, and their effectiveness depends on the specific scenario encountered. Consequently, they may struggle to address the heterogeneous scenarios anticipated in future wireless networks. In response to this challenge, a unified cluster-free SIC-based MIMO-NOMA framework has been proposed \cite[Fig. 9]{liu2022evolution}.

For clarity, we consider a $K$-user MISO channel, where the BS is equipped with $N_{\mathsf{BS}}$ antennas. Under the framework of cluster-free SIC, the transmitted signal remains as given in \eqref{MISO_NOMA_Bemformer_Signal}. The $K$ users are categorized into $N_{\mathsf{Cluster}}$ clusters, where $N_{\mathsf{Cluster}}\in[1,K]$ is an optimizable integer. Let ${\mathcal{G}}_c$ denote the set of users in cluster $c=1,\ldots,N_{\mathsf{cluster}}$ subject to $\bigcup_{c=1}^{N_{\mathsf{cluster}}}{\mathcal{G}}_c={\mathcal{K}}$. Let the binary variable $\alpha_{k,k'}^{c}\in\{0,1\}$, $\forall k\ne k'\in{\mathcal{G}}_c$, $c=1,\ldots,N_{\mathsf{Cluster}}$, specify the SIC decoding order of users $k$ and $k'$ in cluster $c$. $\alpha_{k,k'}^{c}=0$ means that for cluster $c$, user $k$ will carry out SIC to first decode user $k'$'s signal; otherwise, $\alpha_{k,k'}^{c}=1$. Following the same steps as for deriving \eqref{NOMA_Cluster_User_k_Rate}, the achievable rate of user $k$ that is assumed to be grouped into cluster $c$ is given by
\begin{equation}\label{NGMA_Cluster_User_k_Rate}
\begin{split}
&{\mathsf{R}_k}=\log_2(1\\
&+\frac{\lvert{\mathbf{h}}_{k}^{\mathsf{H}}{\mathbf{w}}_{k}\rvert^2p_k}{\sum\limits_{k'\in{\mathcal{G}}_c,k'\ne k}\lvert{\mathbf{h}}_{k}^{\mathsf{H}}{\mathbf{w}}_{k'}\rvert^2\alpha_{k,k'}^{c}
p_{k'}\!+\!\sum\limits_{k'\in {\mathcal{K}}\backslash{\mathcal{G}}_c}\lvert{\mathbf{h}}_{k}^{\mathsf{H}}{\mathbf{w}}_{k'}\rvert^2p_{k'}\!+\!\sigma_{k}^2}\bigg).
\end{split}
\end{equation} 
We comment that successful SIC is assured under a rate condition akin to \eqref{SIC_Ordering_Condition}.
\begin{enumerate}
  \item[\romannumeral1)] When $N_{\mathsf{Cluster}}=K$, cluster-free NOMA reduces to SDMA ({\figurename} {\ref{Figure: SDMA_System}}), where each cluster contains only one user, and NOMA is not employed. In such instances, the transmission rate of each user is determined by \eqref{SDMA_User_k_Rate}. Note that SDMA is applicable only in the underloaded/critically loaded regime (i.e., $K\leq N_{\mathsf{BS}}$), where there are sufficient spatial degrees-of-freedom (DoFs) to design beamformers for suppressing IUI.   
  \item[\romannumeral2)] When $N_{\mathsf{Cluster}}=1$, cluster-free NOMA degenerates to beamformer-based NOMA ({\figurename} {\ref{Figure: BB_NOMA_System}}), where all users are grouped into a single cluster, and NOMA is employed. The transmission rate of each user is then given by \eqref{NOMA_User_k_Rate}. Beamformer-based NOMA is particularly beneficial in underloaded/critically loaded systems with strongly correlated channels and moderately overloaded systems (e.g., $N_{\mathsf{BS}}<K<2N_{\mathsf{BS}}$).    
  \item[\romannumeral3)] When $1<N_{\mathsf{Cluster}}<K$, cluster-free NOMA transforms into cluster-based NOMA ({\figurename} {\ref{Figure: CB_NOMA_System}}), where all users within the same cluster are allocated the same beamformer. The transmission rate of each user is given by \eqref{NOMA_Cluster_User_k_Rate}. Cluster-based NOMA is suitable for moderately overloaded systems with strongly correlated channels within the clusters and extremely overloaded systems (i.e., $K\gg N_{\mathsf{BS}}$). In this scenario, the number of clusters $N_{\mathsf{BS}}$, user clustering $\{{\mathcal{G}}_c\}$, SIC ordering $\{\alpha_{k,k'}^{c}\}$, power allocation $\{p_{c,k}\}$, and transmit beamforming $\{{\mathbf{w}}_c\}$ can be jointly designed to enhance system performance, such as the sum-rate; refer to \cite{xu2023cluster,xu2023distributed} for further details.
\end{enumerate}
The cluster-free NOMA framework encompasses existing transmission schemes as special cases, offering enhanced DoFs by combining the advantages of both multiple antennas and NOMA techniques. Regardless of whether the system is underloaded/critically loaded or overloaded, this framework enables a flexible transmission strategy by appropriately selecting $N_{\mathsf{Cluster}}$ and the associated user clustering, SIC ordering, and transmit beamforming. Further details and mathematical derivations on this framework are provided in \cite{liu2022evolution}. A compassion of SDMA, beamformer-based MIMO-NOMA, cluster-based MIMO-NOMA, and cluster-free MIMO-NOMA is presented in Table \ref{tab: Section: From Conventional NOMA to Cluster Free NOMA}.

\begin{table*}[!t]
\centering
\caption{Comparison of SDMA, beamformer-based MIMO-NOMA, and cluster-based MIMO-NOMA.}
\label{tab: Section: From Conventional NOMA to Cluster Free NOMA}
\resizebox{0.99\textwidth}{!}{
\begin{tabular}{|l|llll|}
\hline
                 & \multicolumn{1}{l|}{SDMA} & \multicolumn{1}{l|}{Beamformer-Based NOMA} & \multicolumn{1}{l|}{Cluster-Based NOMA} & Cluster Free NOMA \\ \hline
Feature          & \multicolumn{1}{l|}{$N_{\mathsf{Cluster}}=K$}    & \multicolumn{1}{l|}{$N_{\mathsf{Cluster}}=K$}                     & \multicolumn{1}{l|}{$1<N_{\mathsf{Cluster}}<K$}                  & $N_{\mathsf{Cluster}}\leq K$                 \\ \hline
Beamforming $\{{\mathbf{w}}_c\}$     & \multicolumn{1}{l|}{$\surd$}    & \multicolumn{1}{l|}{$\surd$}                     & \multicolumn{1}{l|}{$\surd$}                  & $\surd$                 \\ \hline
Power Allocation $\{p_{c,k}\}$& \multicolumn{1}{l|}{$\surd$}    & \multicolumn{1}{l|}{$\surd$}                     & \multicolumn{1}{l|}{$\surd$}                  & $\surd$                 \\ \hline
SIC Ordering $\{\alpha_{k,k'}^{c}\}$    & \multicolumn{1}{l|}{$\times$}    & \multicolumn{1}{l|}{$\surd$}                     & \multicolumn{1}{l|}{$\surd$}                  & $\surd$                 \\ \hline
User Clustering $\{{\mathcal{G}}_c\}$ & \multicolumn{1}{l|}{$\times$}    & \multicolumn{1}{l|}{$\times$}                     & \multicolumn{1}{l|}{$\surd$}                  & $\surd$                 \\ \hline
%Relationship     & ${\mathscr{C}}_{\mathsf{SDMA}}\subseteq{\mathscr{C}}_{\mathsf{BNOMA}}$ & ${\mathscr{C}}_{\mathsf{BNOMA}}$ & $\subseteq{\mathscr{M}}_{\mathsf{CNOMA}}$   & $\subseteq{\mathscr{M}}_{\mathsf{CNOMA}}$                                                                                                            \\ \hline
\end{tabular}}
\end{table*}
\subsection{Discussions and Outlook}
For both scalar and uplink vector channels, power-domain NOMA has been demonstrated to achieve capacity, approaching the limits via SPC-SIC. The research trajectory in downlink MIMO-NOMA, as discussed in Section \ref{Section: MIMO-NOMA}, has been motivated by the pursuit of low-complexity NOMA transmission frameworks that approach the performance of DPC. Simultaneously, novel NOMA transmission protocols are being designed to cater to heterogeneous applications, user quality-of-service (QoS), and diverse propagation environments. NOMA has played a pivotal role in inspiring and empowering various emerging wireless applications. These include unmanned aerial vehicle communications \cite{liu2019uav}, robotic communications \cite{liu2021robotic}, massive and critical machine-type communications \cite{elbayoumi2020noma}, as well as space/air/ground communications \cite{jaafar2020multiple,mu2020non}, among others. Additionally, NOMA has been co-designed with cutting-edge techniques, such as physical layer security \cite{pakravan2023physical}, short-package transmissions \cite{chen2023xurllc}, spatial modulation \cite{zhong2018spatial}, THz/mmWave technology \cite{ding2022design,ding2023joint}, grant-free connectivity \cite{ding2019simple,ding2021new}, visible light communications \cite{bawazir2018multiple}, random access \cite{choi2017noma}, and more. Comprehensive discussions on these topics can be found in several overview papers published in recent years \cite{maraqa2020survey,liu2022evolution,liu2017nonorthogonal}. Looking ahead, the evolution of power-domain NOMA is envisioned to involve integration with new requirements, tools, techniques, and applications. This integration will demand more efficient NOMA transmission protocols. For more insights, interested readers are directed to recent overview papers (e.g., \cite{liu2022developing}) covering this evolving landscape.

To delve deeper into the conceptual underpinnings of NOMA, it is essential to recognize it as a paradigm that fundamentally addresses interference management. Broadly, there exist four quintessential interference management approaches \cite{cadambe2008interference,goldsmith2003capacity}. 
\begin{enumerate}
  \item[\romannumeral1)] Orthogonalization: When the strength of interference is comparable to the desired signal, interference is avoided by orthogonalizing the channel access. This is the basis for OMA.  
  \item[\romannumeral2)] Decoding: When interference is stronger than the desired signal, the interfering signal can be decoded along with the desired signal, as seen in SIC decoding.    
  \item[\romannumeral3)] Treating interference as noise (TIN): When interference is weaker than the desired signal, the interfering signal can be treated as noise and single user encoding/decoding suffices.
  \item[\romannumeral4)] Encoding: When interference is non-causally known at the BS, the interference signal can be pre-subtracted through proper encoding techniques, such as DPC.
\end{enumerate}
A comparison of these four approaches is shown in Table \ref{tab: Section: Summary of the interference management approaches}.

\begin{table*}[!t]
\centering
\caption{Summary of the interference management approaches.}
\label{tab: Section: Summary of the interference management approaches}
\resizebox{0.99\textwidth}{!}{
\begin{tabular}{|l|l|l|l|l|l|}
\hline
                & \begin{tabular}[c]{@{}l@{}}\textbf{Interference}\\\textbf{Levels}\end{tabular} & \textbf{Characteristic}                                                     & \textbf{Advantages}                                                 & \textbf{Disadvantages}                                                             & \textbf{Examples}  \\ \hline
\textbf{Orthogonalization} & Medium              & \begin{tabular}[c]{@{}l@{}}Interference is \emph{avoided} by orthogonalizing \\ the channel access\end{tabular}      & Low-complexity                                             & Low resource efficiency and limited DoFs                                  & OMA       \\ \hline
\textbf{Decoding}        & High                & \begin{tabular}[c]{@{}l@{}}Interference is \emph{decoded} along with the \\ desired signal\end{tabular}              & Improving the rate of the desired signal                 & \begin{tabular}[c]{@{}l@{}}The decodability of the interfering signals \\ limits the other users' rates\end{tabular} & NOMA      \\ \hline
\textbf{TIN}             & Low                 & Interference is \emph{treated as noise}                                   & Requiring only single-user encoding/decoding               & \begin{tabular}[c]{@{}l@{}}Non-robust in environments with medium and \\ high intereference levels\end{tabular}      & NOMA,SDMA \\ \hline
\textbf{Encoding}        & Arbitrary           & \begin{tabular}[c]{@{}l@{}}Interference is \emph{non-causally known} \\ and \emph{pre-subtracted} via encoding\end{tabular} & Capacity-achieving for Gaussian vector BCs & Computationally intensive                                                 & DPC       \\ \hline
\end{tabular}}
\end{table*}

\begin{figure}[!t]
  \centering
  \includegraphics[width=0.45\textwidth]{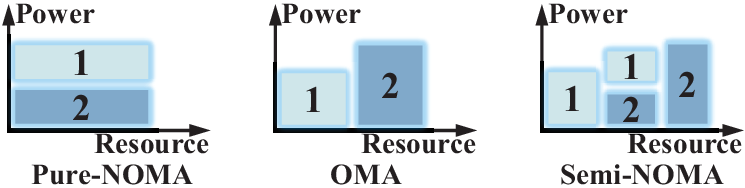}
  \caption{Illustration of hybrid NOMA for a two-user (user 1 and user 2) system.}
  \label{Figure: Semi_NOMA}
\end{figure}

By applying one or amalgamating two or more of these methods, various existing transmission frameworks come into existence. The individual application of OMA and DPC leads to TDMA/FDMA/CDMA/OFDMA and capacity-achieving MU-MIMO, respectively. Power-domain NOMA can be viewed as the combination of SIC decoding and TIN, addressing the disadvantages of these two parent technologies. The amalgamation of NOMA and OMA results in hybrid NOMA or semi-NOMA \cite{ding2022noma,ding2022hybrid}, as depicited in {\figurename} {\ref{Figure: Semi_NOMA}}. Hybrid NOMA allows for the coexistence of OMA- and pure-NOMA-based schemes, not only treating them as special cases but also providing a more flexible resource and interference management paradigm. Another flexible non-orthogonal interference management approach for multiple-antenna multiuser channels is termed RSMA, which combines SIC decoding, TIN, and even DPC \cite{clerckx2016rate,mao2022rate}. Specifically, RSMA splits part of each user's message (i.e., the private message) into a common message intended for all served users. Each user treats the private messages of all users as interference when decoding the common message and subtracts the common message from the received signal (i.e., SIC) before decoding his own private message.
 
Hybrid NOMA, SDMA, and cluster-free MIMO-NOMA are promising flexible interference managements approaches, which are strong candidates for NGMA, and more research is needed on these topics. In the context of future ultra-dense heterogeneous networks, the potential arises to establish a unified MA framework by combining all four interference management strategies. Additionally, novel tools, such as ML, can be leveraged to dynamically determine the most advantageous transmission protocol based on the real-time system layout. This approach facilitates a flexible transmission strategy through judicious allocation of the limited resources.

\section{Multiple Access in Spatial Domain}\label{Section: Spatial Domain}
One significant advancement in next-generation wireless networks is the deployment of multiple antennas at the transceivers \cite{zhang2020prospective,bjornson2024towards}. This enhancement allows for the exploitation of additional spatial DoFs, enabling the simultaneous servicing of multiple users/devices within the same time/frequency/code domain while distinguishing them in the spatial domain \cite{winters1984optimum,winters1994impact,suard1998uplink}.

\begin{figure}[!t]
  \centering
  \includegraphics[width=0.45\textwidth]{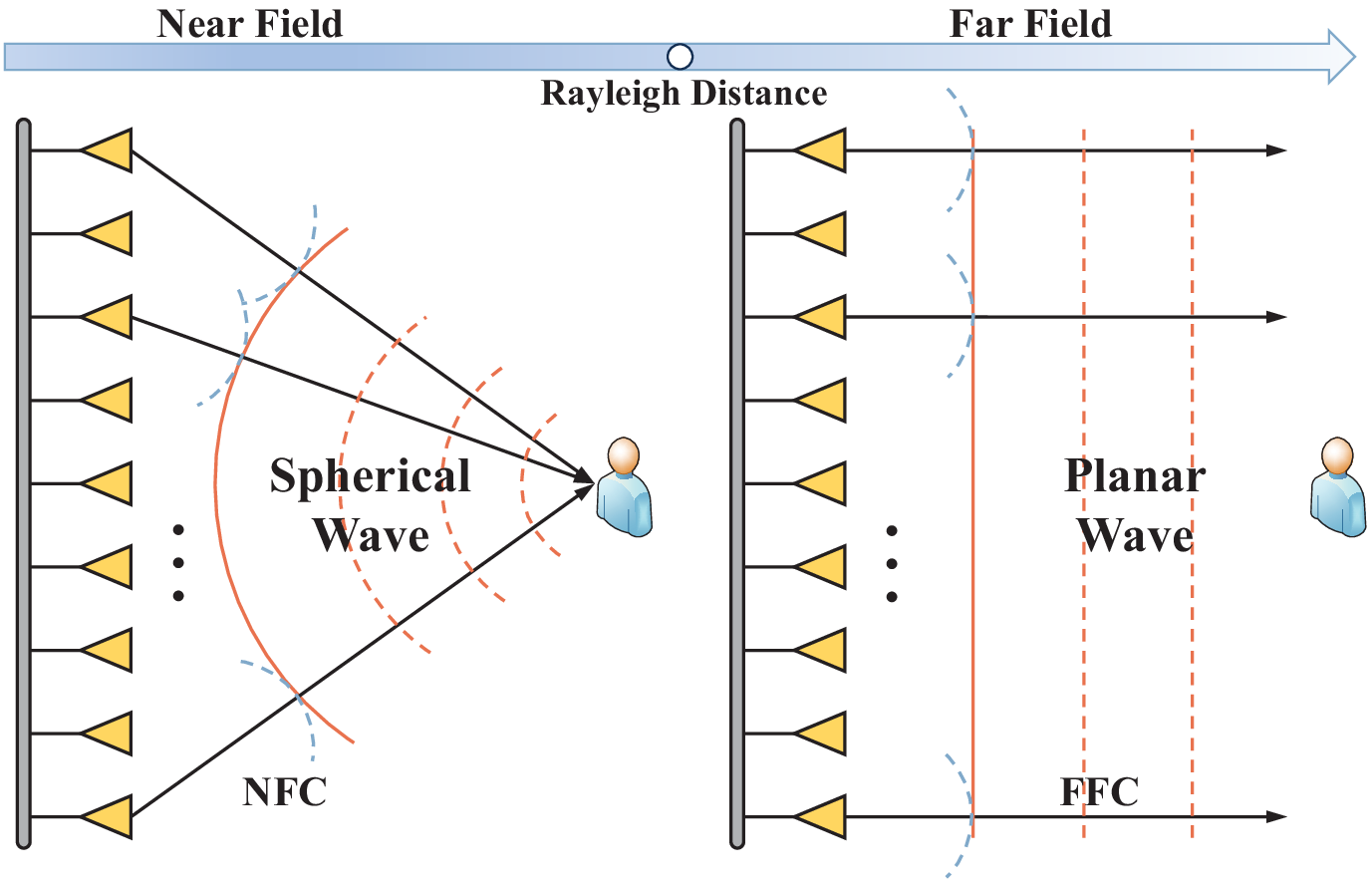}
  \caption{NFC versus FFC.}
  \label{Figure: FFC_vs_NFC}
\end{figure}

In Section \ref{Section: MIMO-NOMA}, we explored the foundational principles of MIMO-NOMA, alongside its comparison with SDMA. However, as Section \ref{Section: MIMO-NOMA} only provided a basic overview, we now embark on a more in-depth exploration of MA techniques in the spatial domain. This includes an extensive review of various SDMA and MIMO-NOMA variants, tailored to specific scenarios. Our investigation will be divided into three parts. In Section \ref{Section: Foundational Principles of SDMA}, we review the foundational principles of SDMA by focusing on commonly used linear beamforming methods. Section \ref{Section: MA in Far-Field Communications (FFC)} is dedicated to MA technologies designed for \emph{far-field communications (FFC)}, where channel modeling heavily relies on planar-wave propagations \cite{liu2023nearfield}; see the right side of {\figurename} {\ref{Figure: FFC_vs_NFC}}. Moving forward to Section \ref{Section: Near-Field SDMA and NOMA}, we shift our focus to recent innovations in spatial-domain MA technologies. Here, we highlight developments in MA for \emph{near-field communications (NFC)}, particularly in scenarios featuring extremely large-scale antenna arrays at the BS. Such setups necessitate channel modeling based on spherical-wave propagations, offering intriguing insights into the evolving landscape of wireless communication paradigms \cite{liu2023nearfield}; see the left side of {\figurename} {\ref{Figure: FFC_vs_NFC}}. For clarity and to provide deeper insights, we narrow our focus to the investigation of MU-MISO channels.
\subsection{Foundational Principles of Space-Division Multiple Access}\label{Section: Foundational Principles of SDMA}
In this subsection, we review classical linear beamforming strategies for both uplink and downlink multiple-antenna systems. 
\subsubsection{Linear Receive Beamforming for Uplink SDMA}
We start by examining an uplink single-cell MU-MISO system with $K$ single-antenna users. The channel between the $N_{\mathsf{BS}}$-antenna BS and user $k$ is denoted by ${\mathbf{h}}_k\in{\mathbbmss{C}}^{N_{\mathsf{BS}}\times1}$. The received signal ${\mathbf{y}}\in{\mathbbmss{C}}^{N_{\mathsf{BS}}\times1}$ is modeled as follows:
\begin{align}
{\mathbf{y}}=\sum\nolimits_{k=1}^{K}{\mathbf{h}}_k \sqrt{p_k}s_k+ {\mathbf{n}},
\end{align}
where $s_k\in{\mathbbmss{C}}$ is the normalized data signal transmitted by user $k$, $p_k$ is user $k$'s transmit power, and ${\mathbf{n}}\sim{\mathcal{CN}}({\mathbf{0}},\sigma^2{\mathbf{I}}_{N_{\mathsf{BS}}})$ denotes the AWGN vector with noise variance $\sigma^2$. In the context of SDMA, the BS decodes each user's message in parallel using a linear beamformer/combiner/detector/equalizer ${\mathbf{v}}_k\in{\mathbbmss{C}}^{N_{\mathsf{BS}}\times1}$ to extract $s_k$ from ${\mathbf{v}}_k^{\mathsf{H}}{\mathbf{y}}$, treating other users as interference. By treating IUI as noise \cite{hassibi2003much}, the transmission rate for user $k$ becomes:
\begin{equation}\label{SDMA_Uplink_User_Rate_MISO_General_Case}
{\mathsf{R}}_k=\log_2\left(1+\frac{p_k\lvert{\mathbf{v}}_k^{\mathsf{H}}{\mathbf{h}}_k\rvert^2}{\sum_{k'\ne k}p_{k'}\lvert{\mathbf{v}}_k^{\mathsf{H}}{\mathbf{h}}_{k'}\rvert^2+\sigma^2\lVert{\mathbf{v}}_k\rVert^2}\right).
\end{equation}
This rate is maximized by initially using a matrix ${\mathbf{V}}_k=(\sum\nolimits_{k'\ne k}{\mathbf{h}}_{k'} {\mathbf{h}}_{k'}^{\mathsf{H}}p_{k'}+ \sigma^2{\mathbf{I}}_{N_{\mathsf{BS}}})^{-\frac{1}{2}}\in{{\mathbbmss{C}}^{N_{\mathsf{BS}}\times N_{\mathsf{BS}}}}$ to whiten the interference-plus-noise term $\sum\nolimits_{k'\ne k}{\mathbf{h}}_{k'} x_{k'}+ {\mathbf{n}}$ into an AWGN vector. Subsequently, maximal-ratio combining (MRC) is applied to the resulting MISO channel ${\mathbf{V}}_k{\mathbf{h}}_kx_k + {\mathbf{n}}_k$ with ${\mathbf{n}}_k\sim{\mathcal{CN}}({\mathbf{0}},{\mathbf{I}}_{N_{\mathsf{BS}}})$. The optimal combining vector is thus given by
\begin{align}
{\mathbf{v}}_k^{\star}&={\mathbf{V}}_k^2{\mathbf{h}}_k=\Big(\sum\nolimits_{k'\ne k}{\mathbf{h}}_{k'} {\mathbf{h}}_{k'}^{\mathsf{H}}p_{k'}+ \sigma^2{\mathbf{I}}_{N_{\mathsf{BS}}}\Big)^{-1}{\mathbf{h}}_k.
\end{align}
Using Woodbury matrix identity, it is found that ${\mathbf{v}}_k^{\star}$ is also proportional to
\begin{align}
{\mathbf{v}}_k^{\star}&\propto\Big(\sum\nolimits_{k=1}^{K}{\mathbf{h}}_{k} {\mathbf{h}}_{k}^{\mathsf{H}}p_{k}+ \sigma^2{\mathbf{I}}_{N_{\mathsf{BS}}}\Big)^{-1}{\mathbf{h}}_k\\
&=({\mathbf{H}}{\mathbf{P}}{\mathbf{H}}^{\mathsf{H}}+ \sigma^2{\mathbf{I}}_{N_{\mathsf{BS}}})^{-1}{\mathbf{h}}_k\triangleq {\mathbf{v}}_k^{\mathsf{LMMSE}}
\label{SDMA_Uplink_User_Rate_MISO_LMMSE_Filter}
\end{align}
with ${\mathbf{H}}=[{\mathbf{h}}_1\ldots{\mathbf{h}}_K]\in{\mathbbmss{C}}^{N_{\mathsf{BS}}\times K}$ and ${\mathbf{P}}=\mathsf{diag}\{p_1,\ldots,p_K\}$. This vector ${\mathbf{v}}_k^{\mathsf{LMMSE}}$ serves as the LMMSE estimator for recovering $s_k$, which is also referred to as the Wiener filter \cite{joham2005linear,tse2005fundamentals}. The resulting transmission rate is
\begin{equation}\label{SDMA_Uplink_User_Rate_MISO_LMMSE_Case}
{\mathsf{R}}_k=\log_2\Big(1+p_k{\mathbf{h}}_k^{\mathsf{H}}\Big(\sum\nolimits_{k'\ne k}{\mathbf{h}}_{k'} {\mathbf{h}}_{k'}^{\mathsf{H}}p_{k'}+ \sigma^2{\mathbf{I}}_{N_{\mathsf{BS}}}\Big)^{-1}{\mathbf{h}}_k\Big).
\end{equation}
The LMMSE combining scheme strikes a balance between amplifying the signal power by aligning the receiver with $\mathbf{h}_k$ and mitigating interference by spatially whitening the received signal using the inverse of the interference-plus-noise covariance matrix \cite{tse2005fundamentals,joham2005linear}.

When $N_{\mathsf{BS}}\geq K$, ZF-based combining schemes can also be utilized to completely eliminate IUI. Specifically, the linear detector for user $k$ satisfies:
\begin{align}
{\mathbf{h}}_{k}^{\mathsf{H}}\mathbf{v}_k^{\mathsf{ZF}}\ne 0,~{\mathbf{h}}_{k'}^{\mathsf{H}}\mathbf{v}_k^{\mathsf{ZF}}=0,~k\ne k',
\end{align}
where $\mathbf{v}_k^{\mathsf{ZF}}$ can be set as the $k$th column of matrix ${\mathbf{H}}({\mathbf{H}}^{\mathsf{H}}{\mathbf{H}})^{-1}\in{\mathbbmss{C}}^{N_{\mathsf{BS}}\times K}$. Consequently, the rate for user $k$ is given by
\begin{equation}\label{SDMA_Uplink_User_Rate_MISO_ZF_Case}
{\mathsf{R}}_k=\log_2\left(1+\frac{p_k}{\sigma^2[({\mathbf{H}}^{\mathsf{H}}{\mathbf{H}})^{-1}]_{k,k}}\right).
\end{equation}
It is noteworthy that ZF combining fails in the overloaded regime when $N_{\mathsf{BS}}< K$. While the use of ZF and LMMSE combining require matrix inversion, a low-complexity combining scheme is MRC, which employs ${\mathbf{v}}_k={\mathbf{h}}_k\triangleq{\mathbf{v}}_k^{\mathsf{MRC}}$, resulting in
\begin{equation}\label{SDMA_Uplink_User_Rate_MISO_MRC_Case}
{\mathsf{R}}_k=\log_2\left(1+\frac{p_k\lVert{\mathbf{h}}_k\rVert^4}{\sum_{k'\ne k}p_{k'}\lvert{\mathbf{h}}_k^{\mathsf{H}}{\mathbf{h}}_{k'}\rvert^2+\sigma^2\lVert{\mathbf{h}}_k\rVert^2}\right).
\end{equation}

%Upon defining ${\mathbf{V}}=[{\mathbf{v}}_1\ldots{\mathbf{v}}_K]\in{\mathbbmss{C}}^{N_{\mathsf{BS}}\times K}$, we obtain
%\begin{equation}
%{\mathbf{V}}=\left\{
%\begin{array}{ll}
%({\mathbf{H}}{\mathbf{P}}{\mathbf{H}}^{\mathsf{H}}+ \sigma^2{\mathbf{I}}_{N_{\mathsf{BS}}})^{-1}{\mathbf{H}}             & {\mathsf{for~LMMSE}}\\
%{\mathbf{H}}({\mathbf{H}}^{\mathsf{H}}{\mathbf{H}})^{-1}           & {\mathsf{for~ZF}}\\
%{\mathbf{H}}          & {\mathsf{for~MRC}}
%\end{array} \right..
%\end{equation}
We note that while ZF combining focuses on canceling interference, MRC combining solely amplifies signal power. However, LMMSE combining considers both signal amplification and interference cancellation, thus outperforming both ZF and MRC. 
\subsubsection{Linear Transmit Beamforming for Downlink SDMA}
By reversing the uplink MU-MISO channel, its downlink counterpart is obtained, and the rate of user $k$ can be expressed as follows:
\begin{align}\label{SDMA_User_k_Rate}
{\mathsf{R}_k}=\log_2\left(1+\frac{p_k\lvert{\mathbf{h}}_k^{\mathsf{H}}{\mathbf{g}}_k\rvert^2}{\sum_{k'\ne k}p_{k'}\lvert{\mathbf{h}}_k^{\mathsf{H}}{\mathbf{g}}_{k'}\rvert^2+\sigma_k^2}\right),
\end{align}
where ${\mathbf{g}}_k\in{\mathbbmss{C}}^{N_{\mathsf{BS}}\times1}$ represents the beamforming vector for user $k$ at the BS, $p_k$ denotes the associated transmit power subjected to the power budget $\sum_{k=1}^{K}{p}_k\lVert{\mathbf{g}}_k\rVert^2\leq p$, and $\sigma_k^2$ denotes the noise power. When $N_{\mathsf{BS}}\geq K$, ZF-based beamforming can be employed to eliminate IUI \cite{spencer2004zero}, resulting in 
\begin{align}
{\mathbf{h}}_{k}^{\mathsf{H}}\mathbf{g}_k^{\mathsf{ZF}}\ne 0,~{\mathbf{h}}_{k'}^{\mathsf{H}}\mathbf{g}_k^{\mathsf{ZF}}=0,~k\ne k',
\end{align}
where $\mathbf{g}_k^{\mathsf{ZF}}$ can be aligned with the $k$th column of the matrix ${\mathbf{H}}({\mathbf{H}}^{\mathsf{H}}{\mathbf{H}})^{-1}\in{\mathbbmss{C}}^{N_{\mathsf{BS}}\times K}$. In practical systems, ZF beamforming provides a high performance primarily for severely underloaded conditions, where $N_{\mathsf{BS}}\gg K$, and when the user channels are sufficiently distinct \cite{krishnamoorthy2022downlinkm}. However, in situations of system overload ($N_{\mathsf{BS}}< K$) or when the user channels are strongly correlated, ZF beamforming tends to exhibit diminished performance, see, e.g., \cite{krishnamoorthy2022downlink}. In these cases, regularized ZF (RZF) beamforming can be utilized \cite{peel2005vector}, where $\mathbf{g}_k^{\mathsf{RZF}}$ aligns with the $k$th column of the matrix ${\mathbf{H}}(\alpha{\mathbf{I}}_{K}+{\mathbf{H}}^{\mathsf{H}}{\mathbf{H}})^{-1}\in{\mathbbmss{C}}^{N_{\mathsf{BS}}\times K}$, with regularization parameter $\alpha>0$ designed to overcome the challenges arising from the inversion of ill-conditioned matrices \cite{krishnamoorthy2022downlinkm}. Alternatively, maximal-ratio transmission (MRT) beamforming can be employed, which adopts ${\mathbf{g}}_k^{\mathsf{MRT}}=\frac{\mathbf{h}_k}{\lVert\mathbf{h}_k\rVert}$. 

ZF beamforming primarily focuses on IUI cancellation, whereas MRT beamforming solely amplifies signal power. Inspired by the concept of simultaneously enhancing signal power and mitigating IUI, a signal-to-leakage-plus-noise ratio (SLNR)-based beamforming approach can be employed. This approach aims to maximize the SLNR for each user $k$ by designing the beamforming vector ${\mathbf{g}}_k^{\mathsf{SLNR}}$ as follows \cite{sadek2007leakage}:
\begin{align}\label{SLNR_Maximization_Prob}
{\mathbf{g}}_k^{\mathsf{SLNR}}=\argmax\nolimits_{\lVert{\mathbf{g}}_k\rVert^2=1}\underbrace{\frac{p_k\lvert{\mathbf{h}}_k^{\mathsf{H}}{\mathbf{g}}_{k}\rvert^2}{\sum_{k'\ne k}p_k\lvert{\mathbf{h}}_{k'}^{\mathsf{H}}{\mathbf{g}}_{k}\rvert^2+\sigma_k^2}}_{\mathsf{SLNR}}.
\end{align}
Here, $p_k\lvert{\mathbf{h}}_k^{\mathsf{H}}{\mathbf{g}}_{k}\rvert^2$ represents the power of the desired signal component for user $k$, while $p_k\lvert{\mathbf{h}}_{k'}^{\mathsf{H}}{\mathbf{g}}_{k}\rvert^2$ denotes the interference power caused by user $k$ on the signal received by some other user $k'$. The aggregate power leaked from user $k$ to all other users, referred to as ``\emph{leakage}'', is given by $\sum_{k'\ne k}p_k\lvert{\mathbf{h}}_{k'}^{\mathsf{H}}{\mathbf{g}}_{k}\rvert^2$. Optimization problem \eqref{SLNR_Maximization_Prob} involves a Rayleigh quotient, and its optimal solution is attained by the normalized principal eigenvector of the matrix $(\sum_{k'\ne k}{\mathbf{h}}_{k'}{\mathbf{h}}_{k'}+\frac{\sigma_k^2}{p_k} {\mathbf{I}}_{N_{\mathsf{BS}}})^{-1}{\mathbf{h}}_{k}{\mathbf{h}}_{k}$ \cite{sadek2007active}. Note that $(\sum_{k'\ne k}{\mathbf{h}}_{k'}{\mathbf{h}}_{k'}+\frac{\sigma_k^2}{p_k} {\mathbf{I}}_{N_{\mathsf{BS}}})^{-1}{\mathbf{h}}_{k}{\mathbf{h}}_{k}$ is a rank-$1$ matrix, and its principal eigenvector is proportional to
\begin{equation}\label{SDMA_Downlink_User_Rate_MISO_SLNR_Precoder}
\begin{split}
&\Big(\sum\nolimits_{k'\ne k}p_k{\mathbf{h}}_{k'}{\mathbf{h}}_{k'}+{\sigma_k^2}{\mathbf{I}}_{N_{\mathsf{BS}}}\Big)^{-1}{\mathbf{h}}_{k}\\
&\propto
(p_k{\mathbf{H}}{\mathbf{H}}^{\mathsf{H}}+{\sigma_k^2}{\mathbf{I}}_{N_{\mathsf{BS}}})^{-1}{\mathbf{h}}_{k}.
\end{split}
\end{equation}
Similarly, LMMSE beamforming can be applied in the downlink channel, which yields \cite{vojcic1998transmitter,joham2005linear}
\begin{align}\label{SDMA_Downlink_User_Rate_MISO_LMMSE_Precoder}
{\mathbf{g}}_k^{\mathsf{LMMSE}}=({\mathbf{H}}{\mathbf{P}}{\mathbf{H}}^{\mathsf{H}}+ \sigma_k^2{\mathbf{I}}_{N_{\mathsf{BS}}})^{-1}{\mathbf{h}}_k.
\end{align}

%Upon defining ${\mathbf{G}}=[{\mathbf{g}}_1\ldots{\mathbf{g}}_K]\in{\mathbbmss{C}}^{N_{\mathsf{BS}}\times K}$ and assuming $\sigma_1^2=\ldots=\sigma_K^2=\sigma^2$, we obtain
%\begin{equation}
%{\mathbf{G}}=\left\{
%\begin{array}{ll}
%({\mathbf{H}}{\mathbf{P}}{\mathbf{H}}^{\mathsf{H}}+ \sigma^2{\mathbf{I}}_{N_{\mathsf{BS}}})^{-1}{\mathbf{H}}             & {\mathsf{for~LMMSE}}\\
%{\mathbf{H}}({\mathbf{H}}^{\mathsf{H}}{\mathbf{H}})^{-1}           & {\mathsf{for~ZF}}\\
%{\mathbf{H}}          & {\mathsf{for~MRT}}
%\end{array} \right..
%\end{equation}
While ZF beamforming is surpassed by MRT beamforming in low-SNR regimes, MRT is interference-limited. Therefore, neither scheme is optimal for all SNR values. By comparing \eqref{SDMA_Downlink_User_Rate_MISO_SLNR_Precoder} and \eqref{SDMA_Downlink_User_Rate_MISO_LMMSE_Precoder}, SLNR beamforming is equivalent to LMMSE beamforming for equal power allocation among different users \cite{patcharamaneepakorn2012equivalence}. Numerous numerical results available in current literature indicate that SLNR and LMMSE outperform RZF/ZF/MRC in most SNR ranges; see \cite{sadek2007leakage,lu2022robust,wiesel2005linear}. However, a comparison between LMMSE and SLNR hinges on the design of the power allocation coefficients, which depends on the specific communication scenario. 

The preceding discussion is based on MU-MISO setups, whose extension to MU-MIMO setups is available in \cite{li2016massive,patcharamaneepakorn2013equivalent}. In addition to the linear beamforming schemes mentioned above, practical beamforming design may need to take into account various constraints and QoS requirements. The resulting problem may not be convex in general but can be efficiently solved using methods such as weighted MMSE (WMMSE) \cite{christensen2008weighted,shi2011iteratively,zhao2023rethinking}, fractional programming (FP) \cite{shen2018fractional,shen2018fractional1}, successive convex approximation (SCA) \cite{razaviyayn2014successive} combined with semi-definite programming (SDP) techniques \cite{luo2010semidefinite}, matrix-monotonic optimization \cite{xing2020matrix,xing2020matrix1}, among others. For further details, we refer to the recent overview paper \cite{liu2024survey}.

\subsection{Multiple Access in Far-Field Communications}\label{Section: MA in Far-Field Communications (FFC)} 
Having laid the foundational principles of MIMO-NOMA (Section \ref{Section: MIMO-NOMA}) and SDMA (Section \ref{Section: Foundational Principles of SDMA}), we now explore various SDMA and MIMO-NOMA variants. In this subsection, our focus shifts to these MA approaches in traditional FFC by studying three cases: massive MIMO transmission, hybrid beamforming (HB)-based transmission, and mmWave/THz transmission.
\subsubsection{Massive MIMO Transmission}
The essence of SDMA revolves around the utilization of spatial beamforming to either amplify signal power or mitigate IUI. However, both approaches can lead to significant IUI, particularly in the overloaded regime, where $N_{\mathsf{BS}}<K$. To address this challenge, MIMO-NOMA techniques can be employed, as discussed in Section \ref{Section: MIMO-NOMA}. Alternatively, another strategy involves equipping the BS with a large-scale antenna array, which yields $N_{\mathsf{BS}}\gg K$ and effectively avoids the overloaded regime. This concept is central to massive MIMO \cite{larsson2014massive}.

\begin{figure}[!t]
 \centering
\setlength{\abovecaptionskip}{0pt}
\includegraphics[width=0.45\textwidth]{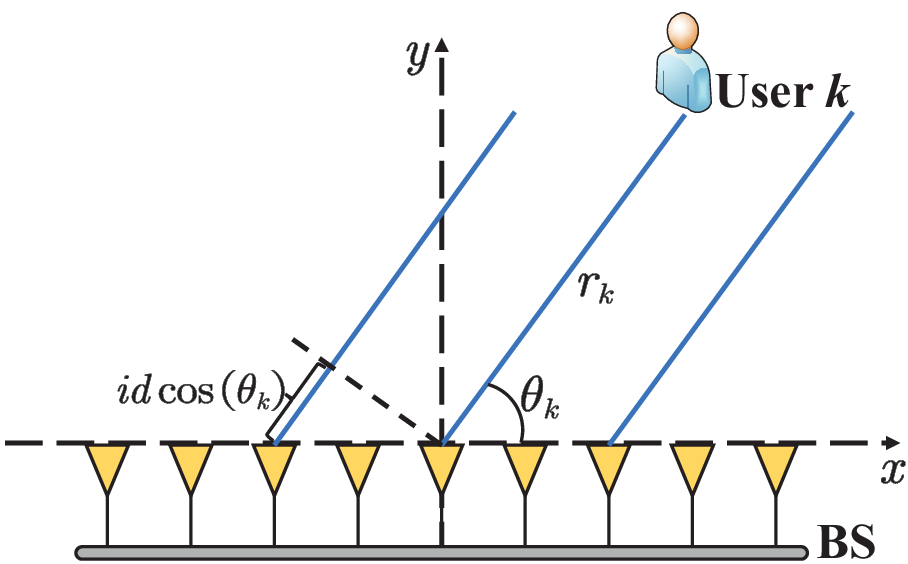}
\caption{Far-field LoS channel model for ULA.}
\label{LoS_2D_Model_NFC}
\end{figure}
Massive MIMO channels exhibit the property of ``favorable propagation'', where the directions of two user channels become asymptotically orthogonal. This property enhances the BS's ability to mitigate IUI between these users, typically improving the spectral efficiency and making it sufficient to employ linear transmit/receive beamforming. To illustrate this, we first consider the line-of-sight (LoS) channel model by assuming the BS is equipped with a uniform linear array (ULA) and the user is located in the \emph{far-field region} of the BS, as depicted in {\figurename} {\ref{LoS_2D_Model_NFC}}. The ULA is situated along the $x$-axis and centered at the origin. Without loss of generality, we assume that $N_{\mathsf{BS}}$ is an odd number with $N_{\mathsf{BS}}=2\widetilde{N}+1$, where $\widetilde{N}$ is a non-negative integer. The indices of the antenna array are denoted by $i\in\{0,\pm1,\ldots,\pm\widetilde{N}\}$. Regarding user $k$'s position, let $r_k$ denote its distance from the center of the antenna array, and $\theta_k$ represent its direction of arrival (DoA). Consequently, user $k$'s location is given by ${\bm{\mathsf{r}}}_k=[r_k\cos(\theta_k),r_k\sin(\theta_k)]^{\mathsf{T}}$ in the $xy$-plane. According to the far-field planar-wave propagation model, the user $k$-to-BS channel vector can be characterized as follows \cite{liu2023near}:
\begin{align}\label{FFC_ULA_LOS_Model}
{\mathbf{h}}_k=\frac{\sqrt{\beta_{\mathsf{r}}}r_{\mathsf{r}}}{r_k}\underbrace{[{\rm{e}}^{-{\rm{j}}\frac{2\pi}{\lambda}(i-1)d\cos(\theta_{k})}]_{i=-\widetilde{N},\ldots,\widetilde{N}}^{\mathsf{T}}}_{
{\mathbf{a}}_{N_{\mathsf{BS}}}(\theta_k)},
\end{align}
where $\lambda$ denotes the wavelength, $d$ is the spacing between adjacent antenna elements, and $\beta_{\mathsf{r}}$ denotes the path loss at the reference distance $r_{\mathsf{r}}=1$ m. From this, we can derive the following theorem.
\vspace{-5pt}
\begin{theorem}\label{Theorem_FFC_Channel_Correlation}
Under the considered far-field model, we have 
\begin{align}
\lim_{N_{\mathsf{BS}}\rightarrow\infty}\frac{\lvert{\mathbf{h}}_k^{\mathsf{H}}{\mathbf{h}}_{k'}\rvert}{\lVert{\mathbf{h}}_k\rVert\lVert{\mathbf{h}}_{k'}\rVert}=
\begin{cases}
0&\theta_k\ne \theta_{k'}\\
1&\theta_k= \theta_{k'}
\end{cases}.
\end{align}
\end{theorem}
\vspace{-5pt}
\begin{IEEEproof}
Please refer to Appendix \ref{Proof_Theorem_FFC_Channel_Correlation} for more details.
\end{IEEEproof}
Next, considering an isotropic scattering channel where 
\begin{align}
{\mathbf{h}}_k=\frac{\sqrt{\beta_{\mathsf{r}}}r_{\mathsf{r}}}{r_k}\overline{\mathbf{h}}_k
\end{align}
with $\overline{\mathbf{h}}_k\sim{\mathcal{CN}}({\mathbf{0}},{\mathbf{I}}_{N_{\mathsf{BS}}})$. Then, the following theorem can be found.
\vspace{-5pt}
\begin{theorem}\label{Theorem_FFC_Channel_Correlation_Scattering}
In FFC with isotropic scattering, it holds that
\begin{align}
\lim_{N_{\mathsf{BS}}\rightarrow\infty}\frac{\lvert{\mathbf{h}}_k^{\mathsf{H}}{\mathbf{h}}_{k'}\rvert}{\lVert{\mathbf{h}}_k\rVert\lVert{\mathbf{h}}_{k'}\rVert}=0,\quad k\ne k'.
\end{align}
\end{theorem}
\vspace{-5pt}
\begin{IEEEproof}
Please refer to Appendix \ref{Proof_Theorem_FFC_Channel_Correlation_Scattering} for more details.
\end{IEEEproof}
The above results suggest that the inner product of the normalized channels asymptotically approaches zero, indicating that the channel directions become asymptotically orthogonal. This property of ``favorable propagation'' is expected to hold in real-world propagation environments, which typically lie between the extreme cases of isotropic and LoS propagation, e.g., correlated Rician fading channels.

This property offers valuable insights for beamforming design in SDMA systems. When the users' channel vectors are non-orthogonal, computationally intensive techniques like DPC are necessary to suppress interference and achieve the sum-rate capacity. Conversely, favorable propagation characterizes an environment where the users' channel vectors become asymptotically orthogonal. In such cases, IUI can be effectively mitigated using simple linear beamforming techniques like MRC/MRT and ZF \cite{ngo2014aspects}.

The application of SDMA in massive MIMO setups has been extensively discussed over the past decade, with detailed information available in well-established textbooks on the topic published in recent years \cite{bjornson2017massive,marzetta2016fundamentals,heath2018foundations}. However, considering the future need to connect an excessive number of users in cellular networks (i.e., the overloaded regime), the spatial DoFs provided by massive MIMO may not be sufficient. In this case, NOMA can help to serve more users via power-domain multiplexing \cite{liu2018gaussian,senel2019role}. Moreover, there may exist communication scenarios where different users have highly correlated channel conditions, in which case massive MIMO can achieve good performance pnly by deploying an extremely large-scale array, which can be costly. In such scenarios, MIMO-NOMA may be a preferable choice to mitigate IUI through user grouping and SIC decoding \cite{senel2019role,ding2016design}. 
\subsubsection{Hybrid Beamforming-Based Transmission}\label{Section: mmWave/THz SDMA and NOMA}
The performance of a multiple-antenna system can be enhanced by increasing the number of antennas. However, it is important to acknowledge that the fully digital implementation of a large-scale antenna array (i.e., massive MIMO), which requires one dedicated radio frequency (RF) chain per antenna element, incurs high hardware costs and energy consumption. To strike a better balance between spectral efficiency and hardware efficiency, researchers are turning their attention to hybrid digital-analog beamforming techniques \cite{el2014spatially}.

\begin{figure}[!t]
 \centering
\setlength{\abovecaptionskip}{0pt}
\includegraphics[width=0.45\textwidth]{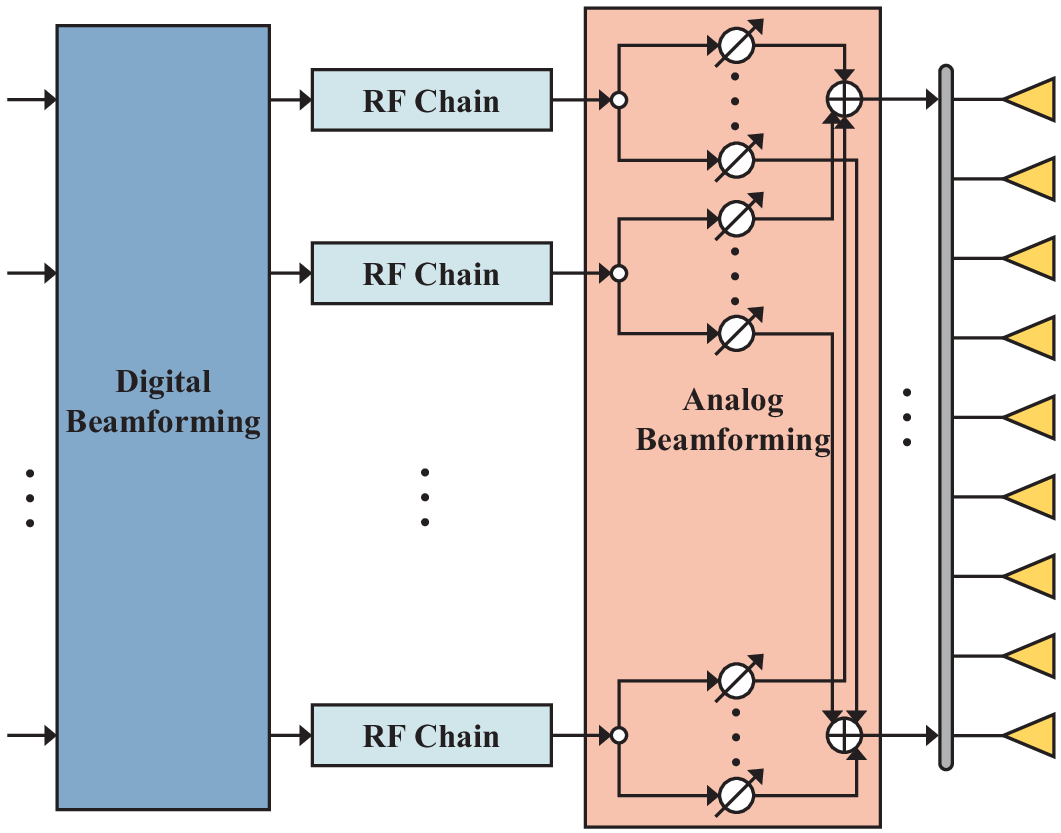}
\caption{Illustration of the architecture for HB.}
\label{HP_Architecture}
\end{figure}

Consider a narrowband downlink single-cell MU-MISO system, in which a BS equipped with $N_{\mathsf{BS}}$ antennas and $N_{\mathsf{RF}}$ RF chains serves $K$ single-antenna users, each supporting single-stream transmission. The system operates under the assumption that $N_{\mathsf{BS}}\gg \max\{N_{\mathsf{RF}},K\}$. A hybrid digital and analog beamforming architecture is employed, where each RF chain is connected to all antenna elements through phase shifters, as illustrated in {\figurename} {\ref{HP_Architecture}}. Initially, the BS processes the data streams digitally in baseband using a digital beamformer. Subsequently, the signals are up-converted to the carrier frequency through $N_{\mathsf{RF}}$ RF chains. The BS then utilizes an analog beamformer, implemented with analog phase shifters, to construct the final transmitted signal.
The resulting received signal at user $k$ can be modeled as follows:
\begin{equation}\label{Received_Signal}
y_k={\mathbf{h}}_k^{\mathsf{H}}{\mathbf{F}}{\mathbf{w}}_ks_k+\sum\nolimits_{k'\ne k}
{\mathbf{h}}_k^{\mathsf{H}}{\mathbf{F}}{\mathbf{w}}_{k'}s_{k'}+n_k,
\end{equation}
where $s_k\in{\mathbbmss{C}}$ is the data signal for user $k$, ${\mathbf{w}}_k\in{\mathbbmss{C}}^{N_{\mathsf{RF}}\times1}$ is the associated digital beamformer, ${\mathbf{h}}_k\in{\mathbbmss{C}}^{N_{\mathsf{BS}}\times1}$ is the channel vector from user $k$ to the BS, and $n_k\sim{\mathcal{CN}}(0,\sigma_k^2)$ represents AWGN with noise power $\sigma_k^2$. Here, ${\mathbf{F}}\in{\mathbbmss{C}}^{N_{\mathsf{BS}}\times N_{\mathsf{RF}}}$ is the analog beamforming matrix with $\lvert[{\mathbf{F}}]_{i,j}\rvert=1/\sqrt{N_{\mathsf{RF}}N_{\mathsf{BS}}}$ and $\angle{[{\mathbf{F}}]_{i,j}}$ denotes the phase shift introduced by the phase shifter connecting RF chain $j$ with antenna $i$. The total power constraint is enforced by normalizing $\mathbf{W}$ such that ${\mathsf{tr}}({\mathbf{W}}^{\mathsf{H}}{\mathbf{F}}^{\mathsf{H}}{\mathbf{F}}{\mathbf{W}})=p$, where ${\mathbf{W}}=[{\mathbf{w}}_1\ldots {\mathbf{w}}_K]^{\mathsf{T}}\in{\mathbbmss{C}}^{N_{\mathsf{RF}}\times K}$ is the digital beamforming matrix. The received SINR at user $k$ is expressed as follows:
\begin{equation}\label{Received_SINR}
\gamma_k=\frac{\lvert{\mathbf{h}}_k^{\mathsf{H}}{\mathbf{F}}{\mathbf{w}}_{k}\rvert^2}{\sum_{k'\ne k}\lvert{\mathbf{h}}_k^{\mathsf{H}}{\mathbf{F}}{\mathbf{w}}_{k'}\rvert^2+\sigma_k^2}.
\end{equation}
HB represents a novel paradigm of SDMA that has garnered significant research attention over the past decade; for a detailed discussion, we refer to the overview papers \cite{heath2016overview,molisch2017hybrid,ahmed2018survey}.

In HB-SDMA, the number of independent data streams is generally set no larger than the number of RF chains to avoid overloading the system. Given the same BS antenna number, we observe that compared to fully digital SDMA systems, HB-SDMA is more susceptible to entering the overloaded regime due to a lower number of RF chains. In such scenarios, as well as in situations involving highly correlated user channels, HB-MIMO-NOMA can play a pivotal role in mitigating IUI and enhancing system spectral efficiency. Consequently, numerous hybrid beamforming algorithms have been proposed across various application domains, including simultaneous wireless information and power transfer (SWIPT) systems \cite{dai2018hybrid}, cache-enabled THz-MIMO networks \cite{zhang2020energy}, UAV networks \cite{feng2021hybrid}, among others. Moreover, several supplementary algorithms for tasks such as beam management, user grouping, and SIC ordering have been introduced; for more details, we refer to \cite{wei2018multi,zhu2019millimeter,ding2023joint}, and the associated references.
\subsubsection{mmWave/THz Transmission}\label{Section: Beam Division Multiple Access and Beamspace MIMO-NOMA}
Typically, the size and spacing of antennas in an antenna array are inversely proportional to the wavelength. As a result, deploying large antenna arrays becomes feasible at higher frequencies, such as the mmWave/THz spectrum bands \cite{rappaport2013millimeter,sun2014mimo}. Using mmWave/THz frequencies offers several advantages. Firstly, it allows for the utilization of a large bandwidth and hundreds of antenna elements at the BS. Secondly, the array gain provided by MIMO beamforming helps offset the disadvantages of mmWave/THz frequencies, such as high path loss and limited communication coverage. Therefore, mmWave/THz transmission and massive MIMO complement each other. Moreover, it is noteworthy that the majority of HB algorithms in the current literature are specifically tailored for mmWave/THz massive MIMO scenarios \cite{heath2016overview}.

In the mmWave/THz bands, the spatial channels are typically sparsely scattered and predominantly characterized by LoS propagation. This characteristic is encapsulated by the channel model presented in the following \cite{spencer2000modeling,gustafson2013mm}:
\begin{equation}
\begin{split}
{\mathbf{h}}_k&=\frac{\sqrt{\beta_{\mathsf{r}}}r_{\mathsf{r}}}{r_k}\sqrt{\frac{{\mathsf{K}}_k}{1+{\mathsf{K}}_k}}{\mathbf{a}}_{N_{\mathsf{BS}}}(\theta_{k,0})\\
&+\frac{\sqrt{\beta_{\mathsf{r}}}r_{\mathsf{r}}}{r_k}\sqrt{\frac{1}{1+{\mathsf{K}}_k}\frac{1}{L_k}}
\sum_{\ell=1}^{L_k}\mu_{k,\ell}{\mathbf{a}}_{N_{\mathsf{BS}}}(\theta_{k,\ell}).
\end{split}
\end{equation}
Here, $L_k\ll N_{\mathsf{BS}}$ represents the number of resolvable non-LoS (NLoS) paths, $\theta_{k,0}$ denotes the DoA of the LoS path, $\{\theta_{k,\ell}\}_{\ell=1}^{L_k}$ are the DoAs of the NLoS paths, $\mu_{k,\ell}\sim{\mathcal{CN}}(0,1)$ denote the associated complex gains, and ${\mathsf{K}}_k\gg 1$ represents the Rician factor. The conditions of $L_k\ll N_{\mathsf{BS}}$ and ${\mathsf{K}}_k\gg 1$ imply that mmWave/THz massive MIMO channels exhibit sparsity in the angle or beam domain, a characteristic that becomes more pronounced for larger antenna array sizes \cite{brady2013beamspace}. 

The angle sparsity becomes explicitly evident in the discrete Fourier transform (DFT) of the spatial channel vector, i.e., ${\mathbf{U}}{\mathbf{h}}_k$, where $\mathbf{U}\in{\mathbbmss{C}}^{N_{\mathsf{BS}}\times N_{\mathsf{BS}}}$ represents a spatial DFT matrix satisfying ${\mathbf{U}}{\mathbf{U}}^{\mathsf{H}}={\mathbf{I}}_{N_{\mathsf{BS}}}$ \cite{brady2013beamspace}. For clarity, we assume that $d=\frac{\lambda}{2}$. Then, ${\mathbf{U}}$ contains the array steering vectors of $N_{\mathsf{BS}}$ orthogonal directions as follows \cite{brady2013beamspace}:
\begin{align}
{\mathbf{U}}=\frac{1}{\sqrt{N_{\mathsf{BS}}}}[{\mathbf{u}}(\psi_1) \ldots {\mathbf{u}}(\psi_{N_{\mathsf{BS}}})]^{\mathsf{H}},
\end{align}
where ${\mathbf{u}}(\psi)=[{\rm{e}}^{-{\rm{j}}{\pi}(i-1)\psi}]_{i=-\widetilde{N}}^{\widetilde{N}}\in{\mathbbmss{C}}^{N_{\mathsf{BS}}\times1}$ and $\psi_{i}=\frac{2i-N_{\mathsf{BS}}-1}{N_{\mathsf{BS}}}\in(-1,1)$ for $i=1,\ldots,N_{\mathsf{BS}}$. Based on the results in Theorems \ref{Theorem_FFC_Channel_Correlation} and \ref{Theorem_FFC_Channel_Correlation_Scattering}, it follows that when $L_k\ll N_{\mathsf{BS}}$ and ${\mathsf{K}}_k\gg 1$, the norm of vector ${\mathbf{U}}{\mathbf{h}}_k$ is dominated by its $i_k^{\star}$th element, where
\begin{align}
i_k^{\star}=\argmin\nolimits_{i=1,\ldots,N_{\mathsf{BS}}}\lvert \psi_{i} - \cos(\theta_{k,0})\rvert.
\end{align}
This sparsity is attributed to the high angle resolution facilitated by massive antenna arrays. ${\mathbf{U}}{\mathbf{h}}_k$ is also termed the beam-domain/angle-domain/beamspace channel vector of user $k$. The beamspace channel matrix of the whole system is thus given by ${\mathbf{U}}[{\mathbf{h}}_1 \ldots {\mathbf{h}}_K]\triangleq \tilde{\mathbf{H}}\in{\mathbbmss{C}}^{N_{\mathsf{BS}}\times K}$, which has the same dimesions as the spatial-domain channel matrix $[{\mathbf{h}}_1 \ldots {\mathbf{h}}_K]\in{\mathbbmss{C}}^{N_{\mathsf{BS}}\times K}$. Each row of $\tilde{\mathbf{H}}$ corresponds to an analog beam in the beamspace, with its row index also referred to as beam index.
\begin{figure}[!t]
 \centering
\setlength{\abovecaptionskip}{0pt}
\includegraphics[width=0.45\textwidth]{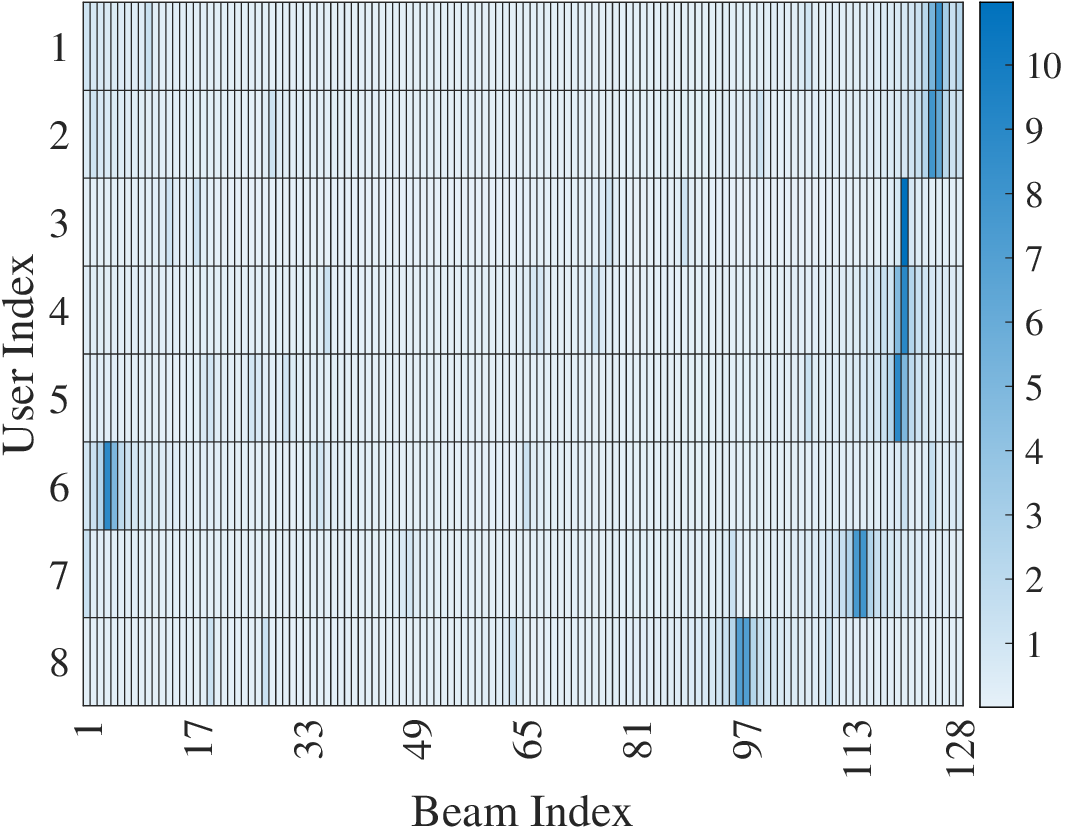}
\caption{Illustration of the beam-domain channel strength for $d=\frac{\lambda}{2}$, $N_{\mathsf{BS}}=128$, $K=8$, $\frac{\sqrt{\beta_{\mathsf{r}}}r_{\mathsf{r}}}{r_k}=1$, ${\mathsf{K}}_k=20$, $L_k=4$, $\mu_{k,\ell}=1$, $\forall k,\ell$. The DoAs for the LoS paths are specified as follows: $\{20^{\circ},20.5^{\circ},-30^{\circ},-30.5^{\circ},-31^{\circ},-160^{\circ},40^{\circ},-60^{\circ}\}$ for users 1 to 8, while the DoAs for the NLoS paths are randomly distributed within the interval $(-180^{\circ},180^{\circ}]$.}
\label{BDMA_Beamspace}
\end{figure}

The illustration of the channel strength of the beamspace channel, viz. the amplitude of each element in matrix $\tilde{\mathbf{H}}$, for an 8-user mmWave/THz MISO system is presented in {\figurename} {\ref{BDMA_Beamspace}}. The figure clearly shows that each user's beamspace channel is primarily characterized by a selected few beam indices within the entire beamspace. Exploiting this observation, a logical approach is to utilize a switch network to identify the dominant beam of each user and connecting these beams with the RF chains. Further suppression of IUI can be achieved by applying digital beamforming within the significantly reduced dimension of the beamspace channel. This concept gives rise to beam division multiple access (BDMA) \cite{sun2015beam,you2017bdma}, where multiple users concurrently share the same time-frequency resources while being discernible in the beam domain \cite{gao2016near,han2018dft}. Notably, BDMA represents a special instance of SDMA tailored for beam-domain-sparse mmWave channels. Studies have shown that beam selection in BDMA can reduce the number of required RF chains without significant performance loss \cite{gao2018low}. The array gains can always be preserved after selecting the dominant beams; thus, even with a simple selection network, satisfactory performance can still be guaranteed \cite{brady2013beamspace}. The transformation of the spatial channel ${\mathbf{h}}_k$ to its beamspace equivalent ${\mathbf{U}}{\mathbf{h}}_k$ is facilitated by a DFT. Practically, this transformation can be realized using a phase-shifter network comprising $N_{\mathsf{BS}}^2$ phase shifters with fixed phase offsets. Implementation-wise, this can be achieved through techniques such as lens antennas \cite{brady2013beamspace}, Butler matrices \cite{chin201025}, or other DFT modules integrated into field-programmable analog arrays \cite{molisch2003dft,suh2010low}.

\begin{figure}[!t]
 \centering
\setlength{\abovecaptionskip}{0pt}
\includegraphics[width=0.45\textwidth]{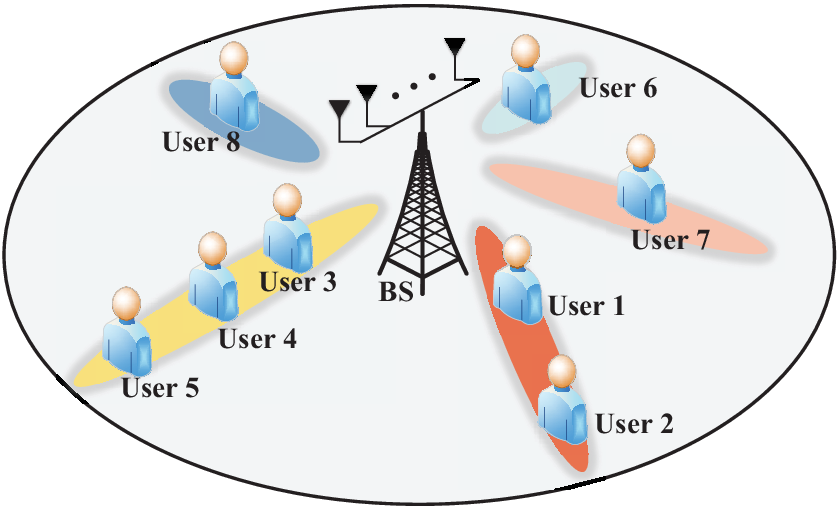}
\caption{Illustration of beamspace NOMA.}
\label{BDMA_NOMA}
\end{figure}

It is important to highlight that in scenarios characterized by either the overload regime or the underloaded regime with highly correlated channels, beam-domain NOMA emerges as a viable solution to further mitigate IUI. Specifically, users sharing the same dominant beam can be effectively grouped together. For instance, as depicted in {\figurename} {\ref{BDMA_Beamspace}}, users 1 and 2, as well as users 3 to 5, exhibit strong correlation in their beamspace channels, as they share the same dominant beam index. In this case, users 1 and 2 as well as users 3 to 5 can be grouped into two NOMA clusters, as illustrated in {\figurename} {\ref{BDMA_NOMA}}. Subsequently, inter-cluster interference can be mitigated through ZF beamforming in the beamspace, while intra-cluster interference can be further alleviated through SC and SIC decoding techniques. Various algorithms have been proposed for beamspace NOMA, encompassing aspects such as beam selection, power allocation, digital beamforming, and user grouping; for a comprehensive review, we refer to \cite{wang2017spectrum,hao2019beamforming,jiao2020max,liu2020multi}.
\subsection{Multiple Access in Near-Field Communications}\label{Section: Near-Field SDMA and NOMA}
Leveraging larger-scale antenna arrays in conjunction with higher carrier frequencies can significantly enhance the spectral efficiency of multiple-antenna multiuser networks. In the past two decades, the evolution of MIMO technologies has progressed towards the development of extremely large aperture arrays (ELAAs) and utilization of extremely high frequencies \cite{cui2022near}. This transition to ELAAs and higher frequency bands signifies not only a quantitative increase in antenna size and carrier frequency but also a qualitative paradigm shift from conventional FFC to NFC \cite{liu2023near}. This subsection explores spatial-domain MA technologies in NFC.
\subsubsection{Near-Field Channel Modeling}
\begin{figure}[!t]
 \centering
\setlength{\abovecaptionskip}{0pt}
\includegraphics[width=0.45\textwidth]{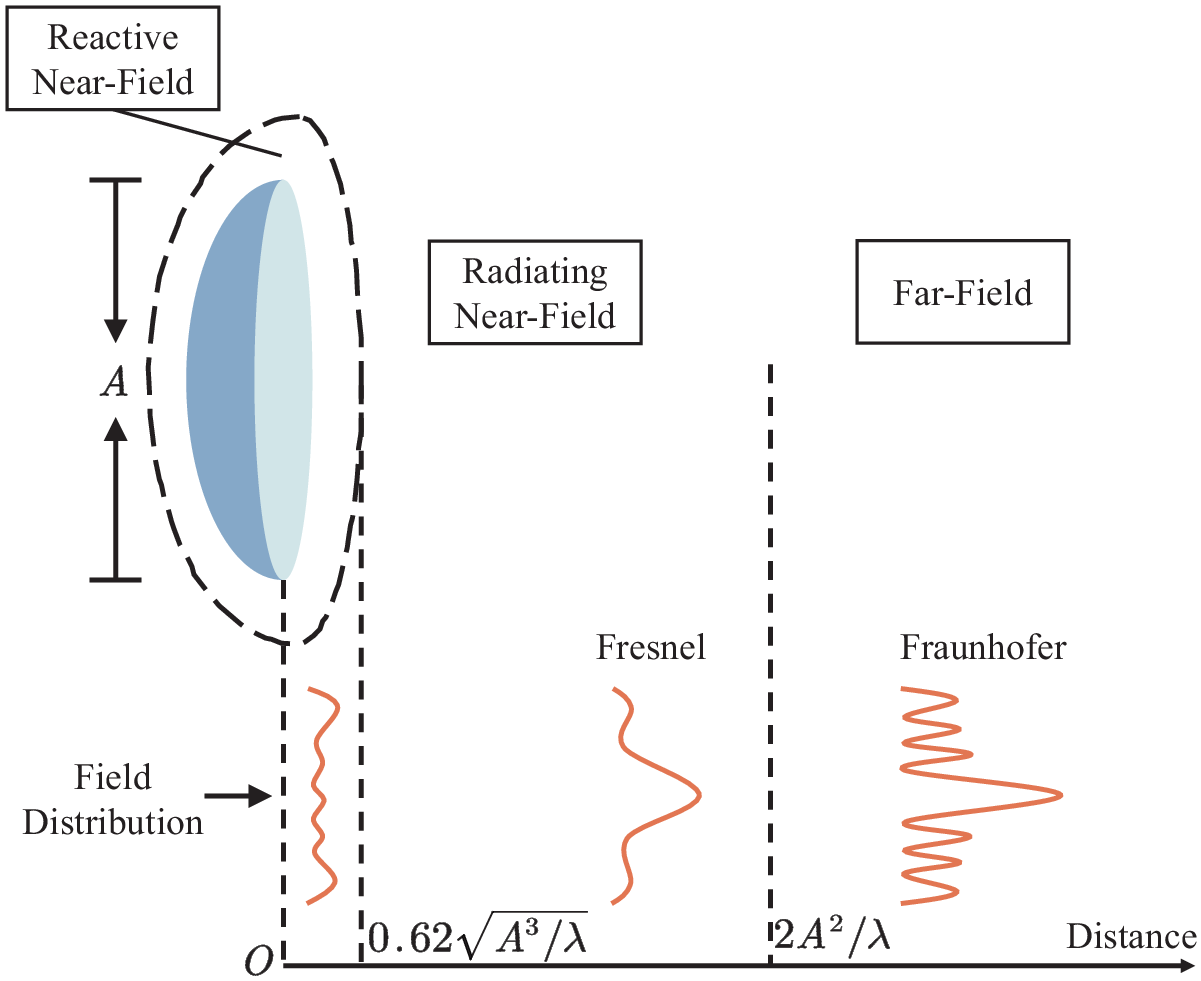}
\caption{Illustration of the EM field regions of an antenna array, where $A$ and $\lambda$ denote the array aperture and signal wavelength, respectively.}
\label{Figure: EM_Field}
\end{figure}

The electromagnetic (EM) field emitted by an antenna can be divided into three regions: the far-field region, the radiating near-field region, and the reactive near-field region, as illustrated in {\figurename} {\ref{Figure: EM_Field}}. Conventionally, the boundary between the reactive and the radiating near–field region is $0.62\sqrt{\frac{A^3}{\lambda}}$, whereas the far–field region corresponds to distances larger than the Rayleigh distance given by $\frac{2A^2}{\lambda}$, where $A$ is the array aperture, i.e., the largest distance between any two antennas of
the array \cite{balanis2016antenna}. Within the reactive near-field region, the energy of the EM field oscillates rather than being permanently removed from the transmitter \cite{liu2023nearfield}. Consequently, this region is dominated by non-propagating fields known as evanescent waves. The propagation of EM waves in the reactive near-field region is intricate and challenging to predict \cite{balanis2016antenna,liu2023nearfield}. Therefore, we focus our attention on the radiating near-field region for the remainder of this discussion. Unless stated otherwise, ``radiating near-field'' is referred to as ``near-field'' for brevity.

\begin{figure}[!t]
 \centering
\setlength{\abovecaptionskip}{0pt}
\includegraphics[width=0.45\textwidth]{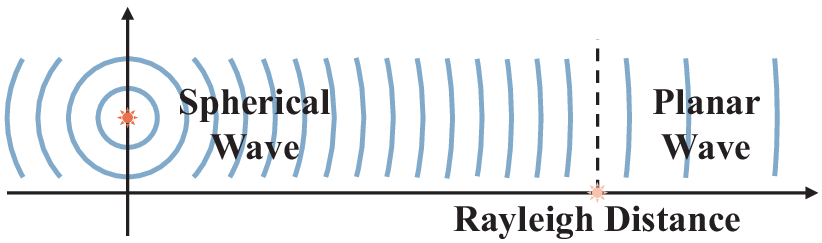}
\caption{Illustration of the flattening of spherical waves with distance.}
\label{NFC_Case}
\end{figure}

Far-field EM propagation is accurately approximated using planar waves, whereas the near-field EM field necessitates precise modeling using spherical waves, as illustrated in {\figurename} {\ref{NFC_Case}} as well as {\figurename} {\ref{Figure: FFC_vs_NFC}}. The deployment of ELAAs (augmented $A$) and the utilization of high-frequency bands (diminished $\lambda$) make NFC relevant over distances spanning hundreds of meters, significantly expanding the previously neglected near-field region of traditional wireless systems. To illustrate, consider an antenna array with a dimension of $A = 4$ m---a size feasible for future conformal arrays such as those deployed on building facades. For signals operating at 3.5 GHz, the Rayleigh distance is calculated to be 373.3 m. This observation underscores how the increased antenna aperture and the shortened wavelength dramatically extend the near-field region. This extension serves as a primary motivation for our exploration of MA schemes in the context of NFC.
 
\begin{figure}[!t]
 \centering
\setlength{\abovecaptionskip}{0pt}
\includegraphics[width=0.45\textwidth]{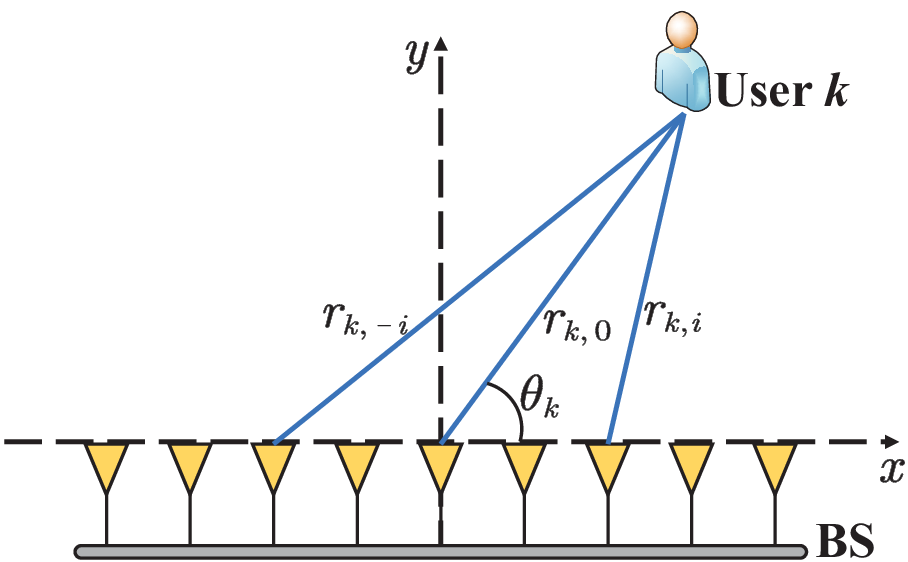}
\caption{Near-field LoS channel model for ULA.}
\label{LoS_2D_Model}
\end{figure}

For clarity, the subsequent discussion is based on a near-field $K$-user MISO model, where the BS is equipped with a ULA containing $N_{\mathsf{BS}}$ antenna elements, as depicted in {\figurename} {\ref{LoS_2D_Model}}. Given that NFC operates at higher frequencies compared to conventional massive MIMO systems, the dominant component of the channels is often LoS propagation. Thus, we focus on the basic free-space LoS propagation scenario. The ULA is situated along the $x$-axis and centered at the origin. To simplify the analysis, we assume that $N_{\mathsf{BS}}=2\widetilde{N}+1$, where $\widetilde{N}$ is a non-negative integer. The indices of the antenna array are represented by $i\in\{0,\pm1,\ldots,\pm\widetilde{N}\}$. Moreover, user $k$'s location is given by ${\bm{\mathsf{r}}}_k=[r_k\cos(\theta_k),r_k\sin(\theta_k)]^{\mathsf{T}}$ in the $xy$-plane, where $r_k$ denotes its distance from the center of the antenna array and $\theta_k$ represents its DoA.

Due to the near-field behavior, it becomes imperative to differentiate the power and phase of different elements when modeling the channel between the BS array and the user. The radiating near-field LoS channel between the BS and user $k$ can be written as follows \cite{liu2023near}:
\begin{equation}\label{NFC_ULA_LOS_Model}
{\mathbf{h}}(r_k,\theta_k)={\sqrt{\beta_{\mathsf{r}}}r_{\mathsf{r}}}
[{1}/{r_{k,i}}{\rm{e}}^{-{\rm{j}}\frac{2\pi}{\lambda}r_{k,i}}]_{i=-\widetilde{N},\ldots,\widetilde{N}}^{\mathsf{T}},
\end{equation}
where $r_{k,i}$ denotes the distance between the user $k$ and the $i$th antenna element with $r_{k,0}=r_k$. Utilizing the spherical-wave propagation model, we obtain \cite{liu2023near}
\begin{equation}
\begin{split}
r_{k,i}&=\sqrt{(r_k\cos(\theta_k)-id)^2+r_k^2\sin^2(\theta_k)}\\
&=r_k\sqrt{1+\frac{1}{r_k^2}(i^2d^2-2idr_k\cos(\theta_k))}.
\end{split}
\end{equation}
Since the array element separation $d$ is typically on the order of a wavelength, in practice, we have $r_k\gg d$ and thus $\frac{d}{r_k}\ll 1$, which leads to the following approximation:
\begin{equation}\label{ULA_F_Appr}
r_{k,i}\approx r_k-id\cos(\theta_k)+\frac{i^2d^2}{2r_k}\sin^2(\theta_k), \forall i.
\end{equation}
The above approximation is due to the fact that $({1+x})^{\frac{1}{2}}\simeq 1+\frac{1}{2}x-\frac{1}{8}x^2$ for $x\rightarrow0$; for further insights, refer to \cite[Remark 1]{liu2023near}. The near-field channel model \eqref{NFC_ULA_LOS_Model} differs from the far-field model \eqref{FFC_ULA_LOS_Model} in the following two aspects: {\romannumeral1}) The far-field model assumes approximately identical angles for the links connecting each array element to the user. Thus, the propagation distance of each link linearly increases with the antenna index, resulting in a linearly varying phase shift for far-field channels. {\romannumeral2}) In the far-field model, variations in channel power along the BS array are considered negligible. In essence, these differences arise from the inclusion of the additional range dimension $\{r_{k,i}\}_{i=-\widetilde{N},\ldots,\widetilde{N}}$ in near-field channels.

Next, we consider strategies pertaining to near-field SDMA and MIMO-NOMA. 
\subsubsection{Near-Field Space-Division Multiple Access}
We commence by investigating whether the property of ``favorable propagation'' holds or approximately holds also for near-field channels. Specifically, the inner product of the normalized near-field channel is given by
\begin{align}
\rho_{k,k'}=\frac{\lvert{\mathbf{h}}^{\mathsf{H}}(r_k,\theta_k){\mathbf{h}}(r_{k'},\theta_{k'})\rvert}
{\lVert{\mathbf{h}}(r_k,\theta_k)\rVert\lVert{\mathbf{h}}(r_{k'},\theta_{k'})\rVert}.
\end{align}
However, this expression is computationally intractable. To facilitate calculation, we adopt certain approximations. Following the principle in \cite{wu2023multiple}, we assume $\frac{\sqrt{\beta_{\mathsf{r}}}r_{\mathsf{r}}}{r_{k,i}}\approx \frac{\sqrt{\beta_{\mathsf{r}}}r_{\mathsf{r}}}{r_{k}}$ ($\forall i$), and we utilize the approximation in \eqref{ULA_F_Appr}. These approximations remain valid when the user is situated beyond the \emph{uniform-power distance}, where the channel gain disparity of each link becomes negligible \cite{liu2024near}. As discussed in \cite{liu2023near,liu2024near}, the uniform-power distance is generally much smaller than the Rayleigh distance. Therefore, these approximations prove effective for the majority of the area within the near-field region. Under these conditions, $\rho_{k,k'}$ can be approximated as follows:
\begin{align}\label{Approximated_Inner_Product_Near_Field_Channels}
\rho_{k,k'}\approx\left\lvert\sum\nolimits_{i=-\tilde{N}}^{\tilde{N}}{\rm{e}}^{{\rm{j}}\delta_1\frac{i}{N_{\mathsf{BS}}}(1+\delta_2\frac{i}{N_{\mathsf{BS}}})}\frac{1}{N_{\mathsf{BS}}}\right\rvert,
\end{align}
where $\delta_1={\mathsf{a}}N_{\mathsf{BS}}$, $\delta_2=\frac{\mathsf{a}}{\mathsf{b}}N_{\mathsf{BS}}$, ${\mathsf{a}}=\frac{2\pi}{\lambda}d^2(\frac{\sin^2(\theta_k)}{2r_k}-\frac{\sin^2(\theta_{k'})}{2r_{k'}})$, and ${\mathsf{b}}=\frac{2\pi}{\lambda}d(\cos(\theta_{k'})-\cos(\theta_{k}))$. Based on the concept of definite integral, we obtain the following theorem.
\vspace{-5pt}
\begin{theorem}\label{Theorem_Favorable_Propagation_LOS_NFC}
(\textbf{Near-Field Favorable Propagation}) Under our considered near-field channels, the inner product of the normalized channel vectors in \eqref{Approximated_Inner_Product_Near_Field_Channels} satisfies
\begin{equation}\label{Favorable_Propagation_LOS_NFC}
\lim_{N_{\mathsf{BS}}\rightarrow\infty}\rho_{k,k'}\approx0.
\end{equation}
\end{theorem}
\vspace{-5pt}
\begin{IEEEproof}
Please refer to Appendix \ref{Proof_Theorem_Favorable_Propagation_LOS_NFC} for more details.
\end{IEEEproof}
Due to the approximations of $\frac{\sqrt{\beta_{\mathsf{r}}}r_{\mathsf{r}}}{r_{k,i}}\approx \frac{\sqrt{\beta_{\mathsf{r}}}r_{\mathsf{r}}}{r_{k}}$ ($\forall i$) and \eqref{ULA_F_Appr}, the result in \eqref{Favorable_Propagation_LOS_NFC} only reveals the approximate asymptotic behaviour of $\rho_{k,k'}$, akin to an approximate ``favorable propagation'' \cite{wu2023multiple,liu2023near}. In simpler terms, we can obtain $\lim_{N_{\mathsf{BS}}\rightarrow\infty}\rho_{k,k'}\ll 1$, which was verified by numerical results in \cite{zhao2023modeling}. This implies that the near-field channels become approximately asymptotically orthogonal for arbitrary users in \emph{different locations}. 
In contrast, as stated in Theorem \ref{Theorem_FFC_Channel_Correlation}, the far-field channels become asymptotically orthogonal only for users in \emph{different directions}. This comparison underscores the broader applicability of the property of ``favorable propagation'' in NFC. This advantageous property stems from the introduction of the range dimension in \eqref{NFC_ULA_LOS_Model} due to the spherical-wave propagation, which can be further leveraged to mitigate IUI and enhance the system's spectral efficiency.
 
In far-field LoS communications, achieving the highest DoFs through SDMA is contingent upon users being situated in different directions. If two users are aligned in the same direction, the resultant IUI becomes too substantial for linear beamforming to mitigate effectively. Consequently, SDMA in FFC essentially transforms into ``angle division multiple access'' (ADMA). In scenarios where two users share the same direction, NOMA becomes imperative to suppress IUI at the receiver. However, employing NOMA may lead to a reduction in the system's multiplexing gain, as the interference at the weaker user's end remains unresolved. Conversely, leveraging the findings in \eqref{Favorable_Propagation_LOS_NFC}, SDMA in NFC can comprehensively eliminate IUI, even when users share the same direction. Consequently, SDMA in NFC fundamentally manifests as ``range division multiple access'' (RDMA), which encompasses ADMA as a special case. This inherent property facilitates the achievement of \emph{massive spatial multiplexing}, enabling the concurrent service of hundreds or even thousands of users with minimal IUI in NFC.

Furthermore, it is important to highlight that, as per Theorem \ref{Theorem_FFC_Channel_Correlation}, FFC can solely support energy focusing in a dedicated direction, commonly referred to as \emph{beamsteering}. In contrast, \eqref{Favorable_Propagation_LOS_NFC} indicates that energy transmitted from the BS can be concentrated on a specific point in NFC, thus characterizing beamforming in NFC as \emph{beamfocusing} \cite{liu2023near}. This distinction underscores the potential for more precise energy localization in NFC compared to FFC. Drawing from this principle, numerous beamforming or beamfocusing algorithms have been developed to enhance the spectral efficiency of near-field SDMA or RDMA systems. For comprehensive reviews of these advancements, readers are encouraged to refer to recent overview papers \cite{wang2024tutorial,lu2023tutorial,liu2024near}.

\begin{figure}[!t]
 \centering
\setlength{\abovecaptionskip}{0pt}
\includegraphics[width=0.45\textwidth]{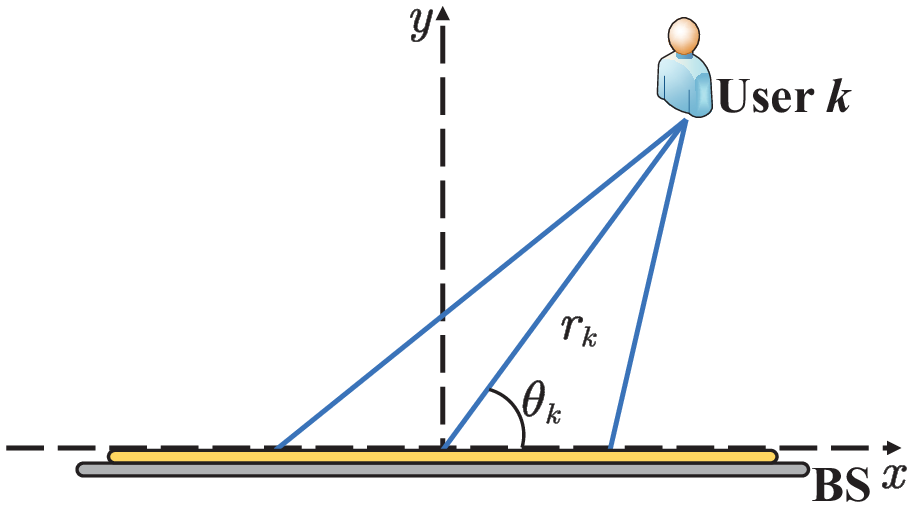}
\caption{Near-field LoS channel model for CAP linear array.}
\label{CAP_ULA_NFC}
\end{figure}

The above discussion on near-field RDMA assumes a scenario where the BS is equipped with a large number of spatially-discrete (SPD) antennas with specific spacing. Since adding more antennas can increase the spatial DoFs, we now turn our attention to a limiting case where the BS is equipped with an infinite number of antennas with infinitesimal spacing, i.e., the BS is equipped with a continuous-aperture (CAP) linear array of length $A=(N_{\mathsf{BS}}-1)d$, as illustrated in {\figurename} {\ref{CAP_ULA_NFC}}.
Unlike the SPD array that delivers finite-dimensional signal vectors, the CAP array supports a continuous distribution of source currents within the transmit aperture. This setup generates an electric radiation field at the receive aperture. The spatial channel impulse response between any two points on the transceiving apertures is described by \emph{Green's function}. This function connects the transmitter’s current distribution and the receiver’s electric field via a spatial integral.

Specifically, let ${\mathsf{j}}({\bm{{\mathsf{a}}}})\in{\mathbbmss{C}}$ denote the continuous distribution of source currents, where ${\bm{{\mathsf{a}}}}=[x,0]^{\mathsf{T}}\in{\mathbbmss{R}}^{2\times1}$ represents the source point within the transmit aperture ${\mathcal{A}}=\{[x,0]^{\mathsf{T}}|x\in[-\frac{A}{2},\frac{A}{2}]\}$. Treating each user as a sizeless point, the electric radiation field $\mathsf{e}({\bm{{\mathsf{r}}}}_k)$ generated at user $k$ is formulated as follows \cite{miller2000communicating}:
\begin{align}
\mathsf{e}({\bm{{\mathsf{r}}}}_k)=\int_{{\mathcal{A}}}{\mathsf{G}}({\bm{{\mathsf{a}}}},{\bm{{\mathsf{r}}}}_k){\mathsf{j}}({\bm{{\mathsf{a}}}}){\rm{d}}{\bm{{\mathsf{a}}}},
\end{align}
where ${\mathsf{G}}({\bm{{\mathsf{a}}}},{\bm{{\mathsf{r}}}}_k)$ denotes Green's function and is given by ${\mathsf{G}}({\bm{{\mathsf{a}}}},{\bm{{\mathsf{r}}}}_k)=\frac{{\rm{e}}^{-{\rm{j}}\frac{2\pi}{\lambda}\lVert {\bm{{\mathsf{r}}}}_k - {\bm{{\mathsf{a}}}}\rVert}}{4\pi \lVert {\bm{{\mathsf{r}}}}_k - {\bm{{\mathsf{a}}}}\rVert}$ in the near-field region. Let ${\mathsf{j}}_k({\bm{{\mathsf{a}}}})\in{\mathbbmss{C}}$ represent the source currents conveying the normalized data symbol $s_k$ to user $k$. Then, the received signal of user $k$ can be written as follows:
\begin{equation}
\begin{split}
y_k&=s_k\int_{{\mathcal{A}}}{\mathsf{G}}({\bm{{\mathsf{a}}}},{\bm{{\mathsf{r}}}}_k){\mathsf{j}}_k({\bm{{\mathsf{a}}}}){\rm{d}}{\bm{{\mathsf{a}}}}\\
&+\sum\nolimits_{k'\ne k}s_{k'}\int_{{\mathcal{A}}}{\mathsf{G}}({\bm{{\mathsf{a}}}},{\bm{{\mathsf{r}}}}_k){\mathsf{j}}_{k'}({\bm{{\mathsf{a}}}}){\rm{d}}{\bm{{\mathsf{a}}}} + n_k,
\end{split}
\end{equation}
where $n_k\sim{\mathcal{CN}}(0,\sigma_k^2)$ denotes AWGN, and the currents are subject to the power budget $\sum_{k=1}^{K}\int_{{\mathcal{A}}}\lvert {\mathsf{j}}_k({\bm{{\mathsf{a}}}}) \rvert^2{\rm{d}}{\bm{{\mathsf{a}}}}\leq p$. The received SINR at user $k$ is expressed as follows:
\begin{equation}\label{Received_SINR_CAP}
\gamma_k=\frac{\left\lvert\int_{{\mathcal{A}}}{\mathsf{G}}({\bm{{\mathsf{a}}}},{\bm{{\mathsf{r}}}}_k){\mathsf{j}}_k({\bm{{\mathsf{a}}}}){\rm{d}}{\bm{{\mathsf{a}}}}\right\rvert^2}{\sum_{k'\ne k}\left\lvert\int_{{\mathcal{A}}}{\mathsf{G}}({\bm{{\mathsf{a}}}},{\bm{{\mathsf{r}}}}_k){\mathsf{j}}_{k'}({\bm{{\mathsf{a}}}}){\rm{d}}{\bm{{\mathsf{a}}}}\right\rvert^2+\sigma_k^2}.
\end{equation}
It can be observed from \eqref{Received_SINR_CAP} that CAP array-based near-field SDMA hinges on the design of the continuous currents $\{{\mathsf{j}}_k({\bm{{\mathsf{a}}}})\}_{k\in{\mathcal{K}}}$, which is in contrast to the discrete beamformer design in SPD array-based SDMA, as shown in \eqref{SDMA_User_k_Rate}.

We now turn our attention to exploring whether the property of ``favorable propagation'' holds or approximately holds for CAP arrays. The inner product of the normalized Green's function is defined as follows:
\begin{align}
\tilde{\rho}_{k,k'}=\frac{\left\lvert\int_{{\mathcal{A}}}{\mathsf{G}}^{*}({\bm{{\mathsf{a}}}},{\bm{{\mathsf{r}}}}_k){\mathsf{G}}({\bm{{\mathsf{a}}}},{\bm{{\mathsf{r}}}}_{k'}){\rm{d}}{\bm{{\mathsf{a}}}}\right\rvert}
{\sqrt{\int_{{\mathcal{A}}}\lvert{\mathsf{G}}({\bm{{\mathsf{a}}}},{\bm{{\mathsf{r}}}}_k)\rvert^2{\rm{d}}{\bm{{\mathsf{a}}}}
\int_{{\mathcal{A}}}\lvert{\mathsf{G}}({\bm{{\mathsf{a}}}},{\bm{{\mathsf{r}}}}_{k'})\rvert^2{\rm{d}}{\bm{{\mathsf{a}}}}}}.
\end{align}
This expression is computationally challenging. To ease computation, we make the assumption that each near-field user is positioned beyond the uniform-power distance. Under this assumption, Green's function can be approximated as follows \cite{miller2000communicating}:
\begin{align}\label{Favorable_Propagation_LOS_NFC_CAP}
{\mathsf{G}}({\bm{{\mathsf{a}}}},{\bm{{\mathsf{r}}}}_k)\approx \frac{{\rm{e}}^{-{\rm{j}}\frac{2\pi}{\lambda}
(r_k-x\cos(\theta_k)+\frac{x^2}{2r_k}\sin^2(\theta_k))}}{4\pi r_k}.
\end{align}
Applying similar derivation steps as in \eqref{Favorable_Propagation_LOS_NFC}, we establish
\begin{align}
\lim\nolimits_{A\rightarrow\infty}\tilde{\rho}_{k,k'}\approx 0,
\end{align}
where $A=(N_{\mathsf{BS}}-1)d$ represents the aperture size of the CAP array used at the BS. This indicates that in CAP array-based NFC, as the array aperture grows infinitely large, the channels tend to become asymptotically orthogonal for users located at arbitrary but different positions. This beneficial characteristic is also a result of spherical-wave propagation, which effectively mitigates IUI and enhances the system's spectral efficiency. Based on the result in \eqref{Favorable_Propagation_LOS_NFC_CAP}, an appropriate design for ${\mathsf{j}}_k({\bm{{\mathsf{a}}}})$ is proposed as follows:
\begin{align}
{\mathsf{j}}_k({\bm{{\mathsf{a}}}})=\sqrt{\frac{p}{K}}\frac{{\mathsf{G}}^{*}({\bm{{\mathsf{a}}}},{\bm{{\mathsf{r}}}}_k)}{\sqrt{\int_{{\mathcal{A}}}\lvert{\mathsf{G}}({\bm{{\mathsf{a}}}},{\bm{{\mathsf{r}}}}_k)\rvert^2{\rm{d}}{\bm{{\mathsf{a}}}}}}
\end{align}
This design aims to achieve asymptotically orthogonal transmission in CAP array-based NFC scenarios.
\subsubsection{Near-Field Hybrid Beamforming-Based Transmission}
Near-field SDMA or RDMA presents the potential for achieving massive spatial multiplexing, but it typically requires a fully digital implementation. However, given the substantial size of ELAAs, practical NFC systems may lean towards a hybrid digital-analog beamforming architecture. In this subsection, we study the performance analysis of near-field HB-SDMA.

Before considering the multiuser scenario, we first analyze the single-user case. To facilitate clarity, we assume that the user is positioned along the $y$-axis, and $N_{\mathsf{RF}}=1$, enabling us to investigate the performance of near-field analog beamforming. By dropping the user index, the analog beamformer can be formulated as ${\mathbf{F}}=[{\rm{e}}^{-{\rm{j}}\frac{2\pi}{\lambda}r_{-\widetilde{N}}},\ldots,{\rm{e}}^{-{\rm{j}}\frac{2\pi}{\lambda}r_{\widetilde{N}}}]$. In this case, the received SNR is calculated in the following lemma.
\vspace{-5pt}
\begin{lemma}\label{Lemma_SU_NFC_Analog_exp2}
Under our considered case, the received SNR in the near-field HB-based system is given by
\begin{align}\label{SU_NFC_Analog_exp2}
\gamma=\frac{p\beta_{\mathsf{r}}}{N_{\mathsf{BS}}\sigma^2}\frac{r_{\mathsf{r}}^2}{d^2}2\ln\left({\tilde{N}}\frac{d}{r}+\sqrt{\left({\tilde{N}}\frac{d}{r}\right)^2+1}\right).
\end{align}
\end{lemma}
\vspace{-5pt}
\begin{IEEEproof}
Please refer to Appendix \ref{Proof_Lemma_SU_NFC_Analog_exp2} for more details.
\end{IEEEproof}
As $N_{\mathsf{BS}}\rightarrow\infty$, we obtain
\begin{align}\label{Theorem_Received_SNR_Analog_SU_Near_Field_Asymptotic}
\lim_{N_{\mathsf{BS}}\rightarrow\infty}\gamma=\frac{2p\beta_{\mathsf{r}}}{\sigma^2}\frac{r_{\mathsf{r}}^2}{d^2}
\lim_{\widetilde{N}\rightarrow\infty}\frac{\ln\left(\frac{{\tilde{N}}d}{r}+\sqrt{\left(\frac{{\tilde{N}}d}{r}\right)^2+1}\right)}{2\widetilde{N}+1}=0.
\end{align} 
The result in \eqref{Theorem_Received_SNR_Analog_SU_Near_Field_Asymptotic} demonstrates the inability of the user to receive any energy from the infinitely large array when employing analog beamforming. This peculiar result can be explained as follows. Equation \eqref{Theorem_Received_SNR_Analog_SU_Near_Field_Asymptotic} is derived under the LoS model shown in \eqref{NFC_ULA_LOS_Model}, which is exact when the user is positioned in the radiating near and far fields. The boundary between the reactive and radiating near fields is determined by $0.62\sqrt{\frac{A^3}{\lambda}}$ \cite{liu2023near}. When the user is positioned at the $y$-axis, the required number of BS antennas to place the user outside the reactive near field is determined as follows:
\begin{equation}\label{SU_ULA_Minimum_Number1}
\begin{split}
r\geq0.62\sqrt{\frac{N_{\mathsf{BS}}^3d^3}{\lambda}}\Leftrightarrow N_{\mathsf{BS}}\leq \frac{\lambda^{\frac{1}{3}}}{d}\left(\frac{r}{0.62}\right)^{2/3}\triangleq N_{\mathsf{rad}},
\end{split}
\end{equation}
which means that \eqref{SU_NFC_Analog_exp2} holds for $N_{\mathsf{BS}}\leq N_{\mathsf{rad}}$. We note that when $N_{\mathsf{BS}}\rightarrow\infty$, the result derived in \eqref{SU_NFC_Analog_exp2} is not exact, and thus the asymptotic result derived in \eqref{Theorem_Received_SNR_Analog_SU_Near_Field_Asymptotic} is also not exact, since an exact characterization of the asymptotic SNR achieved by analog beamforming would require the modeling of the reactive near-field channel, which is a promising avenue for future exploration.

\begin{figure}[!t]
 \centering
\setlength{\abovecaptionskip}{0pt}
\includegraphics[width=0.45\textwidth]{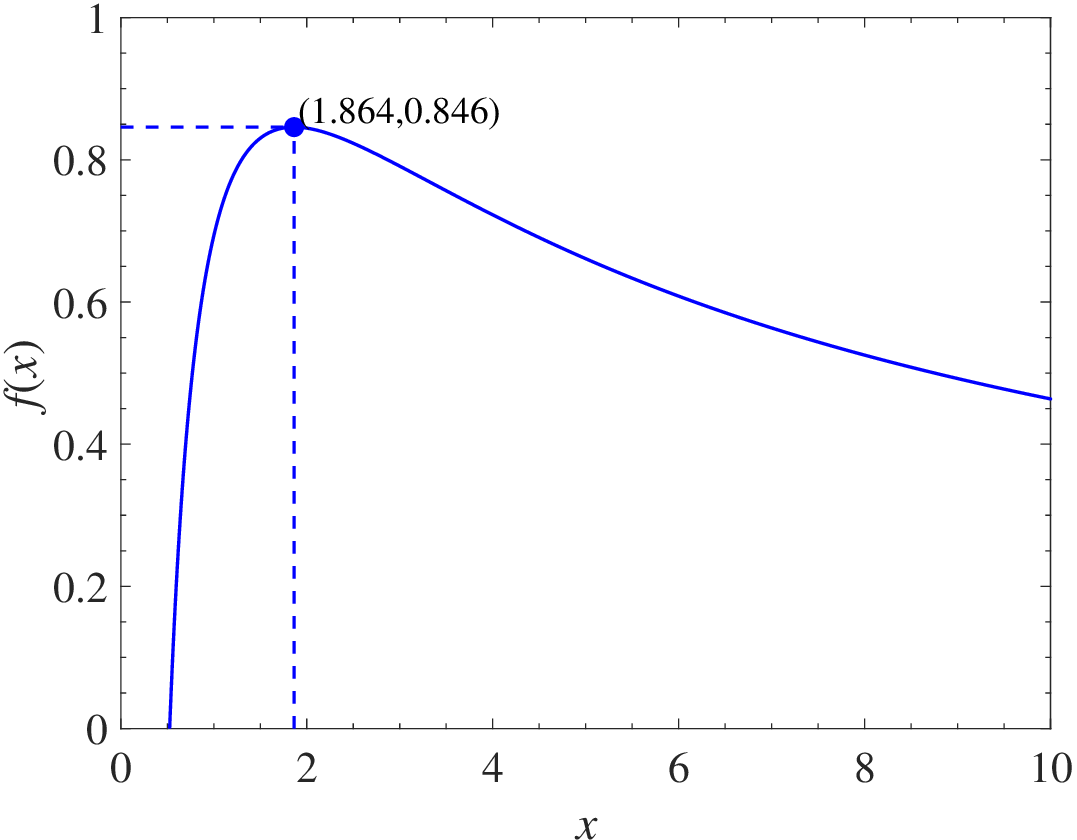}
\caption{Illustration of $f(x)=\frac{\ln(x+\sqrt{x^2+1})}{x}$.}
\label{SU_2_F_1}
\end{figure}

{\figurename} {\ref{SU_2_F_1}} illustrates the values of $f(x)=\frac{\ln(x+\sqrt{x^2+1})}{x}$ versus $x$. From the graph, it is evident that as $x$ increases, $f(x)$ initially rises and then declines. The optimal value of $x$ that maximizes $f(x)$ is found to be $x^{\star}=1.864$. These findings imply that the received SNR in \eqref{SU_NFC_Analog_exp2} monotonically increases with $N_{\mathsf{BS}}$ when $N_{\mathsf{BS}}<\frac{2x^{\star}}{d/r}\frac{3.728 r}{d}\triangleq N_{\mathsf{BS}}^{\star}$, and decreases with $N_{\mathsf{BS}}$ when $N_{\mathsf{BS}}>N_{\mathsf{BS}}^{\star}$. Assuming $d=\frac{\lambda}{2}$, we have $N_{\mathsf{rad}}=\left(\frac{\sqrt{2}r}{0.62d}\right)^{\frac{2}{3}}$. Given the condition $r\gg d$, we have $\frac{r}{d}\gg1$, leading to
\begin{equation}\label{Array_Element_Compare}
N_{\mathsf{rad}}<N_{\mathsf{BS}}^{\star}.
\end{equation}
The results in \eqref{Array_Element_Compare} suggest that, in the radiating near-field region, the received SNR achieved by analog beamforming monotonically increases with the number of antennas. These results also emphasize \emph{the effectiveness of analog beamforming in the radiating near field}.

\begin{figure}[!t]
 \centering
\setlength{\abovecaptionskip}{0pt}
\includegraphics[width=0.45\textwidth]{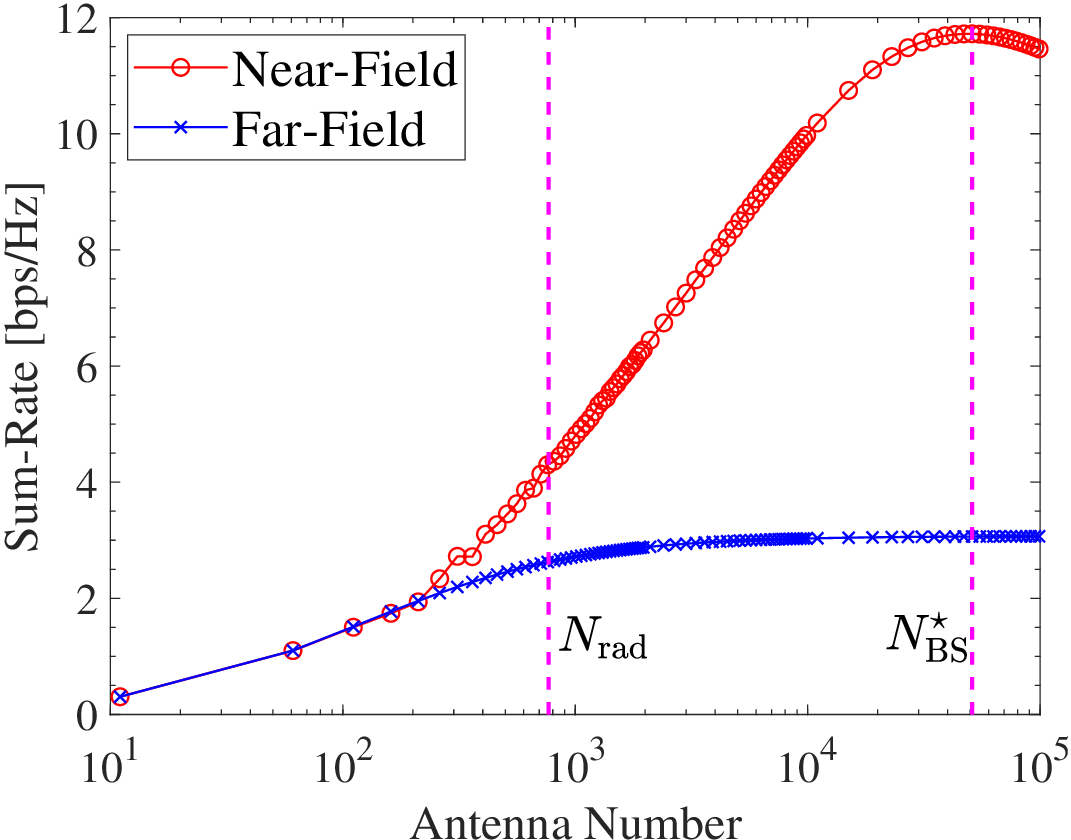}
\caption{Illustration of the sum-rate achieved by HB-SDMA. The system is operating at a frequency of 28 GHz. $K=N_{\mathsf{RF}}=2$, $d=\frac{\lambda}{2}$, $\lambda=0.01$ m, $\frac{p{{\beta_{\mathsf{r}}}}}{\sigma_1^2}=\frac{p{{\beta_{\mathsf{r}}}}}{\sigma_2^2}=10$ dB, $r_1=50$ m, $\theta_1=90^{\circ}$, $r_2=20$ m, and $\theta_2=90^{\circ}$. The transmit power is equally allocated to all users.}
\label{SDMA_NF}
\end{figure}

We extend the above findings to the multiuser scenario by considering a near-field HB-SDMA system. For the sake of clarity of presentation, we assume that  $N_{\mathsf{RF}}=K$. Under this scenario, we first assign to each user $k$ an analog beamformer aligned with its channel. Subsequently, we employ digital MRT beamforming. {\figurename} {\ref{SDMA_NF}} compares the sum-rate of a two-user HB-MISO system under the far-field and near-field channel models. As can be observed, the sum-rate under the near-field model initially increases and then declines as $N_{\mathsf{BS}}$ increases, indicating the existence of an optimal value of $N_{\mathsf{BS}}$, i.e., $N_{\mathsf{BS}}^{\star}$. Additionally, for reference, we present the antenna number required to position the users within the radiated region, labeled as $N_{\mathsf{rad}}$. As can be seen from this graph, $N_{\mathsf{rad}}<N_{\mathsf{BS}}^{\star}$, which indicates that the sum-rate achieved by HB increases monotonically with $N_{\mathsf{BS}}$ when users are positioned in the radiating near-field region. Furthermore, from the graph, we observe that near-field HB-SDMA attains a higher sum-rate compared to its far-field counterpart.
\subsubsection{Near-Field MIMO-NOMA}
Due to the relevance of the range dimension in near-field channels, near-field SDMA or RDMA has the capability to suppress IUI even when two users are positioned in the same direction. This characteristic not only increases the spatial DoFs but also enhances spatial multiplexing gains. However, in scenarios such as massive connectivity where the number of RF chains is less than the number of users, MIMO-NOMA emerges as a promising technology for mitigating IUI and offering flexible resource allocation. The distinctive attributes of NFC introduce novel design opportunities for facilitating near-field NOMA.

In FFC, users in similar angular directions are grouped into clusters and served by a common beamsteering vector. Within each cluster, users are decoded in a \emph{near-to-far} SIC order based on their effective channel gains, which are determined by their distances from the BS. NFC, benefiting from its spherical wavefronts, enables the creation of focused beams in specific spatial regions. This feature introduces additional DoFs in both the angle and range domains, enabling precise signal enhancement and co-channel interference mitigation. By leveraging near-field beamfocusing, signal energy can be concentrated on users situated farther from the BS, resulting in higher effective channel gains for distant users compared to those nearer to the BS. Consequently, a \emph{far-to-near} SIC decoding order among users becomes feasible in NFC, enhancing system transmission flexibility. This operational approach proves advantageous, particularly when the far user requires higher communication capacity than the near user. Building on this foundation, researchers in \cite{zuo2023non} explored resource allocation in near-field NOMA systems, demonstrating that near-field NOMA can achieve higher spectral efficiency compared to conventional far-field NOMA. 

NOMA has found application in downlink legacy networks as well, where preconfigured beamformers intended for legacy near-field users can be repurposed to serve additional far-field NOMA users \cite{ding2023noma}, near-field NOMA users \cite{ding2024resolution}, and semi-NOMA far-/near-field users \cite{ding2023utilizing,ding2024design} by utilizing the preconfigured beams. Theoretical analyses and numerical results indicate that NOMA can effectively enhance connectivity and system throughput.
\subsection{Discussions and Outlook}
In this section, we have presented a concise tutorial treatment on MA technologies exploiting the spatial domain, focusing primarily on simple configurations such as ULAs, MU-MISO systems, and narrowband transmission. Extending these findings to more general cases is relatively straightforward, as recent overview papers in this domain, such as \cite{wang2024tutorial,lu2023tutorial,liu2024near}, offer comprehensive insights. Over the past decades, the utilization of antenna arrays to mitigate IUI has been extensively studied. Despite the existence of various system configurations, multiple-antenna systems are evolving towards ELAAs and higher frequencies, necessitating the exploration into NFC. Presently, research on MA in NFC is still in its infancy, with numerous open research challenges awaiting investigation.

Understanding the information-theoretic aspects of near-field MA (SDMA, NOMA, and others) is crucial. Traditional MA information theory often relies on assumptions that are not entirely physically consistent, such as far-field, discretized, and monochromatic EM fields impaired by white noise. These assumptions can lead to mismatches with the actual nature of the EM fields supporting near-field MA systems. While some results on the information-theoretic limits of NFC can be directly extended from their far-field counterparts, many require revision to account for the characteristics of near-field EM propagation, such as the spatial DoFs \cite{ouyang2023near}. In limiting scenarios where both BS and users are equipped with CAP arrays, characterizing the capacity region and associated coding schemes remains an open problem in the context of near-field MA due to the intractability of EM channel modeling and EM noise modeling \cite{zhu2024electromagnetic}. Further research efforts are needed in this direction to address these challenges.

In practical deployments, NFC will be utilized in multi-cell environments. As the density of wireless networks increases, inter-cell interference emerges as a significant obstacle to realizing the benefits of NFC. Therefore, analyzing the performance of near-field MA at the network level and uncovering system design insights concerning interference management and resource allocation are essential. Multi-cell environments introduce more complex wireless propagation scenarios, where the near-field of one BS may overlap with another BS's far field or near field. Designing efficient SDMA, NOMA, and other MA strategies for such intricate communication scenarios poses a significant challenge and represents an important direction for future research.
\section{Multiple Access: From Communications to Sensing }\label{Section: MA for ISAC}
The future landscape of wireless networks is anticipated to play a pivotal role in shaping a connected, smart, and intelligent wireless world \cite{saad2019vision}. This necessitates a paradigm shift for the design of future networks to support high-quality wireless connectivity alongside high-accuracy sensing capabilities. The preceding sections have offered an overview of MA technologies designed specifically for communications-centric (C-C) tasks. In the sequel, we transition from communications to sensing, exploring MA technologies tailored for \emph{dual-function sensing and communications (DFSAC) tasks}.
\subsection{Overview of Integrated Sensing and Communications}
To meet the dual-functional requirement of sensing and communications (S\&C) in future networks, various technologies have been introduced, with integrated S\&C (ISAC) emerging as a prominent approach to achieve DFSAC using a unified time-frequency-power-hardware resource block \cite{cui2021integrating}. In contrast to conventional orthogonal S\&C (OSAC) approaches, where S\&C functionalities share isolated resources, ISAC is envisioned to be more spectral-, energy-, and hardware-efficient \cite{liu2023seventy}. The term ``\emph{ISAC}'' was first coined in \cite{cui2021integrating}, and its significance was underscored at the 44th meeting in Geneva on June 22, 2023, where ITU-R WP 5D successfully drafted a new recommendation for IMT-2030 (6G) and designated ISAC as one of the six usage scenarios in 6G \cite{dong2024sensing}. Numerous research efforts have been dedicated to this subject; for a detailed overview, we refer to \cite{liu2020joint,zhang2021overview,zhang2021enabling,liu2022integrated,lu2023integrated}.

However, despite its promise for 6G, striking a balance between the two functionalities (i.e., S\&C) poses a significant challenge in the design of ISAC. This challenge is inherently complex due to the potential ``\emph{double non-orthogonality}'' which arises from both the MA of the S\&C functionalities to the same resource, as well as the MA of different communication users (CUs) to the same resource. This ``double non-orthogonality'' results in both \emph{inter-functionality interference (IFI) and IUI}, which underscores the need for the development of efficient interference mitigation and resource management approaches. In conventional communication networks, these approaches are typically facilitated by MA techniques \cite{liu2022developing,liu2022evolution,liu2017nonorthogonal}. Following the same path, we explore the development of ISAC from the MA perspective, with a specific focus on NOMA \cite{mu2023noma}. This emphasis is motivated by two key factors. Firstly, ISAC can be heuristically conceptualized as the non-orthogonal access of S\&C to the same resource block. Secondly, since both IFI and IUI may coexist in ISAC, NOMA emerges as a promising interference management tool \cite{liu2017nonorthogonal}.
\begin{figure}[!t]
    \centering
    \subfigure[Downlink ISAC.]{
        \includegraphics[width=0.4\textwidth]{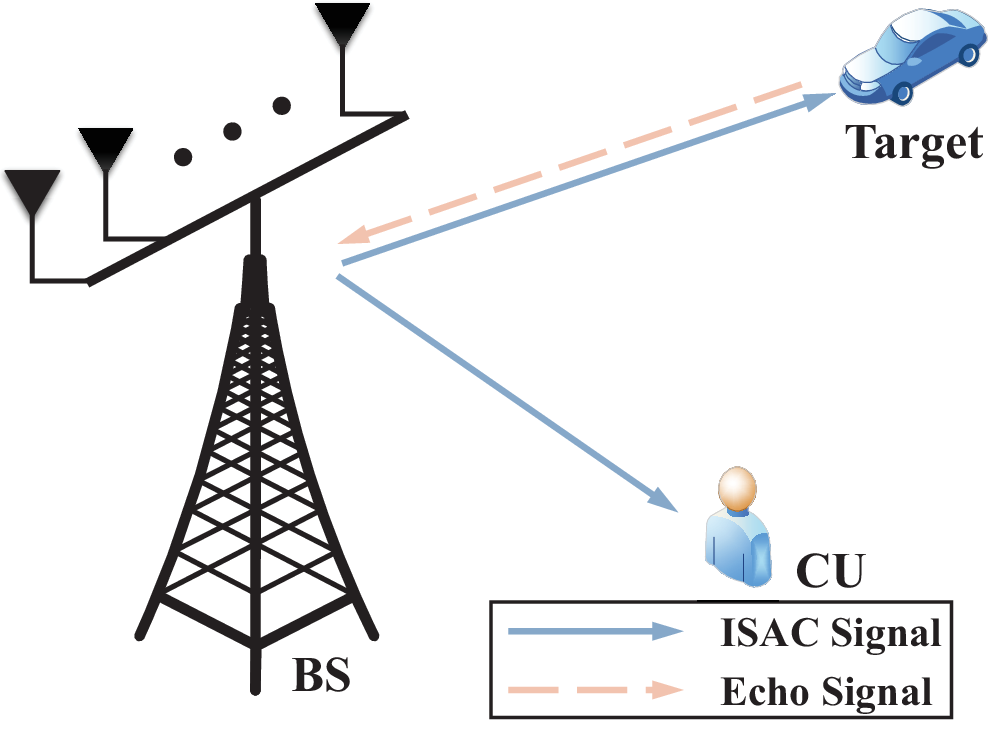}
        \label{Figure: DISAC_System}
    }
    \subfigure[Uplink ISAC.]{
        \includegraphics[width=0.4\textwidth]{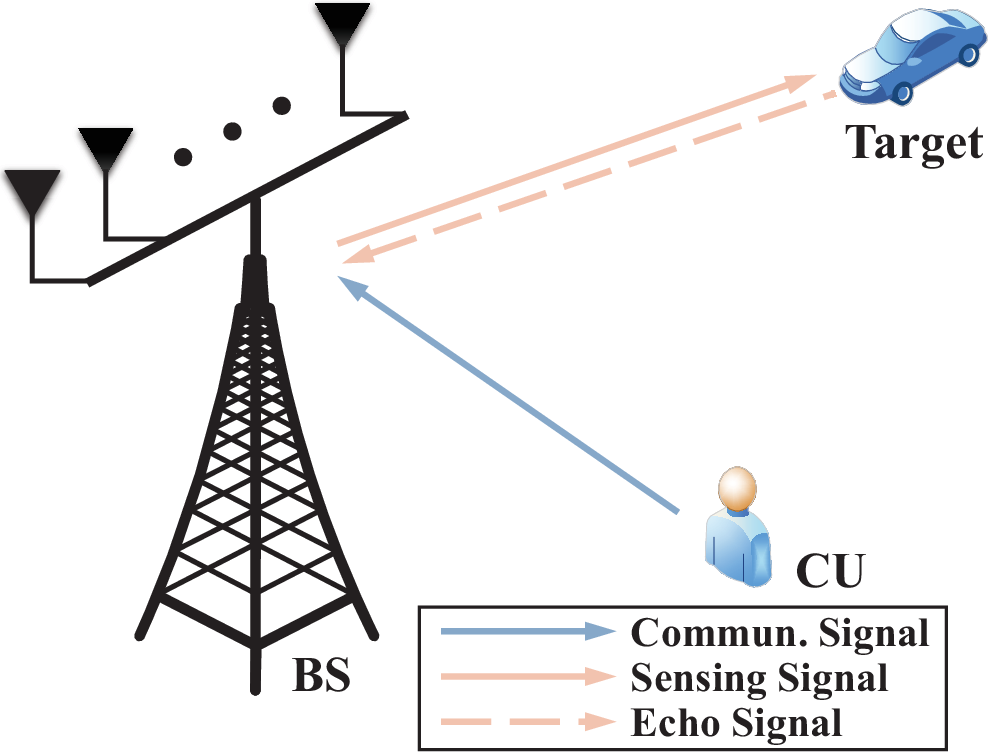}
        \label{Figure: UISAC_System}	
    }
    \caption{Illustration of a downlink ISAC system and an uplink ISAC system. In each system, there is one ISAC BS, one sensing target, and one CU.}
\end{figure}
\subsection{NOMA-Based ISAC Frameworks}\label{Section: NOMA for ISAC}
By harnessing SPC-SIC, NOMA enables the concurrent service of multiple CUs over the same radio resources. This approach not only significantly enhances connectivity but also improves resource efficiency. However, integrating NOMA into ISAC is not a straightforward process and demands a thoughtful redesign. This complexity arises from the fact that NOMA, in its current usage, is tailored for serving CUs. As a result, there is a lack of clarity on how NOMA can be effectively employed to support DFSAC within ISAC \cite{mu2023noma}. In the subsequent sections, we consider various NOMA-based ISAC frameworks.
\subsubsection{NOMA for Downlink Integrated Sensing and Communications}
We commence our exploration with a downlink ISAC system, where the DFSAC BS aims to communicate with a downlink CU while simultaneously sensing the target in the nearby environment, as depicted in {\figurename} \ref{Figure: DISAC_System}. In this scenario, the DFSAC BS transmits the communication signal to the CU. Concurrently, the DFSAC BS emits a sensing probing signal, which is reflected by the target and returned as the sensing echo. By analyzing the received sensing echo, the DFSAC BS can estimate relevant sensing parameters. 

The time interval of the sensing signal may extend beyond the round-trip time of the signal traveling between the BS and the targets. Consequently, the two stages of sensing (sending the probing signal and receiving the echo signal) could be coupled in time. This means that the DFSAC BS should work in the full-duplex mode \cite{rahman2019framework}. To mitigate the resulting self-interference, we assume the BS is equipped with two sets of spatially well-separated antennas. Other methods for self-interference mitigation include RF suppression and baseband suppression, among others \cite{zhang2015full}. To characterize the maximum possible performance, we assume the self-interference can be completely eliminated.

One primary challenge in downlink ISAC is the design of DFSAC waveforms to combat potential IFI. Additionally, managing typical inter-user communication interference becomes more challenging in downlink ISAC due to the added requirements imposed by sensing support. To address these challenges, we explore the potential applications of NOMA to enhance the performance of downlink ISAC. We investigate two novel designs: \romannumeral1) NOMA-empowered downlink ISAC and \romannumeral2) NOMA-inspired downlink ISAC. The key distinction between the two designs lies in whether NOMA is employed within the communication functionality or between the two functionalities in ISAC.

\begin{figure}[!t]
  \centering
  \includegraphics[width=0.45\textwidth]{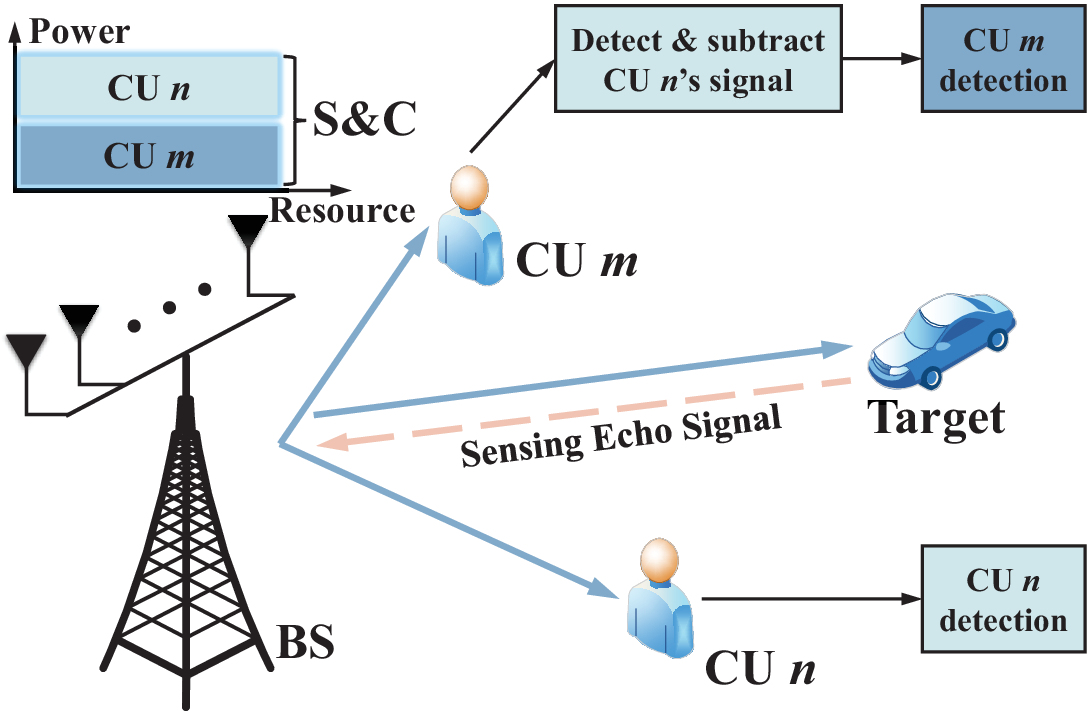}
  \caption{Illustration of the NOMA-empowered downlink ISAC design.}
  \label{Figure: E_NOMA_DISAC}
\end{figure}

$\bullet$ \emph{NOMA-Empowered Downlink Integrated Sensing and Communications}: 
In NOMA-empowered downlink ISAC \cite{wang2022noma}, SPC and SIC techniques are employed for transmitting and detecting the communication signal of each CU, as illustrated in {\figurename} \ref{Figure: E_NOMA_DISAC}. The SPC-based communication signal is also leveraged for target sensing, effectively eliminating IFI. This constitutes a direct extension of NOMA from conventional communication systems to the ISAC framework, thereby empowering the communication functionality of ISAC with NOMA. The NOMA paradigm utilized here can be based on the beamformer-based, cluster-based, or cluster-free NOMA structures as outlined in Section \ref{Section: MIMO-NOMA}.

Notably, the benefits offered by NOMA in the NOMA-empowered downlink ISAC design are twofold. Firstly, in overloaded regimes or underloaded systems with strongly correlated communication channels, NOMA provides additional DoFs for IUI (i.e., inter-CU interference) mitigation through SIC decoding, achieving superior communication performance compared to conventional downlink ISAC designs. Grounded in the potential enhancement of communication performance with NOMA, DFSAC waveforms can be designed to predominantly favour sensing, consequently improving the sensing performance. Secondly, for the sensing-centric (S-C) design of ISAC, DFSAC waveforms need to be constructed based on specific beam patterns to ensure high-quality sensing, which may limit the spatial DoFs available for IUI mitigation. NOMA compensates for this limitation by providing additional DoFs through SIC decoding. Numerical results have affirmed the effectiveness of this NOMA-empowered downlink ISAC design, particularly in overloaded regimes \cite{wang2022noma}. The NOMA-empowered ISAC framework has garnered significant research attention and has been explored alongside unicast/multicast transmission \cite{mu2022noma}, secure transmission \cite{yang2022secure}, RIS-aided transmission \cite{zuo2023exploiting,lyu2023hybrid}, and STAR-RIS-aided transmission \cite{xue2024noma}, among other applications.

\begin{figure}[!t]
  \centering
  \includegraphics[width=0.45\textwidth]{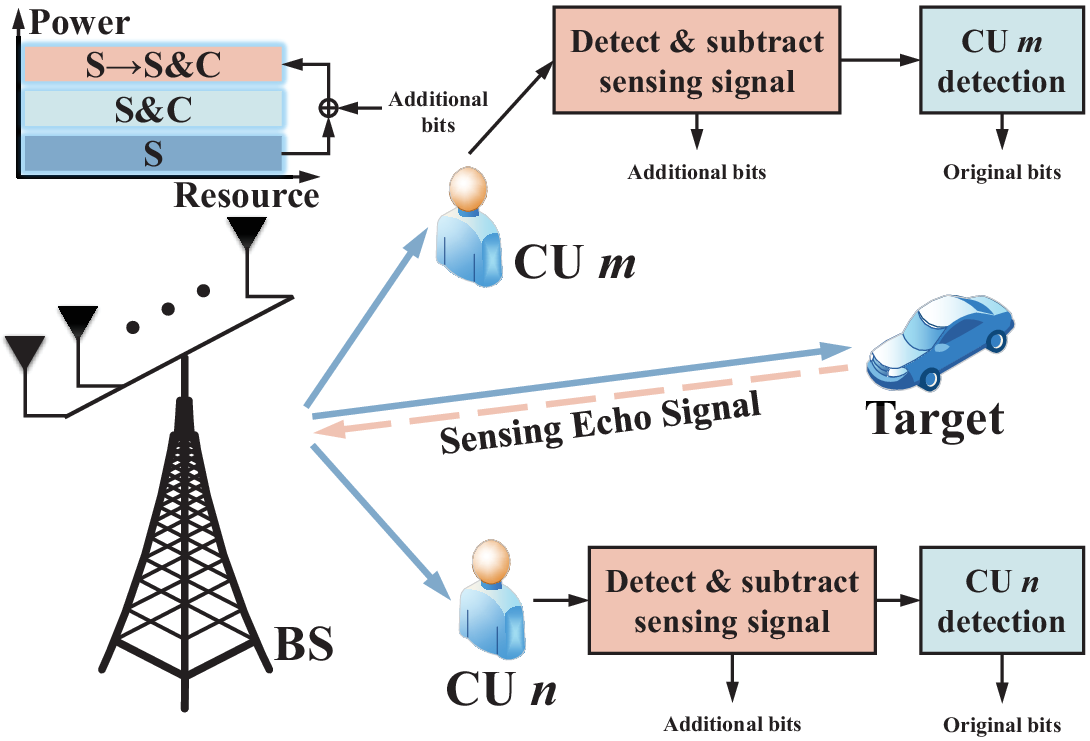}
  \caption{Illustration of the NOMA-inspired downlink ISAC design, where S and C refer to sensing and communications, respectively.}
  \label{Figure: I_NOMA_DISAC}
\end{figure}

$\bullet$ \emph{NOMA-Inspired Downlink Integrated Sensing and Communications}:
In the NOMA-empowered downlink ISAC design, the complete DFSAC waveforms are harnessed to concurrently support communications and sensing without IFI, and NOMA is exclusively employed to address inter-CU interference. As emphasized in \cite{hua2023optimal,liu2020joint}, additional sensing waveforms are often essential for downlink ISAC to achieve high-quality sensing, but they may introduce harmful interference to communications. Consequently, it becomes crucial to address the sensing-to-interference issue in downlink ISAC, leading to the development of the NOMA-inspired framework \cite{wang2022nomai,wang2022exploiting}. This ISAC design offers a practical approach to mitigating sensing interference, which can be further adapted to enhance communications.

As illustrated in {\figurename} \ref{Figure: I_NOMA_DISAC}, the core concept of NOMA-inspired downlink ISAC involves allowing part or all of the additional sensing waveforms to carry information bits, transforming them into additional DFSAC waveforms. Consequently, these extra DFSAC waveforms can be considered as intended for virtual CUs, effectively paired with real CUs through NOMA. At each real CU, the signals of the virtual CUs (i.e., the additional DFSAC waveforms) are initially detected and removed via SIC, thus mitigating sensing-to-communications interference. In comparison to NOMA-empowered ISAC, NOMA-inspired downlink ISAC can be seen as employing NOMA between the sensing and communication functionalities for interference mitigation. Furthermore, as a significant advancement, the information bits conveyed in the additional DFSAC waveforms can also be exploited by the CUs. For instance, these waveforms can be employed to convey public messages intended for all CUs, benefiting both sensing and communications \cite{wang2022nomai,wang2022exploiting}.
\subsubsection{NOMA for Uplink Integrated Sensing and Communications}
After discussing the application of NOMA in downlink ISAC, we transition to uplink ISAC. In uplink ISAC, the DFSAC BS simultaneously receives communication signals from CUs and sensing echo signals reflected by nearby targets, as depicted in {\figurename} \ref{Figure: UISAC_System}. Similar to the downlink case, the sensing probing signal's time interval may be longer than the round-trip time of the sensing signal traveling between the BS and the targets, and thus the BS has to operate in the full-duplex mode. Given the separate transmission of sensing and communication signals, there is no necessity to design DFSAC waveforms. Upon observing the superposed sensing and communication signals, the BS endeavors to retrieve both the data information conveyed by the CUs and the environmental information embedded in the echo signal. Since these two types of information are not known \emph{a priori} to the BS, the primary design challenge in uplink ISAC lies in mitigating the mutual interference between sensing and communications during the processing of the received superposed signals.

In the following sections, we elaborate on two designs for uplink ISAC: \romannumeral1) pure-NOMA-based uplink ISAC and \romannumeral2) semi-NOMA-based uplink ISAC. The primary distinction between these two designs is whether the radio resources are fully or partially shared between sensing and communications.
\begin{figure}[!t]
  \centering
  \includegraphics[width=0.45\textwidth]{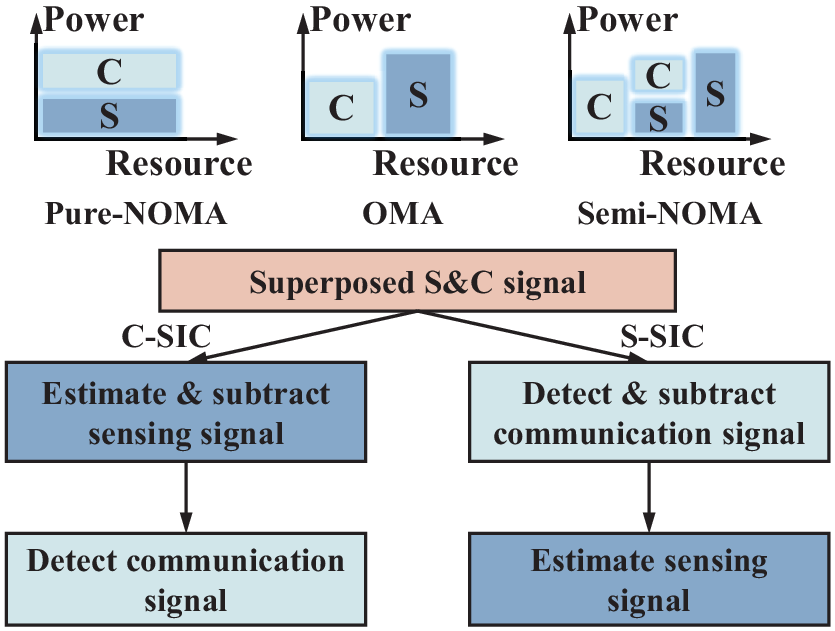}
  \caption{Illustration of an uplink ISAC system with pure-NOMA, OMA, and semi-NOMA schemes.}
  \label{Figure: SIC_NOMA_UISAC}
\end{figure}

$\bullet$ \emph{Pure-NOMA-Based Uplink Integrated Sensing and Communications}:
For clarity, we study the pure-NOMA-based uplink ISAC based on the fundamental uplink ISAC model depicted in {\figurename} \ref{Figure: UISAC_System} as an example. In conventional OMA-based uplink ISAC, dedicated radio resources are assigned to the sensing echo signal and communication signal to separate their functionalities, as illustrated in the top-middle of {\figurename} \ref{Figure: SIC_NOMA_UISAC}. Although the OMA-based approach eliminates IFI, it inherently suffers from resource inefficiency. This inefficiency prompts the development of the pure-NOMA-based uplink ISAC scheme shown in the top-left of {\figurename} \ref{Figure: SIC_NOMA_UISAC}, where the radio resources are fully shared between sensing and communications.

To tackle the mutual interference arising from the superposed sensing and communication signals, a SIC-based framework \cite{ouyang2023revealing} proves to be instrumental. This framework encompasses two SIC schemes: S-C SIC (S-SIC) and C-C SIC (C-SIC), corresponding to two distinct SIC orders, as depicted in {\figurename} \ref{Figure: SIC_NOMA_UISAC}. In the S-SIC scheme, the BS initially decodes the communication signal by treating the sensing signal as interference. Subsequently, the communication signal is subtracted from the superposed signal, and the remaining portion is utilized to sense the parameters of interest. Conversely, in C-SIC, the BS first estimates the target parameters by treating the communication signal as interference. The sensing echo signal is then subtracted from the superposed signal, enabling the recovery of the data information sent by the CUs. While S-SIC enhances sensing performance, C-SIC excels in communication performance.

In practical ISAC systems, S-SIC is often preferred for two primary reasons. Firstly, the nature of the sensing signal differs from that of the communication signal. Typically, the communication signal contains information bits, facilitating straightforward subtraction after decoding. Conversely, the relationship between the entire sensing echo signal and the target parameters may be intricate, posing challenges in subtracting the sensing echo signal from the superposed signal. Secondly, the sensing signal undergoes round-trip path loss during its BS-target-BS transmission, resulting in potentially lower signal strength compared to the communication signal. In terms of fairness, it is more logical to leverage SIC to decode the ostensibly ``stronger'' communication signal before addressing the sensing task. Consequently, numerous studies have studied the performance of uplink ISAC under the S-SIC scheme; for details, we refer to \cite{chiriyath2015inner,ouyang2022on,ouyang2022performance,zhang2023semi,dong2023joint}. In contrast to these works, the authors in \cite{ouyang2023revealing} consider both S-SIC and C-SIC, providing quantitative analyses regarding the influence of SIC ordering.

$\bullet$ \emph{Semi-NOMA-Based Uplink Integrated Sensing and Communications}:
By amalgamating the OMA-based and pure-NOMA-based uplink ISAC designs, a semi-NOMA-based uplink ISAC framework was introduced in \cite{zhang2023semi}, as illustrated in the top-right of {\figurename} \ref{Figure: SIC_NOMA_UISAC}. This framework allows for the coexistence of the OMA- and pure-NOMA-based schemes, not only treating them as special cases but also providing a more flexible ISAC operation capable of satisfying various objectives, such as S-C design, C-C design, and sensing-versus-communications tradeoff design.

To elaborate, the total available radio resources are partitioned into three orthogonal components: the sensing-only resource block, the communications-only resource block, and the DFSAC resource block. In the DFSAC block, the two SIC schemes from the pure-NOMA-based framework can be applied. Simultaneously, the BS processes the pure sensing echo/communication signal received from the single-functionality block in an interference-free manner, akin to the OMA-based uplink ISAC design. Subsequently, the ISAC BS combines the sensing/communication results obtained from the non-orthogonal and orthogonal resource blocks. The semi-NOMA-based uplink ISAC design provides a flexible resource allocation framework for uplink ISAC, accommodating diverse qualities of sensing and communication services. Through meticulous optimization of resource allocation among the three resource blocks, the semi-NOMA-based uplink ISAC design achieves an improved sensing-versus-communication tradeoff \cite{zhang2023semi,mu2023noma}.

\subsection{Performance Analysis of NOMA-ISAC}
After discussing the NOMA-based ISAC frameworks, we proceed to analyze the performance limits of NOMA-ISAC. Before delving into this analysis, it is essential to establish the metrics for performance evaluation. In the subsequent sections, we introduce a framework based on mutual information (MI) for ISAC and utilize it to gauge the system's performance \cite{ouyang2023integrated}.
\subsubsection{Information-Theoretic Limits of Communications and Sensing}
Regarding communications, the information theoretic limits have been expounded in Section \ref{Section: Foundational Principles of Power-Domain NOMA}, where the communication MI is considered, defining the maximum error-free information transmission rate or coding rate. In contrast, sensing performance limits typically depend on the specific sensing task, with performance metrics rooted in estimation and detection theory, such as mean squared error (MSE) \cite{yang2007mimo}, detection probability \cite{kay2009fundamentals}, matching error \cite{li2007mimo}, and Cramér-Rao bound (CRB) \cite{kay1993fundamentals}. These metrics explicitly characterize estimation accuracy or detection reliability. However, it is crucial to note that these sensing metrics may lack compatibility with communication metrics for the following reasons:
\begin{enumerate}
  \item[\romannumeral1)] These sensing metrics involve different units of measurement, mathematical properties, and physical interpretations compared to communication MI.
  \item[\romannumeral2)] These sensing metrics are contingent on specific estimation algorithms. For example, detection probability relies on a statistical hypothesis, and CRB depends on an unbiased maximum likelihood estimator. None of them provide a universal lower bound for distortion metrics in sensing. In contrast, communication MI is rooted in a joint typical decoder, offering a supremum for the efficiency of communications.
  \item[\romannumeral3)] MI has been studied for over 70 years, particularly familiar to researchers in our community. Conversely, many sensing metrics originated from radar theory, potentially less familiar to newcomers in our community.
\end{enumerate}
Motivated by these considerations, it is logical to strive for the establishment of a unified framework for ISAC. This framework would evaluate both communication and sensing performance using metrics with consistent units of measurement, mathematical properties, and physical interpretation, providing a universal bound for the respective tasks. This unified approach facilitates both performance evaluation and system optimization, including waveform design and resource allocation. Given that the essence of sensing is to extract environmental information from the sensing echo signal, we propose using MI, the cornerstone of information theory, to assess sensing performance, as detailed below.

$\bullet$ \emph{Communication Mutual Information}:
Consider the fundamental communication model represented as follows:
\begin{align}
\mathbf{Y}_{\rm{c}} = \mathbf{H}_{\rm{c}}\mathbf{X}_{\rm{c}} + \mathbf{N}_{\rm{c}},
\end{align}
where $\mathbf{X}_{\rm{c}}\in{\mathbbmss{C}}^{N_{\mathsf{BS}}^{\mathsf{t}}\times L_{\rm{c}}}$, $\mathbf{H}_{\rm{c}}\in{\mathbbmss{C}}^{N_{\mathsf{U}}\times N_{\mathsf{BS}}^{\mathsf{t}}}$, and $\mathbf{N}_{\rm{c}}\in{\mathbbmss{C}}^{N_{\mathsf{U}}\times L_{\rm{c}}}$ denote the data symbol matrix, communication channel matrix, and AWGN matrix, respectively. Here, $N_{\mathsf{U}}$ is the number of receive antennas of the CU, $N_{\mathsf{BS}}^{\mathsf{t}}$ is the number of transmit antennas of the BS, and $L_{\rm{c}}$ denotes the length of the communication frame. The communication MI is expressed as $I(\mathbf{X}_{\rm{c}};\mathbf{Y}_{\rm{c}}|\mathbf{H}_{\rm{c}})$, commonly referred to as Shannon information \cite{cover1999elements}. The communication rate (CR) is derived by normalizing the MI by the time interval and bandwidth.

$\bullet$ \emph{Sensing Mutual Information}:
The sensing task involves three key steps: \romannumeral1) designing the sensing waveform, \romannumeral2) transmitting the sensing signal and receiving the echo signal reflected by the target, and \romannumeral3) estimating the target parameters from the echo signal. Most sensing models are characterized by the following equation:
\begin{align}\label{Sensing_Model_Sensing MI}
\mathbf{Y}_{\rm{s}} = \mathbf{H}_{\rm{s}}(\bm\eta)\mathbf{X}_{\rm{s}} + \mathbf{N}_{\rm{s}},
\end{align}
where $\mathbf{X}_{\rm{s}}\in{\mathbbmss{C}}^{N_{\mathsf{BS}}^{\mathsf{t}}\times L_{\rm{s}}}$, $\mathbf{H}_{\rm{s}}(\bm\eta)\in{\mathbbmss{C}}^{N_{\mathsf{BS}}^{\mathsf{r}}\times N_{\mathsf{BS}}^{\mathsf{t}}}$, $\mathbf{Y}_{\rm{s}}\in{\mathbbmss{C}}^{N_{\mathsf{BS}}^{\mathsf{r}}\times L_{\rm{s}}}$, $\mathbf{N}_{\rm{s}}\in{\mathbbmss{C}}^{N_{\mathsf{BS}}^{\mathsf{r}}\times L_{\rm{s}}}$, and ${\bm\eta}\in{\mathbbmss{C}}^{N_{\mathsf{PARA}}\times1}$ represent the predesigned sensing probing signal matrix, target response matrix, echo signal matrix, AWGN matrix, and parameter vector of interest, respectively. Here, $N_{\mathsf{BS}}^{\mathsf{r}}$ is the number of receive antennas at the BS, $L_{\rm{s}}$ denotes the length of the sensing pulse, and $N_{\mathsf{PARA}}$ is the number of parameters of interest. The target response $\mathbf{H}_{\rm{s}}(\bm\eta)$ is a deterministic function of the target parameter $\bm\eta$ to be estimated, interpreted as the sensing channel. Considering the randomness of $\bm\eta$, the sensing MI is defined as the environmental information contained in $\bm\eta$, extractable from $\mathbf{Y}_{\rm{s}}$ when $\mathbf{X}_{\rm{s}}$ is known. Therefore, the sensing MI is calculated as $I({\bm\eta};\mathbf{Y}_{\rm{s}}|\mathbf{X}_{\rm{s}})$. If $\mathbf{H}_{\rm{s}}(\bm\eta)$ itself is the parameter of interest, i.e., ${\bm\eta}=\mathsf{vec}(\mathbf{H}_{\rm{s}}(\bm\eta))$ and $N_{\mathsf{PARA}}=N_{\mathsf{BS}}^{\mathsf{r}}N_{\mathsf{BS}}^{\mathsf{t}}$, then the sensing MI is given by $I(\mathbf{H}_{\rm{s}}(\bm\eta);\mathbf{Y}_{\rm{s}}|\mathbf{X}_{\rm{s}})$. The sensing rate (SR) is obtained by normalizing the MI by the time interval and bandwidth.

The sensing task can be represented through a Markov chain, as follows \cite{ahmadipour2022information}:
\begin{align}
{\bm\eta}\rightarrow{\mathbf{H}}_{\rm{s}}(\bm\eta)\rightarrow\mathbf{Y}_{\rm{s}} = {\mathbf{H}}_{\rm{s}}(\bm\eta)\mathbf{X}_{\rm{s}} + \mathbf{N}_{\rm{s}}\rightarrow\hat{\bm\eta}=f_{\rm{es}}(\mathbf{Y}_{\rm{s}}),
\end{align}
where $f_{\rm{es}}(\cdot)$ is the estimator. The objective of sensing is to estimate\footnote{Note that the task of target detection can be interpreted as a task of estimating a binary random variable, with ``0'' and ``1'' representing the presence and absences of the target.} $\bm\eta$ from $\mathbf{Y}_{\rm{s}}$, and the estimation error is assessed by the distortion function as follows:
\begin{align}
{\mathbbmss{E}}\{d(\bm\eta,\hat{\bm\eta})\}={\mathbbmss{E}}\{\lVert \bm\eta - \hat{\bm\eta}\rVert^2\}={\mathbbmss{E}}\{\lVert \bm\eta - f_{\rm{es}}(\mathbf{Y}_{\rm{s}})\rVert^2\},
\end{align}
where $d(\bm\eta,\hat{\bm\eta})=\lVert \bm\eta - \hat{\bm\eta}\rVert^2$ is the squared-error distortion measure \cite{cover1999elements}. According to the data processing inequality, it holds that \cite{cover1999elements}:
\begin{align}
I({\bm\eta};\mathbf{Y}_{\rm{s}}|\mathbf{X}_{\rm{s}})\geq I({\bm\eta};\hat{\bm\eta}|\mathbf{X}_{\rm{s}}).
\end{align}
Given the \emph{a priori} distribution of $\bm\eta$, the \emph{rate-distortion function} of $\eta$ is defined as follows \cite{cover1999elements}:
\begin{align}
R_{\bm\eta}(D)=\min_{p(\hat{\bm\eta}|{\bm\eta})}I({\bm\eta};\hat{\bm\eta})\quad{\rm{s.t.}}~{\mathbbmss{E}}\{d(\bm\eta,\hat{\bm\eta})\}\leq D,
\end{align}
where $p(\hat{\bm\eta}|{\bm\eta})$ is the transition probability from $\bm\eta$ to $\hat{\bm\eta}$. By definition, $R_{\bm\eta}(D)$ is solely determined by the probability distribution of $\bm\eta$ and is monotonically decreasing with the distortion $D$ \cite{berger2003rate}. The inverse of the \emph{rate-distortion function} is known as the \emph{distortion-rate function}, which is given by \cite{berger2003rate}
\begin{align}
D_{\bm\eta}(R)=\min_{p(\hat{\bm\eta}|{\bm\eta})}{\mathbbmss{E}}\{d(\bm\eta,\hat{\bm\eta})\}\quad{\rm{s.t.}}~I({\bm\eta};\hat{\bm\eta})\leq R,
\end{align}
and is monotonically decreasing with $R$. According to the \emph{rate-distortion theory}, it follows that \cite{tulino2013support}
\begin{align}
R_{\bm\eta}({\mathbbmss{E}}\{d(\bm\eta,\hat{\bm\eta})\})\leq I({\bm\eta};\hat{\bm\eta}|\mathbf{X}_{\rm{s}})\leq I({\bm\eta};\mathbf{Y}_{\rm{s}}|\mathbf{X}_{\rm{s}}),
\end{align}
which implies
\begin{align}\label{Rate_Distortion_Lower_Bound}
{\mathbbmss{E}}\{d(\bm\eta,\hat{\bm\eta})\}\geq D_{\bm\eta}(I({\bm\eta};\hat{\bm\eta}|\mathbf{X}_{\rm{s}}))
\geq D_{\bm\eta}(I({\bm\eta};\mathbf{Y}_{\rm{s}}|\mathbf{X}_{\rm{s}})).
\end{align}

The aforementioned observation implies that employing the sensing MI as the input of the \emph{rate-distortion function} results in an output that serves as a lower bound for the estimation distortion. This finding, initially recognized by Tulino \emph{et al.} \cite{tulino2013support}, has been reiterated in subsequent works \cite{ouyang2023integrated,dong2023rethinking,liu2023deterministic}. It is well-established that numerous sensing performance metrics, including the Bayesian CRB \cite{xiong2023fundamental}, detection probability, and MSE, can be expressed as a squared-error distortion function. Consequently, the sensing MI offers a universal lower bound for the accuracy of the sensing task \cite{ouyang2023integrated,dong2023rethinking,liu2023deterministic}. Additionally, it has been proved that this lower bound is achievable via an MSE estimator when $\bm\eta$ is Gaussian distributed \cite{ouyang2023integrated}. This insight also sheds light on sensing waveform design. Specifically, by crafting a sensing waveform for maximization of the sensing MI, one can simultaneously minimize the lower bound of the sensing accuracy. This criterion, based on maximizing sensing MI, has found application in the current literature; see \cite{tang2010mimo,tang2018spectrally,tang2021constrained,li2024framework} for more details.

The definition of sensing MI is contingent upon knowledge of the \emph{a priori} distribution of $\bm\eta$. Consequently, sensing MI serves as the performance metric for sensing tasks within a Bayesian framework. However, in a non-Bayesian setting where the a priori distribution is absent, one may choose to either assume a uniform distribution or opt for Fisher information instead of Shannon information for performance evaluation. In such cases, the CRB is commonly utilized. Defining a universal lower bound for the sensing task under a non-Bayesian setting remains an open problem.

\subsubsection{Mutual Information-Based ISAC Framework}
The aforementioned arguments signify that sensing MI not only establishes the information-theoretic limit for extracting environmental information but also delineates the estimation-theoretic limit on the distortion of the target parameters. This dual role positions sensing MI as a promising alternative metric for sensing. Moreover, sensing and communication MI exhibit the following properties:
\begin{itemize}
  \item Sensing MI has a similar physical meaning as communication MI, both measured in bits and characterizing information-theoretic limits.
  \item Sensing MI possesses similar mathematical properties as communication MI, both adhering to fundamental MI properties.
\end{itemize}
The above facts imply that sensing MI is a ``communications-friendly'' performance metric. Consequently, an MI-based ISAC framework can be established by employing sensing/communication MI as performance metric \cite{ouyang2023integrated}. Within this framework, sensing and communication performance metrics possess analogous physical and mathematical properties, along with the same unit of measurement. This contrasts with most existing ISAC frameworks utilizing different types of metrics \cite{ouyang2023integrated}. By leveraging this MI-based framework, the fundamental performance limits of ISAC can be unveiled, providing crucial insights into system design. Additionally, this framework facilitates performance comparisons between ISAC and OSAC techniques. One of the key MI-related performance metrics is the sensing-communication rate region, encompassing all achievable SR-CR pairs and thus defining the information-theoretic boundary of the entire system. In the following subsection, we employ this MI-related framework to scrutinize the performance of ISAC by investigating the rate regions of various typical ISAC models, some of which are based on the NOMA frameworks introduced in Section \ref{Section: NOMA for ISAC}.
\subsubsection{Case Study}
Before delving into further details, we introduce a commonly used model used in target sensing, where the target response matrix is characterized as follows \cite{li2007mimo}:
\begin{align}\label{Target_Response_Matrix_Model}
\mathbf{G}=\sum\nolimits_{q}\alpha_q{\mathbf{a}}_{\mathsf{r}}(\theta_q){\mathbf{b}}_{\mathsf{t}}^{\mathsf{T}}(\theta_q),
\end{align}
where $\alpha_q\in{\mathbbmss{C}}$ is the reflection coefficient of the $q$th target, ${\mathbf{a}}_{\mathsf{r}}(\theta_q)\in{\mathbbmss{C}}^{N_{\mathsf{BS}}^{{\mathsf{r}}}\times 1}$ and ${\mathbf{b}}_{\mathsf{t}}(\theta_q)\in{\mathbbmss{C}}^{N_{\mathsf{BS}}^{{\mathsf{t}}}\times 1}$ are the associated receive and transmit array steering vectors, respectively, and $\theta_q$ is its DoA. The reflection coefficient contains both the round-trip path-loss and the radar cross section (RCS) of the corresponding target. Following the Swerling-I model \cite{richards2005fundamentals}, we assume that the RCS is relatively constant from pulse-to-pulse with \emph{a priori} Rayleigh distributed amplitude, in which case $\alpha_q$ follows a complex Gaussian distribution ${\mathcal{CN}}(0,\varsigma_q^2)$. Here, we have $\varsigma_q^2=\varpi_q\varkappa_q^2$, where $\varkappa_q^2$ is the arithmetic mean of all RCS values of the reflecting object and represents the average reflection strength, and $\varpi_q$ characterizes the round-trip path loss. The objective of sensing is to estimate the target response matrix $\mathbf{G}$, i.e., ${\bm\eta}={\mathsf{vec}}(\mathbf{G})$, based on which the angle of each target and the associated reflection coefficient can be estimated using MUSIC (MUltiple SIgnal Classification) and APES (Amplitude and Phase EStimation) algorithms \cite{schmidt1986multiple,li1996adaptive}. 

Substituting $\mathbf{H}_{\rm{s}}(\bm\eta)=\mathbf{G}$ into \eqref{Sensing_Model_Sensing MI} yields the following sensing model:
\begin{align}\label{Sensing_Model_We_Used}
\mathbf{Y}_{\rm{s}} = {\mathbf{G}}\mathbf{X}_{\rm{s}} + \mathbf{N}_{\rm{s}},
\end{align}
where $\mathbf{N}_{\rm{s}}$ is the AWGN matrix with each entry having zero mean and unit variance. The sensing model in \eqref{Sensing_Model_We_Used} can be rewritten as follows:
\begin{align}\label{Sensing_Model_We_Used1}
\mathsf{vec}(\mathbf{Y}_{\rm{s}})=({\mathbf{I}}_{N_{\mathsf{BS}}^{\mathsf{r}}}\otimes{\mathbf{X}}_{\rm{s}}^{\mathsf{H}}){\mathsf{vec}}(\mathbf{G}^{\mathsf{H}})+{\mathsf{vec}}(\mathsf{\mathbf{N}_{\rm{s}}}),
\end{align}
where ${\mathsf{vec}}(\mathsf{\mathbf{N}_{\rm{s}}})\sim{\mathcal{CN}}({\mathbf{0}},{\mathbf{I}}_{L_{\rm{s}}N_{\mathsf{BS}}^{\mathsf{r}}})$ and ${\mathsf{vec}}(\mathbf{G}^{\mathsf{H}})\sim{\mathcal{CN}}({\mathbf{0}},{\mathbf{R}}_{\mathbf{G}})$ with
\begin{align}
{\mathbf{R}}_{\mathbf{G}}=\sum\nolimits_{q}\varsigma_q^2({\mathbf{b}}(\theta_q)\otimes{\mathbf{a}}(\theta_q))({\mathbf{b}}(\theta_q)\otimes{\mathbf{a}}(\theta_q))^{\mathsf{H}}.
\end{align}
As a result, \eqref{Sensing_Model_We_Used1} can be interpreted as the received signal of a hypothetical communication system with ${\mathbf{I}}_{N_{\mathsf{BS}}^{\mathsf{r}}}\otimes{\mathbf{X}}_{\rm{s}}^{\mathsf{H}}\in{\mathbbmss{C}}^{L_{\rm{s}}N_{\mathsf{BS}}^{\mathsf{r}}\times N_{\mathsf{BS}}^{\mathsf{t}}N_{\mathsf{BS}}^{\mathsf{t}}}$ being the equivalent channel matrix. In addition, ${\mathsf{vec}}(\mathbf{G}^{\mathsf{H}})\sim{\mathcal{CN}}({\mathbf{0}},{\mathbf{R}}_{\mathbf{G}})$ and ${\mathsf{vec}}(\mathsf{\mathbf{N}_{\rm{s}}})\sim{\mathcal{CN}}({\mathbf{0}},{\mathbf{I}}_{L_{\rm{s}}N_{\mathsf{BS}}^{\mathsf{r}}})$ are the system input and the thermal noise, respectively. Consequently, the sensing MI can be directly calculated by Shannon's formula, given by \cite{el2011network}
\begin{equation}\label{MI_Sensing_General}
\begin{split}
  &I(\mathbf{G};\mathbf{Y}_{\rm{s}}|\mathbf{X}_{\rm{s}})=I({\mathsf{vec}}(\mathbf{G}^{\mathsf{H}});\mathsf{vec}(\mathbf{Y}_{\rm{s}})|{\mathbf{I}}_{N_{\mathsf{BS}}^{\mathsf{r}}}\otimes{\mathbf{X}}_{\rm{s}}^{\mathsf{H}})\\
  &=\log_2\det({\mathbf{I}}_{L_{\rm{s}}N_{\mathsf{BS}}^{\mathsf{r}}}+({\mathbf{I}}_{N_{\mathsf{BS}}^{\mathsf{r}}}\otimes{\mathbf{X}}_{\rm{s}}^{\mathsf{H}}){\mathbf{R}}_{\mathbf{h}}
  ({\mathbf{I}}_{N_{\mathsf{BS}}^{\mathsf{r}}}\otimes{\mathbf{X}}_{\rm{s}}^{\mathsf{H}})^{\mathsf{H}}).
\end{split}
\end{equation}
Assuming that each sensing waveform symbol lasts $1$ unit time, we write the SR as $L_{\rm{s}}^{-1}I(\mathbf{G};\mathbf{Y}_{\rm{s}}|\mathbf{X}_{\rm{s}})$ \cite{ouyang2023integrated}. For further simplifications, we assume that the BS employs widely separated receive antennas (i.e., MIMO radar) \cite{haimovich2007mimo}, where we can ignore the correlations among the rows of the target response matrix, and the correlation matrix can be simplified as ${\mathbf{R}}_{\mathbf{G}}={\mathbf{I}}_{N_{\mathsf{BS}}^{\mathsf{r}}}\otimes {\mathbf{R}}$ with ${\mathbf{R}}\in{\mathbbmss{C}}^{N_{\mathsf{BS}}^{\mathsf{t}}\times N_{\mathsf{BS}}^{\mathsf{t}}}$ being the transmit correlation matrix \cite{tang2018spectrally}. Under this condition, the sensing MI can be simplified as follows:
\begin{equation}\label{MI_Sensing_General2}
\begin{split}
I(\mathbf{G};\mathbf{Y}_{\rm{s}}|\mathbf{X}_{\rm{s}})=N_{\mathsf{BS}}^{\mathsf{r}}\log_2\det({\mathbf{I}}_{L_{\rm{s}}}+{\mathbf{X}}_{\rm{s}}^{\mathsf{H}}{\mathbf{R}}{\mathbf{X}}_{\rm{s}}).
\end{split}
\end{equation}
Furthermore, it can be proven that when the target response is Gaussian distributed, the sensing waveform that maximizes the sensing MI coincides with the sensing waveform that minimizes the estimation distortion. This implies that the lower bound \eqref{Rate_Distortion_Lower_Bound} provided by \emph{rate-distortion theory} is achievable \cite{yang2007mimo}. 

Next, we will utilize the sensing MI to assess the sensing performance of various typical systems. Unless otherwise specified, we will utilize the MI expression derived in \eqref{MI_Sensing_General2}. Throughout our discussion, we assume that the BS is equipped with $N_{\mathsf{BS}}^{\mathsf{t}}$ transmit antennas and $N_{\mathsf{BS}}^{\mathsf{r}}$ receive antennas. The length of the communication frame/sensing pulse is denoted as $L$.

$\bullet$ \emph{Uplink MISO-NOMA-ISAC}:
Consider an uplink MISO-NOMA-ISAC system where a DFSAC BS serves $K$ uplink single-antenna CUs while simultaneously sensing targets, as illustrated in {\figurename} {\ref{Figure: Uplink-MISO-NOMA-ISAC}}. We employ the \textbf{\emph{pure-NOMA-based uplink ISAC framework}}, where the radio resources are fully shared between sensing and communications. The sensing and communication signals are assumed to be perfectly synchronized at the BS using properly designed synchronization sequences. Consequently, the BS observes the following superposed signal:
\begin{figure}[!t]
  \centering
  \includegraphics[width=0.45\textwidth]{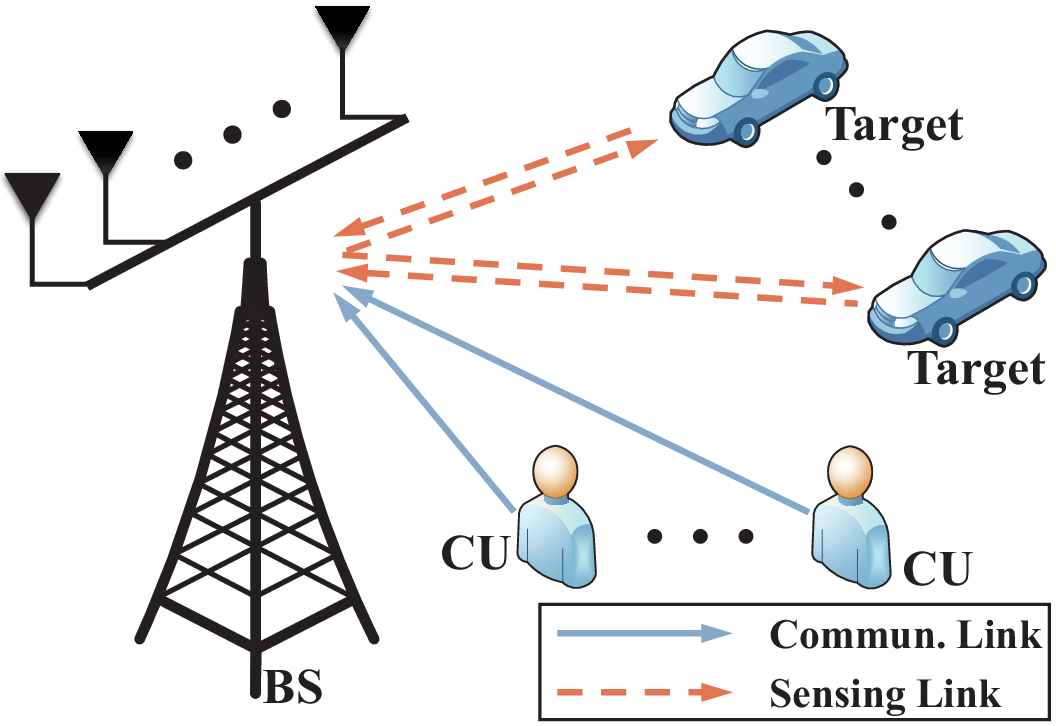}
  \caption{Illustration of an uplink NOMA-ISAC system with $K$ single-antenna uplink CUs.}
  \label{Figure: Uplink-MISO-NOMA-ISAC}
\end{figure}
\begin{align}\label{Uplink_Basic_Model}
{\mathbf{Y}}=\sum\nolimits_{k=1}^{K}{{\mathbf{h}}}_k{\mathbf{x}}_k^{\mathsf{T}}+
{\mathbf{G}}{\mathbf{X}}_{\rm{s}}+{\mathbf{N}},
\end{align}
where $\mathbf{X}_{\rm{s}}\in{\mathbbmss{C}}^{N_{\mathsf{BS}}^{\mathsf{t}}\times L}$ is the sensing signal, $\mathbf{N}\in{\mathbbmss{C}}^{N_{\mathsf{BS}}^{\mathsf{r}}\times L}$ is the AWGN matrix, $\mathbf{G}\in{\mathbbmss{C}}^{N_{\mathsf{BS}}^{\mathsf{r}}\times N_{\mathsf{BS}}^{\mathsf{t}}}$ is the target response matrix defined in \eqref{Target_Response_Matrix_Model}, ${\mathbf{h}}_k\in{\mathbbmss{C}}^{N_{\mathsf{BS}}^{\mathsf{r}}\times 1}$ is the channel vector of CU $k$, and ${\mathbf{x}}_k\in{\mathbbmss{C}}^{L\times 1}$ denotes the message sent by CU $k$.
\begin{figure}[!t]
\centering
\setlength{\abovecaptionskip}{0pt}
\includegraphics[width=0.48\textwidth]{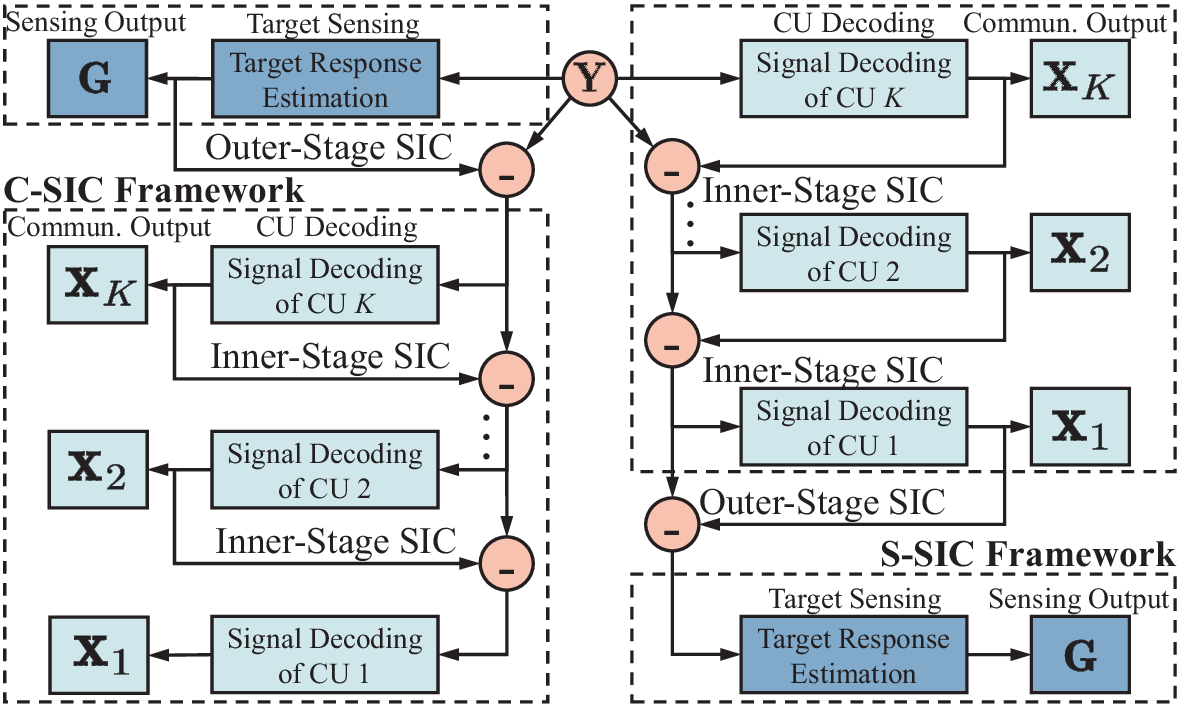}
\caption{Block diagram of the two-stage SIC-based-based framework.}
\label{SIC_Model}
\end{figure}

\begin{figure}[!t]
\centering
\setlength{\abovecaptionskip}{0pt}
\includegraphics[width=0.45\textwidth]{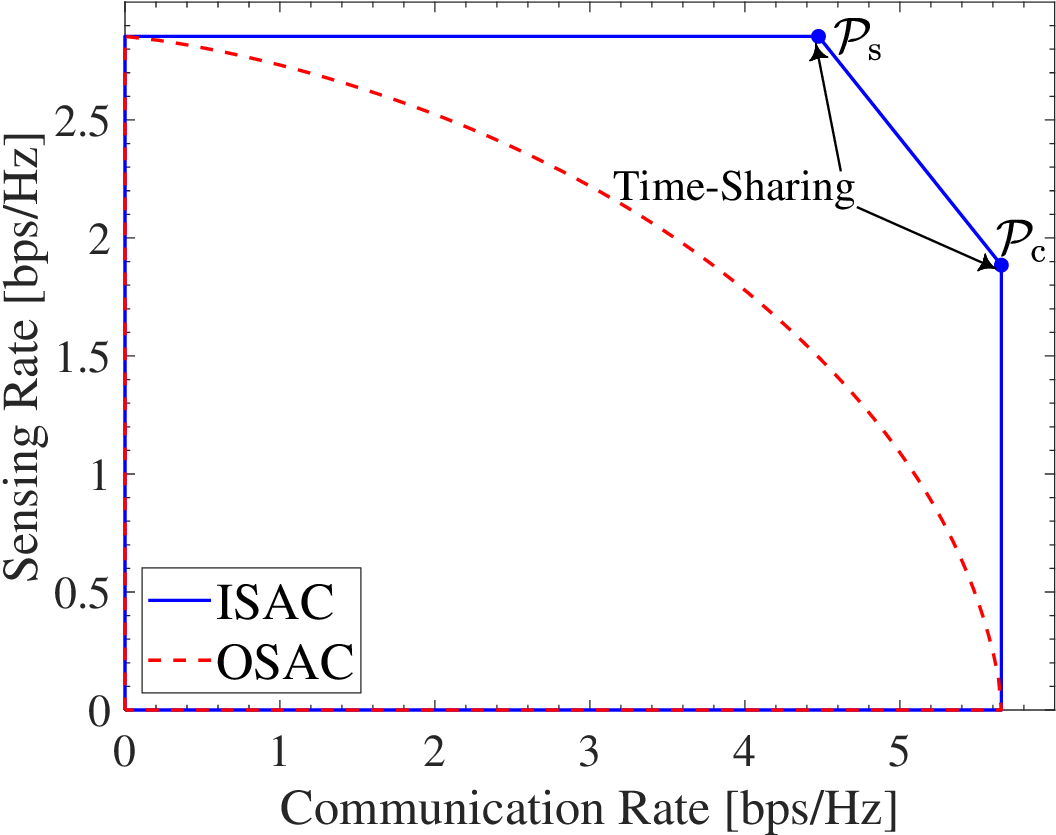}
\caption{Rate region of the uplink NOMA-ISAC system. The system parameter setting can be found in \cite[Section V]{ouyang2023revealing}.}
\label{Figure: Uplink-MISO-NOMA-ISAC-Rate_Region}
\end{figure}

After receiving the superposed signal matrix ${\mathbf{Y}}$, the BS aims to decode the data information contained in the communication signal, $\left\{{\mathbf{x}}_k\right\}_{k=1}^{K}$, as well as extracting the environmental information contained in the target response, $\mathbf{G}$. To mitigate mutual interference, a two-stage SIC-based framework can be utilized \cite{ouyang2023revealing}, in which the inner-stage SIC deals with inter-CU interference, while the outer-stage SIC addresses IFI, as detailed in {\figurename} {\ref{SIC_Model}}. Without loss of generality, we assume that the CUs are arranged in ascending order, i.e., $\lVert{\mathbf h}_K\rVert\geq\cdots\geq\lVert{\mathbf h}_1\rVert$. As shown in {\figurename} {\ref{SIC_Model}}, CUs with better channel conditions are decoded earlier in the inner-stage SIC. Regarding the outer-stage SIC, we consider two SIC orders. Under the first SIC order, the BS first senses the target response $\mathbf{G}$ by treating the communication signal as interference. Then ${\mathbf{G}}{\mathbf{X}}_{\rm{s}}$ is subtracted from $\mathbf{Y}$, and the remaining part is used for detecting the communication signal. Under the second SIC order, the BS first detects the communication signal, $\left\{{\mathbf{x}}_k\right\}_{k=1}^{K}$, by treating the sensing signal as interference. Then, $\sum\nolimits_{k=1}^{K}{{\mathbf{h}}}_k{\mathbf{x}}_k^{\mathsf{T}}$ is subtracted from $\mathbf{Y}$, and the remaining part is used for sensing the target response. Clearly, the first SIC order yields better communication performance (C-SIC), while the second one yields better sensing performance (S-SIC). {\figurename} {\ref{Figure: Uplink-MISO-NOMA-ISAC-Rate_Region}}  compares the SR-sum CR regions achieved by ISAC and OSAC. In OSAC, the communication and sensing functionalities exploit separated frequency-hardware resources. For ISAC, the points ${\mathcal{P}}_{\rm{s}}$ and ${\mathcal{P}}_{\rm{c}}$ are achieved by S-SIC and C-SIC, respectively. The line segment connecting ${\mathcal{P}}_{\rm{c}}$ and ${\mathcal{P}}_{\rm{s}}$ is achieved with the time-sharing strategy. We observe that the rate region of OSAC is completely contained in that of ISAC, highlighting the superiority of ISAC.

\begin{figure}[!t]
    \centering
    \subfigure[System layout.]{
        \includegraphics[width=0.4\textwidth]{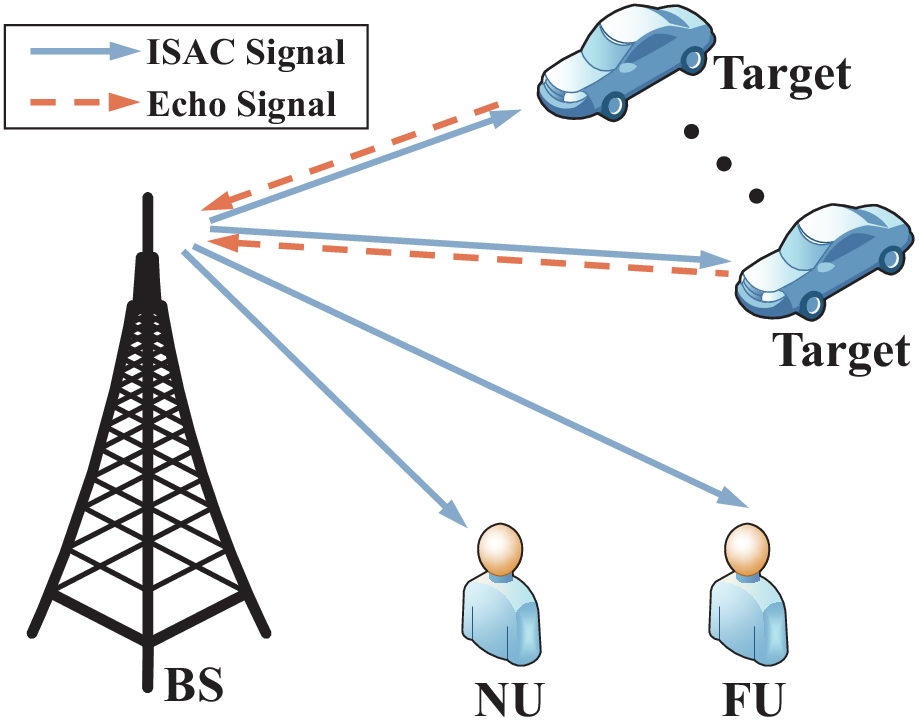}
        \label{Figure: Downlink-SISO-NOMA-ISAC}
    }
    \subfigure[Rate region.]{
        \includegraphics[width=0.45\textwidth]{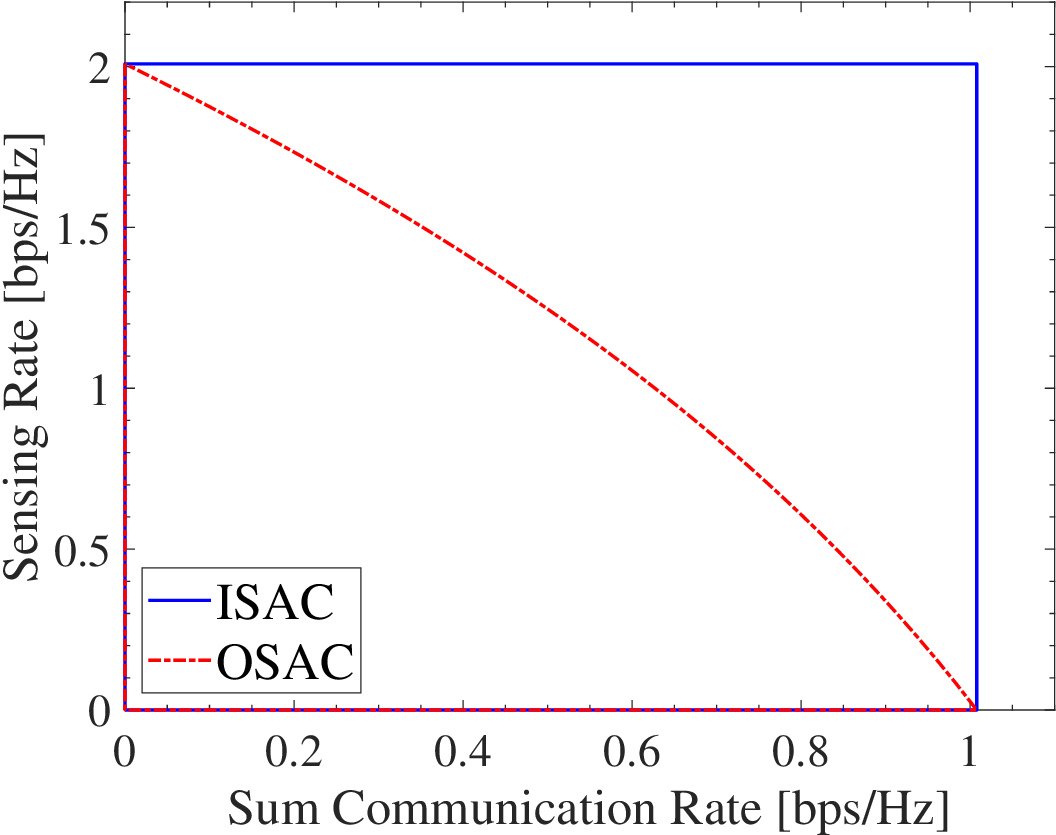}
        \label{Figure: Downlink-SISO-NOMA-ISAC_Rate_Region}	
    }
    \caption{Illustration of (a) downlink SISO-NOMA-ISAC system and (b) the corresponding SR-CR region. The system parameter setting can be found in \cite[Section V]{ouyang2023performance}. }
\end{figure}
$\bullet$ \emph{Downlink SISO-NOMA-ISAC}:
We proceed to examine a downlink SISO-NOMA-ISAC system, where the BS serves a pair of single-antenna CUs, namely one near user (NU) and one far user (FU), while concurrently sensing the surrounding environment, as illustrated in {\figurename} {\ref{Figure: Downlink-SISO-NOMA-ISAC}}. The DFSAC BS is equipped with a single antenna for transmission/reception, i.e., $N_{\mathsf{BS}}^{\mathsf{t}}=N_{\mathsf{BS}}^{\mathsf{r}}=1$, and the objective of sensing is to estimate the RCS coefficients of each target. The \textbf{\emph{NOMA-empowered ISAC framework}} is employed to eliminate IFI and mitigate inter-CU interference \cite{ouyang2023performance}. Given that the BS has a single antenna, the optimal performance of both communications and sensing is achieved by transmitting at full power. This implies that there is no sensing-communication performance tradeoff, as evident from the rate region shown in {\figurename} {\ref{Figure: Downlink-SISO-NOMA-ISAC_Rate_Region}}.

$\bullet$ \emph{Downlink SU-MISO-ISAC}:
The performance of ISAC can be enhanced by employing multiple antennas at the DFSAC BS. To illustrate this, let us consider a downlink single-user (SU)-MISO-ISAC system, as depicted in {\figurename} \ref{Figure: DISAC_System}, where a multiple-antenna DFSAC BS serves a single-antenna CU while concurrently sensing a single target. Assuming full sharing of the radio resources between sensing and communications, the CR can be expressed as: 
\begin{align}
{\mathcal{R}}_{\rm{c}}=\log_2(1+p\lvert{\mathbf{w}}^{\mathsf{H}}{\mathbf{h}}_{\rm{c}}\rvert^2),
\end{align}
where ${\mathbf{h}}_{\rm{c}}\in{\mathbbmss{C}}^{N_{\mathbf{BS}}^{\mathsf{t}}\times1}$ represents the communication channel vector, $p$ denotes the power budget, and ${\mathbf{w}}\in{\mathbbmss{C}}^{N_{\mathbf{BS}}^{\mathsf{t}}\times1}$ is the normalized beamforming vector. The target response matrix is modeled as follows:
\begin{align}
\mathbf{G}=\alpha_1{\mathbf{a}}_{\mathsf{r}}(\theta_1){\mathbf{b}}_{\mathsf{t}}^{\mathsf{T}}(\theta_1).
\end{align}
Let us focus on estimating the reflection coefficient $\alpha_1$, assuming perfect tracking of the target angle. The corresponding SR is given by:
\begin{align}
{\mathcal{R}}_{\rm{s}}=L^{-1}\log_2(1+pN_{\mathsf{BS}}^{\mathsf{r}}L\varsigma_1^2\lvert{\mathbf{w}}^{\mathsf{H}}{\mathbf{h}}_{\rm{s}}\rvert^2),
\end{align}
where ${\mathbf{h}}_{\rm{s}}\triangleq{\mathbf{b}}_{\mathsf{t}}(\theta_1)$ represents the sensing channel. Note that both $\mathcal{R}_{\rm{c}}$ and $\mathcal{R}_{\rm{s}}$ are affected by the choice of the beamforming vector $\mathbf{w}$. However, finding an optimal $\mathbf{w}$ that maximizes both $\mathcal{R}_{\rm{s}}$ and $\mathcal{R}_{\rm{c}}$ simultaneously presents a challenging task. Motivated by this challenge, we explore three typical scenarios: the S-C design, which aims at maximizing the SR, the C-C design, which aims at maximizing the CR, and the Pareto optimal design, which characterizes the Pareto boundary of the SR-CR region.
\begin{figure}[!t]
  \centering
  \includegraphics[width=0.45\textwidth]{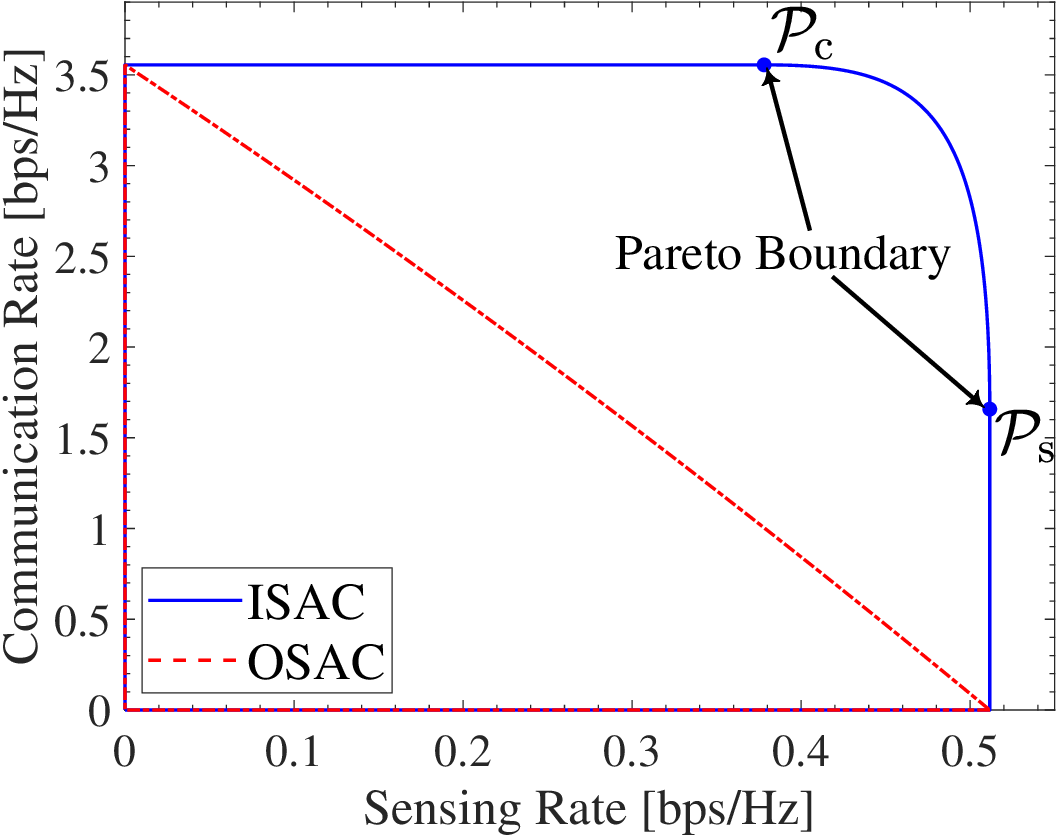}
  \caption{Rate region of the downlink SU-MISO-ISAC system. The system parameter setting can be found in \cite[Section V]{ouyang2023revealingthe}.}
  \label{Figure: Downlink-SU-MISO-ISAC-Rate_Region}
\end{figure}

The S-C beamforming vector is given by ${\mathbf{w}}=\frac{{\mathbf{h}}_{\rm{s}}}{\lVert{\mathbf{h}}_{\rm{s}}\rVert}$, while the C-C beamforming vector is ${\mathbf{w}}=\frac{{\mathbf{h}}_{\rm{c}}}{\lVert{\mathbf{h}}_{\rm{c}}\rVert}$. Moreover, any rate-pair on the Pareto boundary can be obtained using a rate-profile based method \cite{zhang2010cooperative}. This involves solving the following optimization problem:
\begin{equation}\label{Problem_CR_SR_Tradeoff}
\max\nolimits_{{\mathbf{w}},\mathcal{R}}{\mathcal{R}},~{\rm{s.t.}}~{\mathcal{R}}_{\rm{s}}\geq \alpha{\mathcal{R}},{\mathcal{R}}_{\rm{c}}\geq (1-\alpha){\mathcal{R}},\lVert{\mathbf{w}}\rVert^2=1,
\end{equation}
where $\alpha\in[0,1]$ is the rate-profile parameter. The entire Pareto boundary can be obtained by solving this problem with $\alpha$ varying from $0$ to $1$. A closed-form solution to this non-convex problem is provided in \cite{ouyang2023revealingthe}. In {\figurename} {\ref{Figure: Downlink-SU-MISO-ISAC-Rate_Region}}, we compare the CR-SR region achieved by ISAC with that achieved by OSAC. Two distinct points are of particular interest in the context of ISAC: point ${\mathcal{P}}_{\rm{s}}$, attained by the S-C design, and point ${\mathcal{P}}_{\rm{c}}$, achieved by the C-C design. The curve segment connecting ${\mathcal{P}}_{\rm{s}}$ and ${\mathcal{P}}_{\rm{c}}$ represents ISAC's Pareto boundary in terms of the rate region. Significantly, we observe that OSAC's rate region is entirely contained in ISAC's region, and a proof for this is available in \cite{ouyang2023revealingthe}. Similar results were also obtained in \cite{zhao2023modeling,zhao2024performanceana,zhao2024performancea,zhao2024near} for near-field transmission.

\begin{figure}[!t]
    \centering
    \subfigure[System layout.]{
        \includegraphics[width=0.4\textwidth]{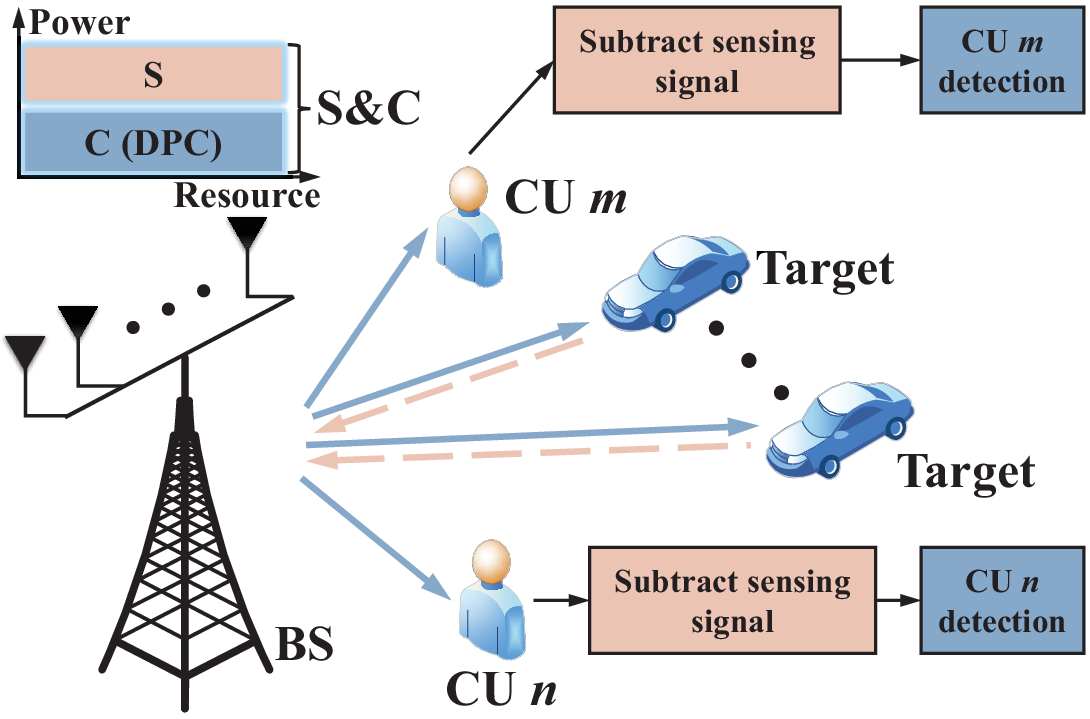}
        \label{Figure: Downlink-MU-MISO-ISAC}
    }
    \subfigure[Rate region.]{
        \includegraphics[width=0.45\textwidth]{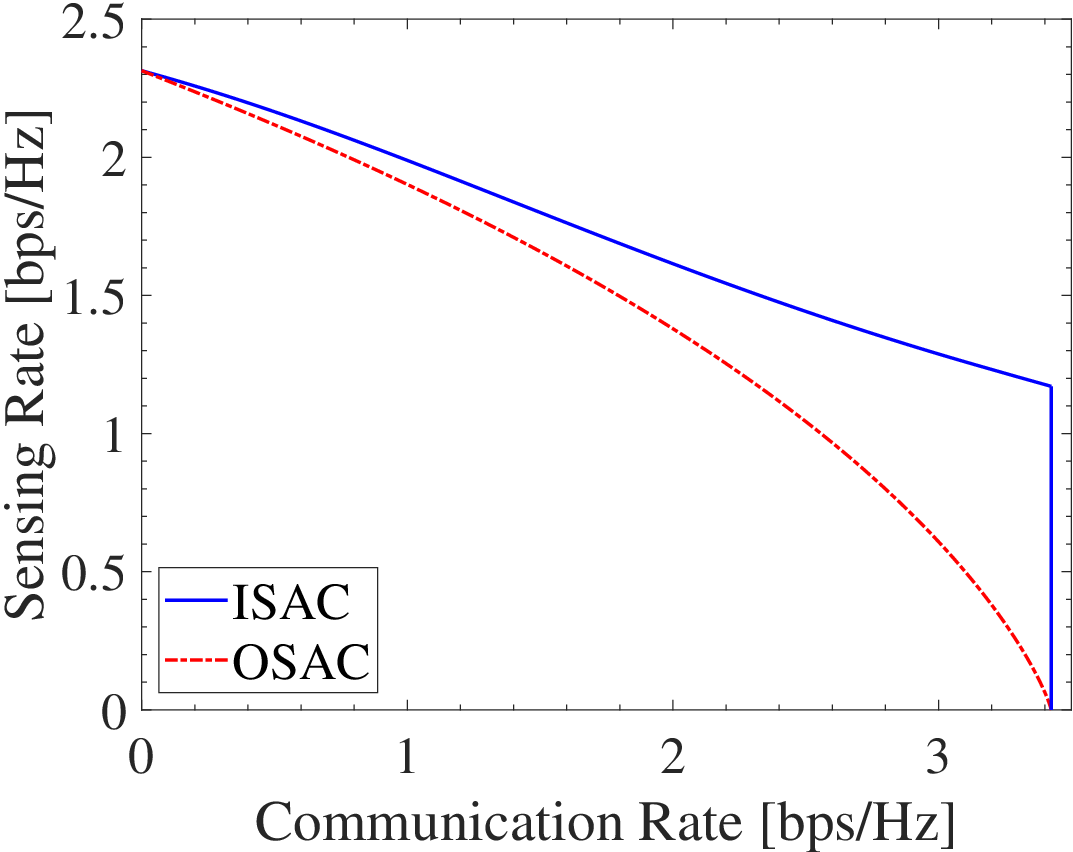}
        \label{Figure: Downlink-MU-MISO-ISAC_Rate_Region}	
    }
    \caption{Illustration of (a) downlink MU-MISO-ISAC system and (b) the corresponding SR-CR region. The system parameter setting can be found in \cite[Section VI]{ouyang2022performance}. }
\end{figure}
$\bullet$ \emph{Downlink MU-MISO-ISAC}:
The downlink SU-MISO-ISAC architecture can be further expanded to accommodate multiple users, as illustrated in {\figurename} {\ref{Figure: Downlink-MU-MISO-ISAC}}. Motivated by the \textbf{\emph{NOMA-inspired downlink NOMA framework}}, data messages designated for individual CUs undergo DPC to mitigate inter-CU interference \cite{ouyang2022performance}. These messages are then superposed with the sensing signal. Notably, the sensing signal design takes into account the statistical characteristics of the communication channels and signals. Given its static nature over prolonged periods \cite{ouyang2022performance}, it is reasonable to inform each CU of this fixed signal. Upon receiving the superposed signal, each CU can then subtract the sensing component from its observation and proceed to decode the data message \cite{ouyang2022performance}. The resulting SR-CR region is depicted in {\figurename} {\ref{Figure: Downlink-SISO-NOMA-ISAC_Rate_Region}}. As depicted, the achievable rate region of OSAC is entirely contained in the achievable rate region of ISAC, underscoring the superiority of ISAC over conventional OSAC methodologies.
\begin{figure}[!t]
    \centering
    \subfigure[System layout.]{
        \includegraphics[width=0.4\textwidth]{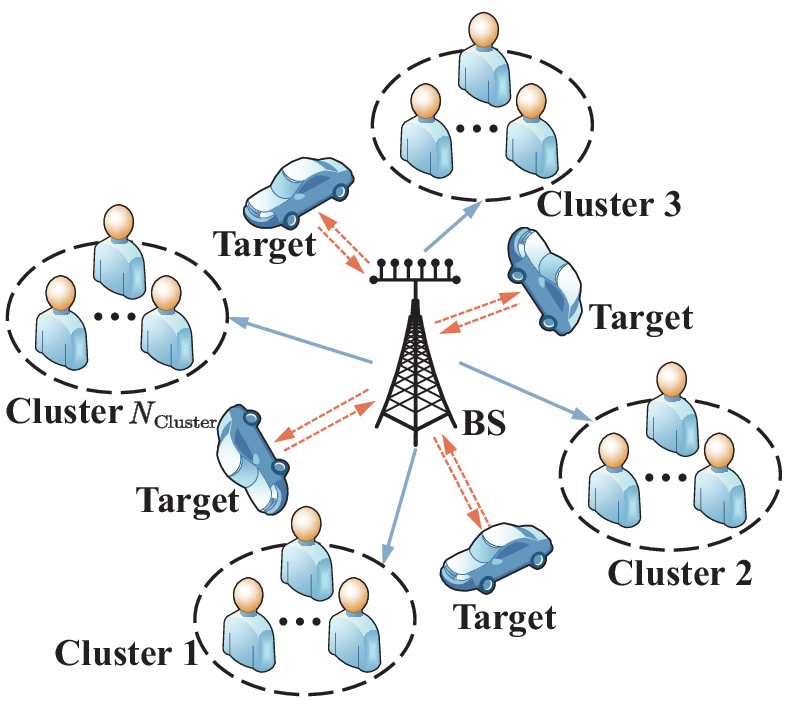}
        \label{Figure: Downlink-Cluster-Based-MIMO-NOMA-ISAC}
    }
    \subfigure[Rate region.]{
        \includegraphics[width=0.45\textwidth]{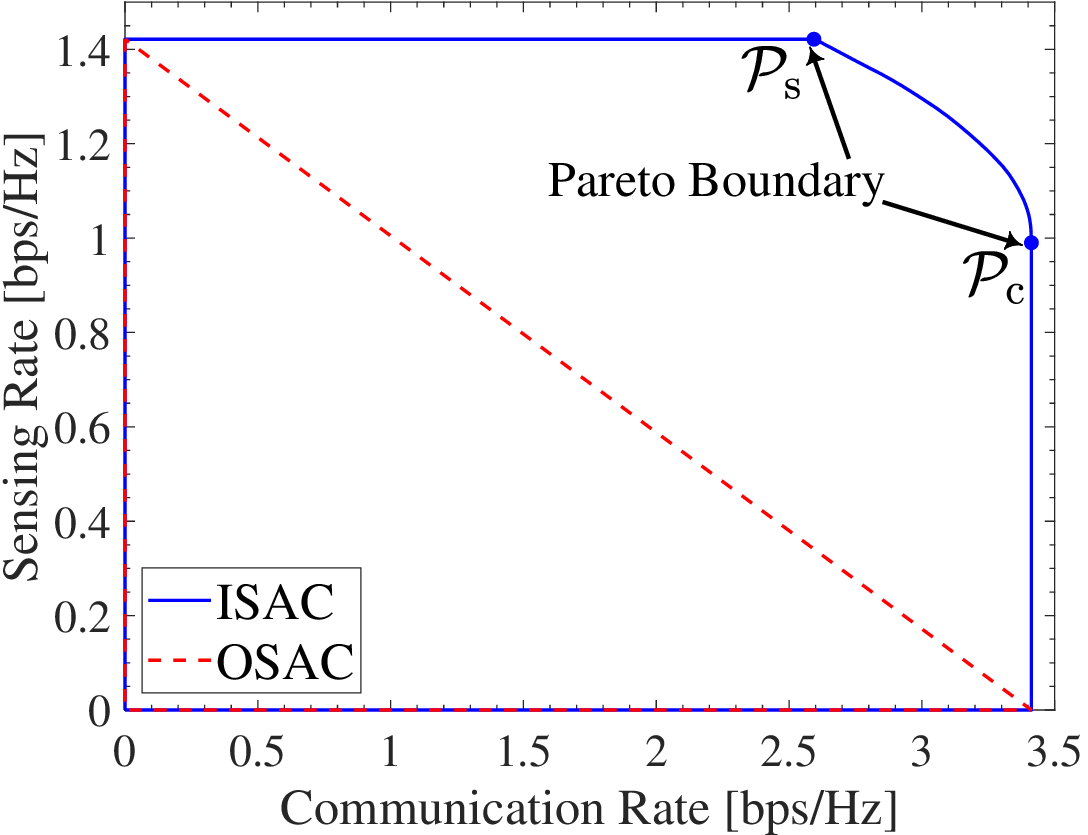}
        \label{Figure: Downlink-Cluster-Based-MIMO-NOMA-ISAC_Rate_Region}	
    }
    \caption{Illustration of (a) downlink cluster-based MIMO-NOMA-ISAC system and (b) the corresponding SR-CR region. The system parameter setting can be found in \cite[Section V]{zhao2024cluster}. }
\end{figure}

$\bullet$ \emph{Downlink Cluster-Based MIMO-NOMA-ISAC}:
Finally, let us consider a downlink cluster-based MIMO-NOMA ISAC system, as depicted in {\figurename} {\ref{Figure: Downlink-Cluster-Based-MIMO-NOMA-ISAC}}, built upon the \textbf{\emph{NOMA-empowered ISAC framework}} \cite{zhao2024cluster}. In this setup, the BS is tasked with communicating with multiple CUs, each equipped with $N_{\mathsf{U}}$ antennas, while simultaneously sensing the surrounding environment for targets. The CUs are randomly grouped into $N_{\mathsf{Cluster}}=N_{\mathsf{BS}}^{\mathsf{t}}$ clusters \cite{ding2015application,zhao2024cluster}, with $K_c$ CUs in each cluster $c=1,\ldots,N_{\mathsf{Cluster}}$. The DFSAC signal can be expressed as follows:
\begin{align}
{\mathbf{X}}=\sum\nolimits_{c=1}^{N_{\mathsf{Cluster}}}{\mathbf{w}}_c\sum\nolimits_{k=1}^{K_c}\sqrt{p_{c,k}}{\mathbf{s}}_{c,k},
\end{align}
where ${\mathbf{w}}_c\in{\mathbbmss{C}}^{N_{\mathsf{BS}}^{\mathsf{t}}\times1}$ is the normalized beamforming vector for cluster $c$, ${\mathbf{s}}_{c,k}\in{\mathbbmss{C}}^{L\times1}$ is the unit-power data stream for CU $k$ in cluster $c$, and $p_{c,k}$ is the associated transmit power. NOMA is utilized to mitigate intra-cluster interference, while ZF detection is employed at each CU to eliminate inter-cluster interference. Let ${\mathbf{H}}_{c,k}\in{\mathbbmss{C}}^{N_{\mathsf{BS}}^{\mathsf{t}}\times N_{\mathsf{U}}}$ and ${\mathbf{v}}_{c,k}\in{\mathbbmss{C}}^{N_{\mathsf{U}}\times1}$ denote the user-to-BS channel matrix and detection vector, respectively. Then, ${\mathbf{v}}_{c,k}$ should be designed as follows:
\begin{equation}
\left\{
\begin{array}{ll}
{\mathbf{v}}_{c,k}^{\mathsf{H}}{\mathbf{H}}_{c,k}^{\mathsf{H}}{\mathbf{w}}_{c}\ne 0             & k=1,\ldots,K_c\\
{\mathbf{v}}_{c,k}^{\mathsf{H}}{\mathbf{H}}_{c,k}^{\mathsf{H}}{\mathbf{w}}_{c'}=0           & c'\ne c,~k=1,\ldots,K_{c'}
\end{array} \right.,
\end{equation}
which can be realized if $N_{\mathsf{U}}\geq N_{\mathsf{Cluster}}$ \cite{ding2015application}. Let $1\rightarrow\ldots\rightarrow K_{c}$ denote the SIC decoding order in cluster $c$. Then, the sum-rate CU is given by
\begin{align}
{\mathcal{R}}_{\rm{c}}=\sum_{c=1}^{N_{\mathsf{Cluster}}}\sum_{k=1}^{K_c}\log_2\left(1+\frac{\lvert{\mathbf{v}}_{c,k}^{\mathsf{H}}{\mathbf{H}}_{c,k}^{\mathsf{H}}{\mathbf{w}}_{c}\rvert^2 p_{c,k}}{
\lvert{\mathbf{v}}_{c,k}^{\mathsf{H}}{\mathbf{H}}_{c,k}^{\mathsf{H}}{\mathbf{w}}_{c}\rvert^2\sum\nolimits_{k'>k}{p_{c,k'}}+\sigma^2}\right),
\end{align}
where $\sigma^2$ is the noise power. On the other hand, when the DFSAC signal $\mathbf{X}$ is used for sensing the targets, the resulting SR is given by 
\begin{align}
{\mathcal{R}}_{\rm{s}}=L^{-1}N_{\mathsf{BS}}^{\mathsf{r}}\log_2\det\left({\mathbf{I}}_{N_{\mathsf{BS}}^{\mathsf{t}}}+L{\mathbf{P}}^{\mathsf{H}}{\mathbf{R}}{\mathbf{P}}\right),
\end{align}
where ${\mathbf{P}}=[\tilde{p}_1{\mathbf{w}}_1,\ldots,\tilde{p}_{N_{\mathsf{Cluster}}}{\mathbf{w}}_{N_{\mathsf{Cluster}}}]\in{\mathbbmss{C}}^{N_{\mathsf{BS}}^{\mathsf{t}}\times N_{\mathsf{Cluster}}}$, $\tilde{p}_c=\sum_{k=1}^{K_c}p_{c,k}$ is the total power allocated to cluster $c=1,\ldots,N_{\mathsf{Cluster}}$, and ${\mathbf{R}}\in{\mathbbmss{C}}^{N_{\mathsf{BS}}^{\mathsf{t}}\times N_{\mathsf{BS}}^{\mathsf{t}}}$ is the transmit correlation matrix of the target response. Both $\mathcal{R}_{\rm{c}}$ and $\mathcal{R}_{\rm{s}}$ are influenced by the choice of $\mathbf{P}$. However, finding an optimal $\mathbf{W}$ that maximizes both $\mathcal{R}_{\rm{s}}$ and $\mathcal{R}_{\rm{c}}$ simultaneously is challenging. To address this challenge, we explore three typical scenarios: the S-C design, aiming to maximize $\mathcal{R}_{\rm{s}}$, the C-C design, aiming to maximize $\mathcal{R}_{\rm{c}}$, and the Pareto optimal design, characterizing the Pareto boundary of the SR-CR region.

In \cite{zhao2024cluster}, it was proven that the C-C design is obtained via water-filling among the communication channels, while the S-C deign is obtained via water-filling among the sensing channels. Furthermore, the Pareto optimal design can be attained by solving a convex rate-profile optimization problem similar to \eqref{Problem_CR_SR_Tradeoff}. In {\figurename} {\ref{Figure: Downlink-Cluster-Based-MIMO-NOMA-ISAC_Rate_Region}}, we compare the SR-CR region achieved by ISAC with that achieved by OSAC. The point ${\mathcal{P}}_{\rm{s}}$ is attained by the S-C design, and the point ${\mathcal{P}}_{\rm{c}}$ is achieved by the C-C design. The curve segment connecting ${\mathcal{P}}_{\rm{s}}$ and ${\mathcal{P}}_{\rm{c}}$ represents ISAC's Pareto boundary in terms of the rate region. It is observed, and further proven in \cite{zhao2024cluster}, that OSAC’s rate region is entirely contained in ISAC's region. These results were initially observed in the case of $K_{c}=1$, i.e., one CU in each cluster \cite{ouyang2023mimo}, and then extended to the case of $K_{\rm{c}}=2$ under a signal alignment-based protocol \cite{zhao2023downlink,zhao2024performance}.
\subsection{Discussions and Outlook}
Resource-efficient ISAC requires the application of MA techniques. Essentially, ISAC combines sensing and communication functionalities within a single time-frequency-power-hardware resource block, embodying the concept of MA. MA further offers efficient interference management tools for ISAC, addressing both IFI and IUI. In this section, we have highlighted the effectiveness of MA in ISAC and proposed an MI-based framework to assess the performance of the NOMA-ISAC paradigm. However, several research challenges remain open.

For instance, while sensing MI provides a lower bound for sensing distortion metrics, its achievability is not always guaranteed \cite{ouyang2023integrated}. Although previous discussions suggest that this lower bound is attainable in cases where the target response follows a Gaussian distribution, further investigation is needed to determine the conditions for achieving it in non-Gaussian scenarios. Additionally, while the NOMA- and MI-based framework has been used to reveal the rate region of many typical channels, the rate region of the most general downlink MU-MIMO-ISAC channel remains unknown. Moreover, the rate region under practical QoS constraints, such as user fairness and hardware impairments, has yet to be reported. Finally, while the NOMA- and MI-based framework holds promise for facilitating ISAC waveform design, this area of research is still in its nascent stages.

Looking ahead, future networks will integrate communication, sensing, and computing (such as mobile edge computing \cite{mao2017survey} or over-the-air computing \cite{yang2020federated}) capabilities. This integration can be realized through an integrated sensing, communications, and computation (ISCC) framework \cite{qi2020integrated}. Here again, the NOMA-based framework remains crucial, as ISCC essentially requires MA for the sensing, communication, and computation functionalities within a single resource block \cite{wang2023noma}. Hence, future research can extend the NOMA-based ISAC frameworks to their ISCC counterparts. Furthermore, in a recent study \cite{jeon2014computation}, the authors proposed leveraging MI to evaluate computing performance, resulting in the computation rate metric \cite{wu2019computation}. Building on these findings, the MI-based framework can be expanded to encompass computation, offering a comprehensive approach to evaluate the performance of ISCC systems.

\section{Next-Generation Multiple Access: The Road Ahead}\label{Section: MA for NGMA}
After discussing the fundamental properties of MA in the power domain, spatial domain, and ISAC, we pivot to explore potential applications of MA, focusing on NOMA, within next-generation wireless networks, referred to as NGMA. In this context, we will delineate the research opportunities and challenges associated with the following three new directions:
\begin{enumerate}
  \item [\romannumeral1)]New application scenarios: Exploring the utilization of NOMA in heterogeneous environments to accommodate emerging 6G applications;
  \item [\romannumeral2)]New techniques: Investigating the integration of NOMA with other cutting-edge 6G physical layer technologies to enhance overall system performance;
  \item [\romannumeral3)]New tools: Harnessing the power of ML to develop adaptive resource allocation strategies tailored to the dynamic nature of wireless networks.
\end{enumerate}
\subsection{New Applications}
This section explores the application of NGMA, focusing on its fundamental principles and their extension to novel scenarios in the context of 6G. Specifically, our primary focus will be on semantic communications as a case study.
\subsubsection{Semantic-NOMA/NGMA}
Recently, semantic communications have garnered significant attention from both industry and academia \cite{tong2022nine,gunduz2022beyond}, emerging as a fundamental challenge in the evolution of 6G wireless networks. Often referred to as task or goal-oriented communications, semantic communications focus on extracting the ``meaning'' of the transmitted information at the transmitter. By leveraging a matched knowledge base between the transmitter and receiver, semantic information can be successfully ``interpreted'' at the receiver \cite{weaver1963mathematical}. In practical terms, the message is first encoded via a semantic encoder at the transmitter before undergoing channel encoding, a departure from the conventional Shannon paradigm, which typically employs a bit source encoder prior to channel coding. By transmitting only essential information relevant to a specific task, semantic communication reduces the demands on energy and wireless resources, fostering a more sustainable communication network.

In the realm of multiuser semantic communications, researchers have explored various approaches. One such method involves leveraging conventional OMA strategies like TDMA/FDMA/OFDMA. For instance, the authors of \cite{yan2022resource} addressed the problem of sub-channel allocation in uplink text semantic communications to enhance semantic spectral efficiency through OMA strategies. However, it is noted in \cite{zhang2023deepma} that CDMA is ill-suited for wireless semantic communication due to its binary data transmission design, rendering it generally unsuitable for continuous symbol transmission inherent to semantic communications. To overcome this limitation, the authors of \cite{zhang2023deepma} proposed employing a set of well-trained semantic encoders to encode input data into mutually orthogonal semantic symbol vectors, akin to the semantic counterpart of CDMA. This approach achieves higher resource efficiency compared to other OMA strategies. Additionally, the authors of \cite{xie2021task} employed ZF-based SDMA to mitigate IUI in an uplink two-user MISO semantic system, demonstrating its robustness to channel variations compared to traditional communication systems. This concept is extended to a general MU-MISO scenario in \cite{xie2022task}. Furthermore, the integration of SPC and SIC decoding, i.e., power-domain NOMA, was explored to reduce IUI in \cite{li2023non}, where semantic NOMA systems achieve a larger rate region compared to semantic OMA systems.

In practical multiuser networks, integrating semantic communications into a multiuser network introduces heterogeneity. New semantic users coexist with existing bit users, and users may employ both semantic and bit communication approaches for information transmission. This necessitates efficient MA and resource management to support heterogeneous semantic and bit multiuser communications. To address this challenge, the authors of \cite{mu2022heterogeneous} proposed a downlink semi-NOMA scheme, where the bit stream is split into two streams: one transmitted with the semantic stream over a shared frequency sub-band, and the other transmitted over a separate orthogonal frequency sub-band. For uplink communications, the authors of \cite{mu2023exploiting} proposes an opportunistic semantic and bit communication approach, allowing the secondary user to participate in NOMA via the most suitable communication method (semantic or bit) at each fading state, achieving a balance between performance and interference.

In summary, these studies highlight the potential of NOMA to enhance system performance in both semantic multiuser communications and heterogeneous semantic and bit multiuser communications. 
\subsubsection{Other Heterogeneous Application Scenarios}
In addition to heterogeneous semantic and bit multiuser communications, there are various heterogeneous application scenarios in next-generation wireless networks, where users with diverse QoS requirements have to coexist. In such scenarios, NOMA is widely regarded as a promising approach for interference management and resource allocation \cite{zhang2018energy}. One typical example is ISAC, where sensing services and communication services coexist. In this context, the NOMA-based frameworks introduced in Section \ref{Section: NOMA for ISAC} provide favorable performance. Another scenario arises in heterogeneous radio access networks, serving a mixture of enhanced mobile broadband (eMBB), massive machine-type communication (mMTC), and ultra-reliable low-latency communication (URLLC) devices within the same time-frequency resource block \cite{popovski20185g}. In this case, heterogeneous-NOMA has been shown to be an efficient strategy, leading to significant gains in performance trade-offs among the three generic services compared to OMA schemes \cite{popovski20185g}. Additionally, there exist numerous other heterogeneous application scenarios; for further examples, we refer to \cite{liu2022developing}.

By now, research regarding the application of NOMA in next-generation heterogeneous scenarios, including semantic-NGMA, is still in its infancy. Existing studies have predominantly focused on simplified communication scenarios such as two-user SISO channels. The integration of the NOMA frameworks discussed in the preceding sections with more intricate system settings remains an open challenge. Considering the large scales and massive connectivity anticipated in future ultra-dense heterogeneous networks, the ensuing resource allocation and interference management challenges in heterogeneous-NGMA are poised to become exceedingly complex and demanding. Further research efforts are imperative in this direction, encompassing practical considerations of resource and service constraints, alongside the utilization of novel tools such as ML. These endeavors will be pivotal in paving the way for the deployment of heterogeneous-NGMA networks in the context of 6G.
\subsection{New Techniques}
In 6G, a plethora of newly emerging techniques for the physical layer will gradually evolve to tackle fundamental challenges in future networks. Here, we explore the integration of NOMA/NGMA with reconfigurable antennas, with a particular focus on reconfigurable intelligent surfaces (RISs) \cite{liu2021reconfigurable,mu2024reconfigurable}.
\subsubsection{Reconfigurable Intelligent Surfaces}
Due to their capability to intelligently reconfigure the wireless propagation environment using passive reflecting elements, RISs offer the potential for significant performance enhancements and interference mitigation by creating a software-defined wireless environment. Furthermore, their low-power and cost-effective nature facilitates seamless deployment into next-generation wireless systems.

The integration of NOMA and RIS is motivated by their mutual benefits. On the one hand, NOMA serves as an efficient MA strategy in multiuser networks, thereby enhancing spectral efficiency and connectivity in RIS-aided systems. On the other hand, RISs confer several advantages to existing NOMA networks. Firstly, the reflected links provided by RISs enhance the performance of NOMA networks by providing additional signal diversity. Secondly, by adjusting the phase shifts of reflecting elements and their positions, RISs can improve or degrade the channel quality of individual users, enabling more feasible settings for SIC ordering, transitioning from ``channel condition-based NOMA'' to ``QoS-based NGMA'' \cite{liu2022reconfigurable}. Thirdly, integrating RISs in MIMO-RIS networks relaxes the strict constraint on the number of antennas at the transceivers in MIMO-NOMA, owing to the multiple reflecting elements of the RISs. Consequently, research on the integration of RIS and NOMA has proliferated in the past years, with numerous algorithms proposed for NOMA-RIS, encompassing beamforming, user grouping, and SIC ordering; for a comprehensive review, we refer to recent overview papers \cite{liu2021reconfigurable,liu2022reconfigurable,ding2022state}.

One limitation of conventional RISs is their ability to only reflect incoming signals, providing a half-space coverage. To address this limitation, the concept of simultaneous transmitting and reflecting (STAR)-RIS was introduced in \cite{xu2021star,liu2021star}, enabling the reflection of incident wireless signals within the half-space on the same side of the STAR-RIS, while also allowing transmission to the other side of the RIS. Consequently, STAR-RISs offer full-space coverage. The integration of STAR-RIS and NGMA has also garnered significant research attention; for detailed insights, we refer to \cite{ahmed2023survey}.
\subsubsection{New Forms of Antennas}
The most promising aspect of RISs/STAR–RISs lies in their ability to reconfigure the propagation of incident wireless signals, thereby creating a ``smart radio environment'' (SRE) through control over both the phase and amplitude of these reconfigurable elements \cite{di2020smart}. Additionally, one of the significant advantages that RISs offer to NGMA is the ability to adjust the effective channel quality of each user feasibly through phase or amplitude control of each RIS element. However, it is important to note that these properties are not exclusive to RISs. 

In recent years, advancements in antenna manufacturing have led to the emergence of novel antenna forms \cite{smith2004metamaterials,smith2017analysis,wang2020metantenna}, such as holographic MIMO \cite{deng2021reconfigurable}, dynamic metasurface antennas \cite{shlezinger2021dynamic}, and fluid antennas \cite{wong2020fluid}, among others. These new antennas have transcended their conventional roles as signal conduits; instead, they have evolved into dynamic, adaptable, and intelligent components. Unlike RISs, which are typically employed as relays, these new antennas find applications both at the BS-side (such as holographic MIMO and dynamic metasurface antennas) and user-side (fluid antennas), enabling the establishment of advanced transceivers. These transceivers have demonstrated hardware- and energy-efficiency, capable of shaping the wireless environment through software control. Specifically, these antennas can generate programmable signal patterns by manipulating their radiated EM waves. Each pattern corresponds to a set of beams directed towards different directions, facilitating precise control over signal direction and coverage. From this standpoint, these novel antennas share a similar lineage with RISs, as they can be programmed to establish an ``SRE'' and can support the deployment of NGMA. Further details and a comparison of these novel antennas can be found in \cite{liu2024near}.

The past five years have witnessed a growing research interest in RISs/STAR-RISs, with a wealth of published findings exploring their integration with NOMA and its variants. These developments provide a solid groundwork for the practical implementation of RISs/STAR-RISs in NGMA systems. Moving forward, the next phase for RISs/STAR-RISs-assisted NGMA systems may involve proof-of-concept designs and verification. This entails focusing on several key practical aspects, such as mitigating error propagation in SIC when employing NOMA, designing efficient channel estimation and signal detection methods, and developing real-world deployment strategies for RISs/STAR-RISs. These challenges warrant further investigation. Regarding other emerging forms of antennas, much of the research on their integration with NOMA/NGMA is still in its nascent stages. Fundamental theoretical questions, including characterizing information-theoretic limits and designing optimal resource allocation methods, remain open. Addressing these challenges requires a comprehensive consideration of how these new antennas influence the EM fields.
\subsection{New Tools}
Given the heterogeneous, dynamic, and complex nature of the communication and sensing scenarios in 6G networks, novel approaches are imperative to address the increasingly challenging problems they present \cite{saad2019vision}. ML-based algorithms have emerged as a promising avenue in this regard.
\subsubsection{Machine Learning for NGMA} 
While NGMA holds great promise, the intricate multi-domain multiplexing poses significant challenges for interference suppression and system optimization. For instance, optimizing the cluster-free NOMA framework, as outlined in Section \ref{Section: From Conventional NOMA to Cluster-Free NOMA}, necessitates the design of transmit/receive beamforming, user grouping, power allocation, and SIC ordering, culminating in a highly complex non-convex mixed-integer nonlinear programming problem. Traditional convex optimization methods may achieve local optima but face critical challenges, including intricate mathematical transformations, sensitivity to initialization parameters, and high computational complexity. Moreover, these methods are often ill-equipped to handle problems with a large number of controllable variables and struggle with efficiently optimizing Markovian problems \cite{xu2023artificial}.

In contrast, ML-based algorithms offer a more versatile framework capable of addressing problems with a large number of controllable variables and temporally correlated events. ML techniques, particularly reinforcement learning (RL) approaches, have gained traction for resource allocation in heterogeneous wireless networks owing to their learning capability, particularly in dynamic and unpredictable network environments. RL models empower BSs to learn from real-time feedback from the environment and mobile users, as well as from historical experiences. By observing the system state at each time slot, RL determines optimization variables (actions) to maximize accumulated discounted rewards. The state space comprises the wireless resources (e.g., power, bandwidth, computing, and caching resources) allocated to each user. RL enables BSs to swiftly adapt their control policy to dynamic environments, enhancing decision-making abilities and offering an efficient approach to tackling resource allocation problems in dynamic heterogeneous networks \cite{luong2019applications}. RL has been employed in NOMA networks to address dynamic user clustering/pairing \cite{ahsan2021resource} and long-term resource allocation \cite{he2019joint,ahsan2022reliable}, consistently outperforming conventional algorithms in dynamic scenarios through interaction with the environment.

In addition to RL, deep learning (DL) has found applications in NOMA to address various challenges. DL models utilize cascaded neural network layers to automatically extract features from input data and make decisions, accurately tracking network states and predicting future evolution. DL has been extensively utilized in NOMA for tasks such as channel acquisition \cite{gui2018deep}, signal constellation design \cite{jiang2019deep}, signal detection \cite{emir2021deep}, and resource allocation \cite{yang2019deep}. For resource allocation, the DL approach can be model-based, leveraging prior knowledge of NOMA systems to reduce the number of parameters requiring learning. Conversely, for tasks like acquiring timely and accurate channel knowledge, the DL approach can be data-driven, optimizing NOMA communication systems based on large training datasets.
\subsubsection{NGMA for Machine Learning}
MA techniques can also enhance the performance of ML, as illustrated by federated learning (FL) \cite{niknam2020federated}. FL is a distributed learning approach enabling multiple devices to collaboratively train deep neural networks while preserving privacy. Each device trains its local FL models using its own data and transmits the trained models (e.g., weights, gradients) to the BS, which aggregates them to generate a global FL model sent back to the devices. In a heterogeneous learning and communication system, FL users and conventional CUs coexist. Communication strives to enhance transmission rates, while FL aims to minimize a loss function \cite{chen2020joint}. To accommodate the distinct QoS requirements of learning and communications, NOMA can be utilized to mitigate IUI, thereby enhancing the performance of both communications and FL \cite{ni2022star,ni2022integrating}. In these studies, an SIC-based framework was implemented at the BS to detect superposed learning and communication signals, with numerical findings suggesting that the proposed NOMA framework achieves a promising balance between communication and learning performance.

Next, we consider computing-heterogeneous learning systems that incorporate both computing-limited centralized learning (CL) devices and computing-powerful FL users. In this scenario, CL devices upload raw data to the BS, while FL devices upload model parameters to the BS \cite{han2024semi}. In this setup, CL users prioritize the transmission rate between the BS and themselves since the training of the entire model is attributed to the BS. However, FL users are primarily concerned about the FL loss. This mirrors the aforementioned heterogeneous learning and communication system, suggesting that NOMA can enhance the performance of both FL and CL.

The preceding discussion highlights the synergistic relationship between ML and NOMA, offering promising avenues for ML applications in NGMA systems. However, within the realm of ML-aided NGMA, several research challenges remain open. For instance, while ML can efficiently predict optimal solutions through low-complexity forward propagation, the training of models using back-propagation often demands extensive datasets and imposes significant computational burdens. Developing high-performance, lightweight models and accelerating the training process remains a crucial yet challenging endeavor, particularly in NGMA applications dealing with delay-sensitive services. Furthermore, conventional ML applications in NOMA typically address single-objective problems. In the context of heterogeneous NGMA networks, the need to solve dynamic multi-objective problems becomes increasingly pertinent. This necessitates the design of efficient multi-task ML methodologies. Regarding NGMA-aided ML, such as the NOMA-based heterogeneous learning and communication framework, research is still in its early stages. Further exploration is warranted, especially towards characterizing the information-theoretic limits of both communication and learning. Additionally, the design of efficient resource allocation strategies catering to diverse communication and learning requirements emerges as a critical frontier requiring thorough investigation.
\subsection{Summary and Discussions}
This section has outlined the road ahead of NOMA technologies towards NGMA along three new directions: new application scenarios, new techniques, and new tools. While these directions may seem distinct, they are inherently intertwined. The seamless integration of NOMA into new applications within future heterogeneous networks necessitates the adoption of cutting-edge tools and innovative techniques. The evolving demands of new applications will, in turn, stimulate further advancements and refinements in these tools and techniques. It is through the multidisciplinary fusion of these directions that the practical realization of NGMA within 6G networks becomes achievable. Moreover, along each direction, the corresponding technologies and scenarios can be harmoniously integrated with the MA frameworks and technologies developed in preceding sections. These include, but are not limited to, cluster-free MIMO-NOMA, NOMA-based ISAC, and near-field MA, among others.
\section{Summary and Conclusions}\label{Section: Conclusions}
This paper provided a comprehensive tutorial review on the MA technologies developed over the past 50 years, focusing on power-domain NOMA, spatial-domain MA, the application of NOMA to ISAC, and the integration of MA with emerging technologies. For power-domain NOMA, we investigated its capacity limits for both scalar and vector channels and provided a unified cluster-free MIMO-NOMA framework. In terms of spatial-domain MA, we reviewed several transmission schemes, including linear beamforming, massive MIMO, and mmWave/THz communications, for both conventional far-field and near-field communication scenarios. Moreover, our review covered current research on applying NOMA to ISAC, introduced corresponding NOMA-based ISAC frameworks, and analyzed their performance using an MI-based framework. Finally, we considered the integration of MA with other emerging technologies, such as semantic communications, RIS, and ML, outlining research opportunities and challenges in these areas. Throughout this comprehensive review, we pinpointed open problems and outline future research directions within the realm of MA and NGMA. 

This comprehensive 50-year tutorial review sheds light on the evolving landscape of MA technologies, showcasing a trajectory aimed at enhancing \emph{flexibility, granularity, and maintainability}. The evolution is evident for both power-domain MA and spatial-domain MA. In power-domain MA, a transition from OMA to NOMA, and further to semi-NOMA and RSMA, reflects a significant stride towards enhanced flexibility. Similarly, advancements in spatial-domain MA, progressing from SDMA to the cluster-free NOMA paradigm, further underscore this trend towards increased flexibility. Moreover, the integration of new physical-layer technologies, such as ISAC and NFC, has refined the granularity of channel access. This refinement translates into heightened efficiency concerning resource allocation and interference management. Notably, the transition from far-field ADMA to near-field RDMA, and from OSAC to NOMA-based ISAC, stand out as prime examples. Lastly, the establishment of unified MA frameworks, incorporating diverse interference management strategies and catering to heterogeneous service requirements, promises enhanced system maintainability. A unified control over all resources ensures streamlined operation and management. With these three core properties---\emph{flexibility, granularity, and maintainability}---MA technologies are poised to illuminate the path towards the realization of 6G networks.

\begin{appendices}
\section{Proof of Theorem \ref{Theorem_FFC_Channel_Correlation}}\label{Proof_Theorem_FFC_Channel_Correlation}
According to \eqref{FFC_ULA_LOS_Model}, we have
\begin{subequations}
\begin{align}
\lvert{\mathbf{h}}_k^{\mathsf{H}}{\mathbf{h}}_{k'}\rvert&=\frac{{\beta_{\mathsf{r}}}r_{\mathsf{r}}^2}{r_kr_{k'}}\left\lvert\sum_{i=-\widetilde{N}}^{\widetilde{N}}{\rm{e}}^{{\rm{j}}\frac{2\pi}{\lambda}(i-1)d(\cos(\theta_{k})-\cos(\theta_{k'}))}\right\rvert\\
&=\frac{{\beta_{\mathsf{r}}}r_{\mathsf{r}}^2}{r_kr_{k'}}\left\lvert\frac{\sin\left(\frac{\pi}{\lambda}N_{\mathsf{BS}}d(\cos(\theta_{k})-\cos(\theta_{k'})\right)}
{\sin\left(\frac{\pi}{\lambda}d(\cos(\theta_{k})-\cos(\theta_{k'})\right)}\right\rvert,
\end{align}
\end{subequations}
which yields
\begin{align}
\lim_{N_{\mathsf{BS}}\rightarrow\infty}\frac{\lvert{\mathbf{h}}_k^{\mathsf{H}}{\mathbf{h}}_{k'}\rvert}{\lVert{\mathbf{h}}_k\rVert\lVert{\mathbf{h}}_{k'}\rVert}
=\lim_{N_{\mathsf{BS}}\rightarrow\infty}\frac{\lvert{\mathbf{h}}_k^{\mathsf{H}}{\mathbf{h}}_{k'}\rvert}{{\beta_{\mathsf{r}}}r_{\mathsf{r}}^2N_{\mathsf{BS}}r_k^{-1}r_{k'}^{-1}}=0
\end{align}
for $\theta_k\ne \theta_{k'}$. Furthermore, when $\theta_k= \theta_{k'}$, we have $\lvert{\mathbf{h}}_k^{\mathsf{H}}{\mathbf{h}}_{k'}\rvert=\lVert{\mathbf{h}}_k\rVert\lVert{\mathbf{h}}_{k'}\rVert$. Consequently, it holds that $\lim_{N_{\mathsf{BS}}\rightarrow\infty}\frac{\lvert{\mathbf{h}}_k^{\mathsf{H}}{\mathbf{h}}_{k'}\rvert}{\lVert{\mathbf{h}}_k\rVert\lVert{\mathbf{h}}_{k'}\rVert}=1$. This completes the proof.
\section{Proof of Theorem \ref{Theorem_FFC_Channel_Correlation_Scattering}}\label{Proof_Theorem_FFC_Channel_Correlation_Scattering}
From the law of large numbers, we have
\begin{align}
\frac{1}{N_{\mathsf{BS}}}{\mathbf{h}}_k^{\mathsf{H}}{\mathbf{h}}_{k'}\rightarrow0,\frac{1}{N_{\mathsf{BS}}}\lVert{\mathbf{h}}_k\rVert^2\rightarrow \frac{{\beta_{\mathsf{r}}}r_{\mathsf{r}}^2}{r_k^2}
\end{align}
almost surely as $N_{\mathsf{BS}}\rightarrow\infty$, which implies
\begin{align}
\frac{\lvert{\mathbf{h}}_k^{\mathsf{H}}{\mathbf{h}}_{k'}\rvert}{\lVert{\mathbf{h}}_k\rVert\lVert{\mathbf{h}}_{k'}\rVert}=
\frac{\lvert{\mathbf{h}}_k^{\mathsf{H}}{\mathbf{h}}_{k'}\rvert/N_{\mathsf{BS}}}{\lVert{\mathbf{h}}_k\rVert\lVert{\mathbf{h}}_{k'}\rVert/N_{\mathsf{BS}}}
\rightarrow0
\end{align}
almost surely as $N_{\mathsf{BS}}\rightarrow\infty$. This completes the proof.
\section{Proof of Theorem \ref{Theorem_Favorable_Propagation_LOS_NFC}}\label{Proof_Theorem_Favorable_Propagation_LOS_NFC}
Based on the concept of definite integral, $\rho_{k,k'}$ can be approximated as follows:
\begin{align}
\rho_{k,k'}&\approx\left\lvert\int_{-\frac{1}{2}}^{\frac{1}{2}}{\rm{e}}^{{\rm{j}}\delta_1x(1+\delta_2x)}{\rm{d}}x\right\rvert\\
&=\!\frac{\pi^{\frac{1}{2}}/2}{N_{\mathsf{BS}}\lvert \mathsf{a}\rvert^{\frac{1}{2}}}\!
\left\lvert{\rm{erf}}\!\left(\!\frac{{\rm e}^{{\rm{j}}\frac{\pi}{4}}(\delta_2\!-\!1)}{2\sqrt{{\lvert \mathsf{a}\rvert}/{\mathsf{b}^2}}}\!\right)
\!+\!{\rm{erf}}\!\left(\!\frac{{\rm e}^{{\rm{j}}\frac{\pi}{4}}(\delta_2\!+\!1)}{2\sqrt{{\lvert \mathsf{a}\rvert}/{\mathsf{b}^2}}}\!\right)\label{Integral_NFC_Corr11}
\right\rvert,
\end{align}
where $\rm{erf}(\cdot)$ is the error function, and where \eqref{Integral_NFC_Corr11} is due to \cite[Eq. (2.33.3)]{gradshteyn2014table}. Furthermore, based on \cite[Eq. (7.1.23)]{abramowitz1948handbook} and \cite{olver1997asymptotics}, we obtain
\begin{equation}
\lim\nolimits_{x\rightarrow\infty}{\rm{erf}}({\rm e}^{{\rm{j}}\frac{\pi}{4}}x)=-\lim\nolimits_{x\rightarrow-\infty}{\rm{erf}}({\rm e}^{{\rm{j}}\frac{\pi}{4}}x)=1.
\end{equation}
Letting $N_{\mathsf{BS}}\rightarrow\infty$ yields $\lim_{N_{\mathsf{BS}}\rightarrow\infty}\lvert\delta_2\rvert=\lim_{N_{\mathsf{BS}}\rightarrow\infty}\lvert\frac{\mathsf{a}}{\mathsf{b}}N_{\mathsf{BS}}\rvert=\infty$, and it follows that
\begin{equation}
\lim_{N_{\mathsf{BS}}\rightarrow\infty}\rho_{k,k'}\approx\lim_{N_{\mathsf{BS}}\rightarrow\infty}\frac{\pi^{\frac{1}{2}}}{N_{\mathsf{BS}}\lvert {\mathsf{a}}\rvert^{\frac{1}{2}}}=0.
\end{equation}
This completes the proof.
\section{Proof of Lemma \ref{Lemma_SU_NFC_Analog_exp2}}\label{Proof_Lemma_SU_NFC_Analog_exp2}
Given that ${\mathbf{F}}=[{\rm{e}}^{-{\rm{j}}\frac{2\pi}{\lambda}r_{-\widetilde{N}}},\ldots,{\rm{e}}^{-{\rm{j}}\frac{2\pi}{\lambda}r_{\widetilde{N}}}]$, the received SNR can be calculated as follows:
\begin{subequations}
\begin{align}
\gamma&=\frac{p{{\beta_{\mathsf{r}}}r_{\mathsf{r}}^2}}{N_{\mathsf{BS}}\sigma^2}\left\lvert\sum\nolimits_{i=-\tilde{N}}^{\tilde{N}}\frac{1}{r_{i}}\right\rvert^2\\
&=\frac{p\beta_{\mathsf{r}}}{N_{\mathsf{BS}}\sigma^2}\frac{r_{\mathsf{r}}^2}{d^2}\left\lvert\sum\nolimits_{i=-\tilde{N}}^{\tilde{N}}\frac{d/r}{\sqrt{1+(id/r)^2}}\right\rvert^2\\
&\approx\frac{p\beta_{\mathsf{r}}}{N_{\mathsf{BS}}\sigma^2}\frac{r_{\mathsf{r}}^2}{d^2}\left\lvert\int_{-\tilde{N}d/r}^{\tilde{N}d/r}\frac{1}{\sqrt{1+x^2}}{\rm{d}}x\right\rvert^2\label{SU_NFC_Analog_exp111}\\
&=\frac{p\beta_{\mathsf{r}}}{N_{\mathsf{BS}}\sigma^2}\frac{r_{\mathsf{r}}^2}{d^2}2\ln\left({{\tilde{N}}d}/{r}+\sqrt{\left({{\tilde{N}}d}/{r}\right)^2+1}\right),\label{SU_NFC_Analog_exp222}
\end{align}
\end{subequations}
where the approximation in \eqref{SU_NFC_Analog_exp111} arises from $\frac{d}{r}\ll 1$ and the concept of definite integral, and \eqref{SU_NFC_Analog_exp222} follows from $\int\frac{{\rm{d}}x}{\sqrt{1+x^2}}=\ln(x+\sqrt{x^2+1})$. This completes the proof.
\end{appendices}

\bibliographystyle{IEEEtran}
\bibliography{ref}
\end{document}